\documentclass[12pt]{article}

\usepackage{graphicx}
\usepackage{graphics}
\usepackage{amssymb}
\usepackage{amsmath}

\usepackage{epsf,amsfonts,hyperref}

\renewcommand{\appendix}[1]{
    \addtocounter{section}{1}
    \setcounter{equation}{0}
    \renewcommand{\thesection}{\Alph{section}}
    \section*{Appendix \thesection\protect\indent #1}
    \addcontentsline{toc}{section}{Appendix \thesection\ \ \ #1}
}
\newcommand\encadremath[1]{\vbox{\hrule\hbox{\vrule\kern8pt 
\vbox{\kern8pt \hbox{$\displaystyle #1$}\kern8pt} 
\kern8pt\vrule}\hrule}}
\def\enca#1{\vbox{\hrule\hbox{
\vrule\kern8pt\vbox{\kern8pt \hbox{$\displaystyle #1$}
\kern8pt} \kern8pt\vrule}\hrule}}

\newcommand\figureframex[3]{
\begin{figure}[bth]
\hrule\hbox{\vrule\kern8pt 
\vbox{\kern8pt \vbox{
\begin{center}
{\mbox{\epsfxsize=#1.truecm\epsfbox{#2}}}
\end{center}
\caption{#3}
}\kern8pt} 
\kern8pt\vrule}\hrule
\end{figure}
}
\newcommand\figureframey[3]{
\begin{figure}[bth]
\hrule\hbox{\vrule\kern8pt 
\vbox{\kern8pt \vbox{
\begin{center}
{\mbox{\epsfysize=#1.truecm\epsfbox{#2}}}
\end{center}
\caption{#3}
}\kern8pt} 
\kern8pt\vrule}\hrule
\end{figure}
}

\makeatletter
\@addtoreset{equation}{section}
\makeatother
\newtheorem{theorem}{Theorem}[section]
\newtheorem{conjecture}{Conjecture}[section]
\newtheorem{remark}{Remark}[section]
\newtheorem{proposition}{Proposition}[section]
\newtheorem{lemma}{Lemma}[section]
\newtheorem{corollary}{Corollary}[section]
\newtheorem{definition}{Definition}[section]
\def\br{\begin{remark}\rm\small}
\def\er{\end{remark}}
\def\bt{\begin{theorem}}
\def\et{\end{theorem}}
\def\bd{\begin{definition}}
\def\ed{\end{definition}}
\def\bp{\begin{proposition}}
\def\ep{\end{proposition}}
\def\bl{\begin{lemma}}
\def\el{\end{lemma}}
\def\bc{\begin{corollary}}
\def\ec{\end{corollary}}
\def\beaq{\begin{eqnarray}}
\def\eeaq{\end{eqnarray}}
\newcommand{\proof}[1]{{\noindent \bf proof:}\par
{#1} $\square$}

\newcommand{\eq}[1]{eq.~(\ref{#1})}

\newcommand{\beq}{\begin{equation}}
\newcommand{\eeq}{\end{equation}}
\newcommand{\bea}{\begin{eqnarray}}
\newcommand{\eea}{\end{eqnarray}}

\renewcommand{\and}{{\qquad {\rm and} \qquad}}

\newcommand{\virg}{{\qquad , \qquad}}

\newcommand{\Res}{\mathop{\,\rm Res\,}}

\newcommand{\td}[1]{{\tilde{#1}}}

\newcommand{\om}{\omega}

\newcommand{\ii}{{\mathrm{i}}}

\newcommand{\ee}[1]{{{\rm e}^{#1}}}

\renewcommand{\d}{{{\partial}}}

\newcommand{\Pint}{{\int\kern -1.em -\kern-.25em}}

\renewcommand{\Re}{{\mathrm{Re}}}

\newcommand{\ovl}{\overline}
\newcommand{\genus}{{\mathfrak g}}

\newcommand{\acycle}{{\cal A}}
\newcommand{\bcycle}{{\cal B}}

\newcommand\spcurve{{\cal S}}
\newcommand\curve{{\cal C}}

\newcommand\arond{{\stackrel{\circ}{a}}}
\newcommand\brond{{\stackrel{\circ}{b}}}

\newcommand\Brond{{\stackrel{\circ}{B}}}

\newcommand\grond{{\stackrel{\circ}{g}}}

\newcommand\Phirond{{\stackrel{\circ}{\Phi}}}

\newcommand\xrond{{\stackrel{\circ}{x}}}
\newcommand\yrond{{\stackrel{\circ}{y}}}
\newcommand\Xrond{{\stackrel{\circ}{X}}}
\newcommand\Yrond{{\stackrel{\circ}{Y}}}
\newcommand\xirond{{\stackrel{\circ}{\xi}}}

\newcommand\spcurverond{{\stackrel{\circ}{\spcurve}}}

\newcommand\fram{{\mathfrak f}}

\newcommand\CL{{{\hat \Lambda}}}
\newcommand\modsp{{\cal M}}

\newcommand\Ber{{{\cal B}}}

\newcommand\bpt{{\mathfrak b}}

\newcommand\CYX{{\mathfrak X}}

\begin{document}
\sloppy


\pagestyle{empty}

\hfill IPHT-T12/030
\addtolength{\baselineskip}{0.20\baselineskip}
\begin{center}
\vspace{26pt}
{\large \bf {Computation of open Gromov-Witten invariants for toric Calabi-Yau 3-folds by topological
recursion, a
proof of the BKMP conjecture}}
\newline
\vspace{26pt}

{\sl B.\ Eynard}\hspace*{0.05cm}\footnote{ E-mail: bertrand.eynard@cea.fr },
{\sl N.\ Orantin}\hspace*{0.05cm}\footnote{ E-mail: norantin@math.ist.utl.pt }\\
\vspace{6pt}
${}^1$ Institut de Physique Th\'{e}orique, IPHT CEA Saclay,\\
F-91191 Gif-sur-Yvette Cedex, France.\\

${}^2$ CAMGSD, Departamento de Matem\'atica,
Instituto Superior T\'ecnico,
Av. Rovisco Pais,
1049-001 Lisboa, Portugal.
\end{center}

\vspace{20pt}
\begin{center}
{\bf Abstract}

The BKMP conjecture (2006-2008), proposed a new method to compute closed and open Gromov-Witten invariants for every toric Calabi-Yau 3-folds, through a topological recursion based on mirror symmetry.
So far, this conjecture had been verified to low genus for several toric CY3folds, and proved to all genus only for $\mathbb C^3$.

In this article we  prove the general case. Our proof is based on the fact that both sides of the conjecture can be naturally written in terms of combinatorial sums of weighted graphs: on the A-model side this is the localization formula, and on the B-model side the graphs encode the recursive algorithm of the topological recursion. One can slightly reorganize the set of graphs obtained in the B-side, so that it coincides with the one obtained by localization in the A-model.Then it suffices to compare the weights of vertices and edges of graphs on each side, which is done in 2 steps: the weights coincide in the large radius limit, due to the fact that the toric graph is the tropical limit of the mirror curve. Then the derivatives with respect to K\"ahler radius coincide due to special geometry property implied by the topological recursion.

\end{center}
%





\vspace{26pt}
\pagestyle{plain}
\setcounter{page}{1}


\section{Introduction}

Topological string theories have raised a lot of interest, because they represent a limit of string theory which is mathematically well defined and where computations can be entirely performed. 
Among the two possible types of topological string theories, the topological A-model string theory is mathematically formulated as Gromov--Witten theory.

For applications in physics, one is often concerned by topological string theories in target spaces which are Calabi--Yau 3-folds.
Not so many examples of Calabi-Yau spaces are known explicitly, but there is a family which is particularly well understood, this is the family of "toric" Calabi--Yau 3-folds. These are particularly well studied thanks to their toric symmetry, which allows to go even deeper in the computations. In addition, these theory lie at the crossroad of many interesting objects studied both in mathematics and physics, some of which we remind now.

\smallskip

There are mainly two types of topological string theories: A-model and B-model, and it was conjectured (and proved in some cases) that the A-model and B-model are dual to each other, through {\bf mirror symmetry} which exchanges the complex and Kh\"ahler structures of the target spaces.
For both A and B theories, the "string amplitudes" enumerate, in some appropriate way, some maps from a Riemann surface of given topology, into the target space (the Calabi-Yau 3-fold $\CYX$ or its mirror $\hat\CYX$). The string amplitudes for the A-model are called "Gromov-Witten" invariants and are well defined and extensively studied mathematical objects.

So, string amplitudes depend on a target space and on the topology (genus and possibly number of boundaries) of a Riemann surface. Closed amplitudes $F_g(\CYX)$ enumerate surfaces without boundaries, they depend only on a genus $g$.
They are encoded into generating functions by making formal series:
$$
F(\CYX,g_s) = \sum_{g={\rm genus}} g_s^{2g-2}\,F_g(\CYX)
$$
where $\CYX$ is our target space (the toric CY 3-folds) and where the formal parameter $g_s$ is traditionally called the "string coupling constant".
The goal is eventually to compute $F_g=F_g(\CYX)$, i.e. the amplitudes corresponding to the enumeration of Riemann surfaces of given genus. For example $F_0(\CYX)$ computes the planar amplitudes, i.e. rational maps from $\mathbb P^1$ into $\CYX$.

\smallskip

{\bf Topological vertex.}
In principle, topological string amplitudes for toric CY 3-folds are entirely known, through the "topological vertex" method \cite{Aganagic2004, LiJun2004, maulik-2008, maulik-2003, maulik-2004}. In that method, one introduces 
$$
q=\ee{-g_s}
$$
and string amplitudes are given in terms of a series in powers of $q$:
$$
Z = \ee{F(g_s)} = \ee{{\displaystyle \sum_g} g_s^{2g-2}\,F_g} = \sum_k q^k C_k
$$
where the coefficients $C_k$ are of combinatoric nature (typically they enumerate 2d or 3d partitions, this is often called a "crystal model"), they are known rather explicitly, or at least they are given by explicit combinatoric sums over partitions.

The problem is to extract from this $q$-series the asymptotic behavior and expansion near $q=1$ in powers of $g_s=-\ln q$. Indeed, the coefficients of this expansion are the ones of physical interest for precision computations in high energy physics whereas the $q$-series is defined as an expansion near $q=0$.
Even computing the leading order i.e. $F_0$, requires an infinite combinatoric sum. This makes its computation not straightforward, and going beyond leading order directly from the combinatorics sum is a very difficult challenge.

\medskip

{\bf Methods for computing Gromov--Witten invariants of fixed genus.}
Many methods have been introduced to compute directly the $g_s$ expansion, most of them are based on solving a differential equation:

- one is the famous "{\bf holomorphic anomaly equations}" \cite{Bershadsky:1993cx, aganagic-2006}. It is based on the observation that topological string amplitudes should be "modular" invariant. This implies a relationship between their holomorphic and  anti-holomorphic parts with respect to the parameters of the target space, which can be translated into a set of EDP satisfied by the string amplitudes. A drawback of the method is that one can compute the amplitudes only up to an unknown holomorphic function, which can be fixed by knowing the answer in some limiting cases. When applicable, the holomorphic anomaly equation method is extremely efficient for actual computations of Gromov-Witten invariants.

- Another is {\bf Givental's} method \cite{givental-2001, givental_sb_2003}. This method translates some geometric relations (like gluing surfaces) into a set of EDP. Those EDP can be formally solved, and the solution can be written as a linear operator acting on a product of Kontsevich integrals (depending on an infinite number of times), i.e. one has to compute derivatives of Kontsevich integrals and at the end set the times to some special values.
This method shows that the generating function of Gromov-Witten invariants is a Tau-function for some integrable hierarchy.

\smallskip

In both methods one has to find the Gromov-Witten invariants of a whole family of Calabi-Yau 3-folds, one can't find the invariants of one manifold directly. 

\smallskip
{\bf The remodeling method}

- In 2006 \cite{Mar1} M. Mari\~no suggested a new method, and then with Bouchard, Klemm and Pasquetti further formalized the statement under the name "{\bf remodeling the B--model}" in 2008 \cite{BKMP}. That method is based on the "topological recursion" of \cite{Eynard2004, EOFg} and on mirror symmetry (this will be described in more details in section \ref{secBmodel} below). This method proposes to compute the amplitudes by recursion on the Euler characteristics, without having to solve a differential equation, and in particular allows to compute the amplitudes of one given manifold without having to study a family of manifolds. Also, they give a recipe to compute "open Gromov-Witten invariants", as well as invariants for orbifold geometries, which were not known before. 
This claim of \cite{BKMP} is often called the {\bf "BKMP conjecture"}, and can be seen as an explicit example of mirror symmetry to all genus. We write it explicitly as conjecture \ref{BKMPconj} in section \ref{secBKMPconj}.

The authors of \cite{BKMP} and many others afterwards checked this conjecture for many examples of target space manifolds, and for low genus Gromov--Witten invariants, but the statement was so far proved to all genus only for the simplest toric Calabi--Yau 3-fold, namely $\CYX=\mathbb C^3$ \cite{ChenLin2009,ZhouJian2009}.

It was also noticed that the topological recursion implies the holomorphic anomaly equations \cite{Eynard2007} as well as some properties very similar to Givental's formalism \cite{OrantinN.2008}, but the converse has not been proved. 

In \cite{ZhouJian2009}, the proof for $\CYX=\mathbb C^3$ was mostly combinatorical, and used the "cut and join" equations of Goulden--Jackson \cite{Goulden2000}.
Unfortunately the tools involved in the combinatorics were very specific to the $\mathbb C^3$ geometry, and not easy to generalize, and has prevented the authors of \cite{ChenLin2009, ZhouJian2009} to extend their proof to other toric Calabi-Yau spaces.

A matrix model's heuristic argument was also presented, using the Chern-Simons matrix model \cite{Mar1}, and using a new matrix model reproducing all toric Calabi-Yau 3-folds \cite{EKM1}, but the saddle point analysis of the matrix model \cite{EKM2} was not proved with mathematical rigor.

\medskip

The goal of the present article is to present a {\bf general rigorous proof} for every toric Calabi-Yau 3-fold, mostly combinatorial although not based on cut and join, but more based on special properties of the topological recursion as well as localization.

\medskip

As a guideline for reading and understanding the proof of the BKMP conjecture, here is a short summary with references to the different steps carried out in the present article.

{\bf Sketch of the proof:}

- on the A-model side, it is known \cite{LiJun2004, Diaco} how to write the Gromov-Witten invariants through a {\bf localization} formula, as a combinatorial sum over {\bf graphs}, with weights associated to edges and vertices. Vertices are labeled by a "genus" and valency $(g,n)$.
Weights of vertices are Gromov-Witten invariants of $\mathbb C^3$ (i.e. the topological vertex) and are given by the Mari\~no--Vafa formula, i.e. they are triple Hodge integrals in $\overline\modsp_{g,n}$. We remind this procedure in theorem \ref{thlocclosedGW} and section \ref{secopenGW}. 

- on the B-model side, the topological recursion can be naturally written as a combinatorial sum over {\bf graphs}, with weights associated to edges and vertices (but not the same graphs and weights as the A-model side). Vertices are also labeled by a genus and valency $(g,n)$. 
Weights of vertices are combinations of residues of meromorphic forms computed at the branchpoints of the mirror curve. We explain it in section \ref{omgngraphs} by recalling some previous results in theorems \ref{WgnCL1bp} and \ref{theoremWngclasses}. We then specialize this result to the spectral curve obtained by mirror symmetry from a toric Calabi-Yau 3-fold in theorem \ref{thWngtdHtdF}.

- contrarily to the A-model side, the graphs of the B-model side have no $(0,1)$ or $(0,2)$ vertices (genus zero, valency 1 or 2). There is a standard graph combinatorial toolkit which allows to relate sums of graphs with or without vertices of valency $1$ or $2$. In other words we can add $(0,1)$ and $(0,2)$ vertices to the B-model sum of graphs, at the price of changing ("renormalizing") the weights of edges and vertices. This is done in section \ref{secrenormalize} through a few intermediate steps.

- after this graph manipulation, so that we have the same graphs on both sides, it remains to check whether the weights of edges and vertices in the A-model and in the B-model coincide.
This is done in two steps:

- In the large radius limit, where all K\"ahler parameters are large $t_j\to +\infty$, i.e. the {\bf tropical limit}, the fact that the tropical limit of the spectral curve is the toric graph of the A-model side, implies that the weights coincide at all $t_j=\infty$. This is shown in connection with the Mari\~no-Vafa formula in theorem \ref{thWngvertex}.

- For finite K\"ahler radius $t_j$, thanks to the topological recursion, the weights of the B-model side satisfy the "{\bf special geometry} property" (similar to Seiberg--Witten relations), i.e. a differential equation with $\d/\d t_j$, which allows to compute their derivatives with respect to $t_j$, and thus show that the weights of the A-model and B-model side, coincide for all $t_j$'s (large enough). This procedure is carried out in lemma \ref{lemmarenorm01}.

Combining all these results together concludes the proof of the BKMP conjecture in the last section of the present article.

\bigskip

This article is organised as follows:
\begin{itemize}
\item Section 2 is a reminder on Toric Calabi-Yau 3-folds geometry and mirror symmetry.

\item Section 3 is devoted to the description of the A-model side reminding the topological vertex formalism through localization.

\item In section 4, we present the BKMP conjecture and prove it by reorganising the graphs involved by the topological recursion into a set of graphs matching the ones rising for the localisation analysis of section 3.

\item Section 5 is a conclusion.

\item The first two appendices are reminder of the topological recursion formalism and its relation to intersection numbers. The other ones present some of the technical proofs of theorems requested for proving the BKMP conjecture.

\end{itemize}

\section{Reminder: geometry of Toric Calabi--Yau 3-folds}

The geometry of toric Calabi-Yau 3-folds is well known and described in many works and review articles \cite{mirrorbook,BouchardToric,Fulton93,LiJun2004, vonk-2005}, it is mostly combinatorial.
The goal of this section is to give a brief description of the geometry and combinatorics of those spaces and introduce notations, which will be useful for the rest of the article. Since these are classical results of toric geometry, we only provide sketches of proofs here and we refer the reader to the aforementioned literature for more details.

\subsection{Construction of toric Calabi--Yau 3-folds\label{secdefXtoric}}

Every toric Calabi-Yau 3-fold $\CYX$ can be constructed as follows.
Let $r$ be a non negative integer, and let $\vec q_1,\dots,\vec q_{r}$ be independent integer vectors called charges $\vec q_i=(q_{i,1},\dots,q_{i,r+3})\in \mathbb Z^{r+3}$ for $i =1,\dots,r$ such that
\beq
\forall\,i=1,\dots,r \quad , \quad \sum_{j=1}^{r+3} q_{i,j} = 0.
\eeq
In addition, these charges will be asked to fulfill additional condtions called
"smoothness condition" which are described later in section \ref{secsmoothness}.

Let $t_1,\dots,t_{r}$ be $r$ positive real numbers $t_i>0$, called K\"ahler parameters, or radii
and
$X_1,\dots, X_{r+3}$ be the  cannonical coordinates  of $\mathbb C^{r+3}$. We write
\beq
X_i = |X_i|\,\ee{i\theta_i} .
\eeq

\smallskip
\bd\label{defCYXtoricrelthetaequiv}
For charges $\left\{\vec q_i \right\}_{i=1}^r$, K\"ahler parameters $\left\{t_i\right\}_{i=1}^r$ and $\left\{X_i\right\}_{i=1}^{r+3}$ as above,
one defines the 6-dimensional real manifold $\CYX$ as follows.
$\CYX$ is the submanifold of $\mathbb C^{r+3}$ defined by the $r$ relations
\beq
\forall\,i=1,\dots,r\, \qquad , \quad \sum_{j=1}^{r+3} q_{i,j}\,|X_j|^2 = t_i,
\eeq
and quotiented by the equivalence relations:
\beq
\forall\, i=1,\dots,r \,,\, \forall \alpha\in \mathbb R \qquad , \quad 
(\theta_{1},\dots,\theta_{r+3}) \equiv  (\theta_{1},\dots,\theta_{r+3})+\alpha\,(q_{i,1},\dots,q_{i,r+3}).
\eeq
\ed

Under a suitable choice of charges $\vec q_i$'s, $\CYX$ is a smooth 6 dimensional manifold. It turns out that $\CYX$ has a complex structure inherited from that of $\mathbb C^{r+3}$ and it has the Calabi-Yau property which is equivalent to
${\displaystyle \sum_{j=1}^{r+3}} q_{i,j} = 0$. 
One can check that the following symplectic form:
\beq
\omega = \frac{1}{2}\,\sum_{i=1}^{r+3} d|X_i|^2\,\wedge\,d\theta_i
\eeq
is a well defined symplectic form on $\CYX$ (it is the reduction of the cannonical symplectic form on $\mathbb C^{r+3}$, and it descends to the equivalence classes).

Let us consider 2 examples which will often illustrate our general method:

$\bullet$ The resolved connifold is defined with $r=1$ and $q=(1,1,-1,-1)$, i.e.
\beq
|X_1|^2+|X_2|^2-|X_3|^2-|X_4|^2=t
\virg
(\theta_1,\theta_2,\theta_3,\theta_4) \equiv (\theta_1+\alpha,\theta_2+\alpha,\theta_3-\alpha,\theta_4-\alpha).
\eeq

$\bullet$ The local $\mathbb P^2$ is defined with $r=1$ and $q=(1,1,-3,1)$, i.e.
\beq
|X_1|^2+|X_2|^2-3|X_3|^2+|X_4|^2=t
\virg
(\theta_1,\theta_2,\theta_3,\theta_4) \equiv (\theta_1+\alpha,\theta_2+\alpha,\theta_3-3\alpha,\theta_4+\alpha).
\eeq


\subsubsection{The toric graph}

\begin{figure}[t]
\centering
$$\includegraphics[height=4.5cm]{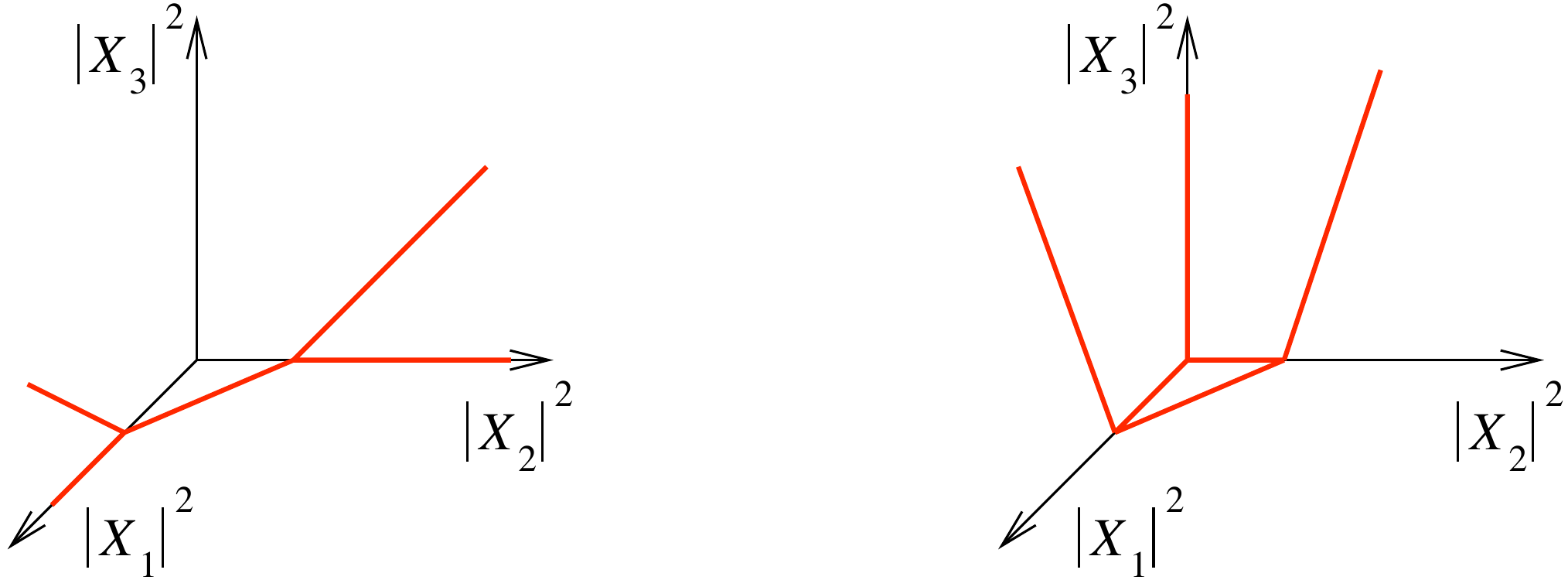}  $$
\caption{The polyhedra of  the resolved connifold $q=(1,1,-1,-1)$ and of local $\mathbb P^2$ $q=(1,1,-3,1)$.
\label{figtoricgraphex1}}
\end{figure}

Let $\pi$ be the moment map
\beq
\begin{array}{cccl}
\pi: & \CYX & \to & \mathbb R^3  \cr
& & \mapsto & (|X_1|^2, |X_2|^2,|X_3|^2)
\end{array} .
\eeq
The image $\pi(\CYX)$ is a convex polyhedral subdomain of $\mathbb R^3$.
Its faces are given by $|X_i|^2=0$ for some $i=1,\dots,3+r$. Its edges are the locci where a pair of $|X_i|^2$ vanish and its vertices are reached when 3 of the $|X_i|^2$ vanish.
See fig.\ref{figtoricgraphex1} for our 2 examples.

\medskip

Let us define the three vectors
\beq
\alpha= (1,0,0,\alpha_4,\dots,\alpha_{r+3}) 
\,\, , \,\,\,
\beta= (0,1,0,\beta_4,\dots,\beta_{r+3})
\,\, , \,\,\,
\gamma= (0,0,1,\gamma_4,\dots,\gamma_{r+3})
\eeq
satisfying
\beq
\sum_{j=1}^{r+3} q_{i,j}\alpha_j=0
\virg
\sum_{j=1}^{r+3} q_{i,j}\beta_j=0
\virg
\sum_{j=1}^{r+3} q_{i,j}\gamma_j=0 
\eeq
for all $i=1,\dots,r$.
Notice that this definition implies that
\beq
\forall\, j=1,\dots, r+3 \quad , \quad \alpha_j+\beta_j+\gamma_j=1.
\eeq
Thus, the 3-dimensional vector $(\alpha_j,\beta_j,\gamma_j)$ is the normal to the $j^{\rm th}$ face define as the image of the set $\left\{ \left. \vec X \right| |X_j|^2=0\right\}$.

We define the graph in $\mathbb R^3$ whose $r+3$ vertices are the points
\beq
(\alpha_j,\beta_j,\gamma_j)
\eeq
and we draw an edge between $(\alpha_i,\beta_i,\gamma_i)$ and $(\alpha_j,\beta_j,\gamma_j)$ iff the face $|X_i|^2=0$ and $|X_j|^2=0$ have a common edge.

This graph is in $\mathbb R^3$, but since it lies on the hyperplane $\alpha+\beta+\gamma=1$ (thanks to the Calabi-Yau condition ${\displaystyle \sum_{j=1}^{r+3}} q_{i,j}=0$), we can actually view it as a graph in $\mathbb R^2$.

We thus define:

\bd\label{defdualgraph}
The dual toric graph $\hat{\Upsilon}_{\CYX}$ of $\CYX$, is the graph in $\mathbb R^2$ whose $r+3$ vertices are the points
\beq
v_j = (\alpha_j,\beta_j)
\eeq
such that two vertices $v_i$ and $v_j$ are linked by an edge if and only if the faces $|X_i|^2=0$ and $|X_j|^2=0$ have a common edge.
$\hat{\Upsilon}_{\CYX}$ is thus the graph whose vertices are the normal vectors to the faces of the polyhedra of $\CYX$.

\ed

\medskip

\bd\label{deftoricgraph}
The toric graph ${\Upsilon}_{\CYX}$ of $\CYX$, is the dual of $\hat{\Upsilon}_{\CYX}$, i.e. a graph whose edges are orthogonal to those of $\hat{\Upsilon}_{\CYX}$, and such that the length of the $\alpha$--projection of the compact edges are equal to the $|X_1|^2$ projection of the corresponding edges in the polyhedra. The lenghts of compact edges of the toric graph, are thus linear combinations of the $t_j$'s.

The coordinates of a vertex $\sigma$ of $\Upsilon_\CYX$ in the $\mathbb R^2$ plane are denoted:
\beq
\sigma = (\arond_\sigma,\brond_\sigma).
\eeq
Thus $\arond_\sigma-\arond_{\sigma'}$ (resp. $\brond_\sigma-\brond_{\sigma'}$) is a linear combination of $t_j$'s.

\ed

\begin{figure}[t]
\centering
$$\includegraphics[height=5.5cm]{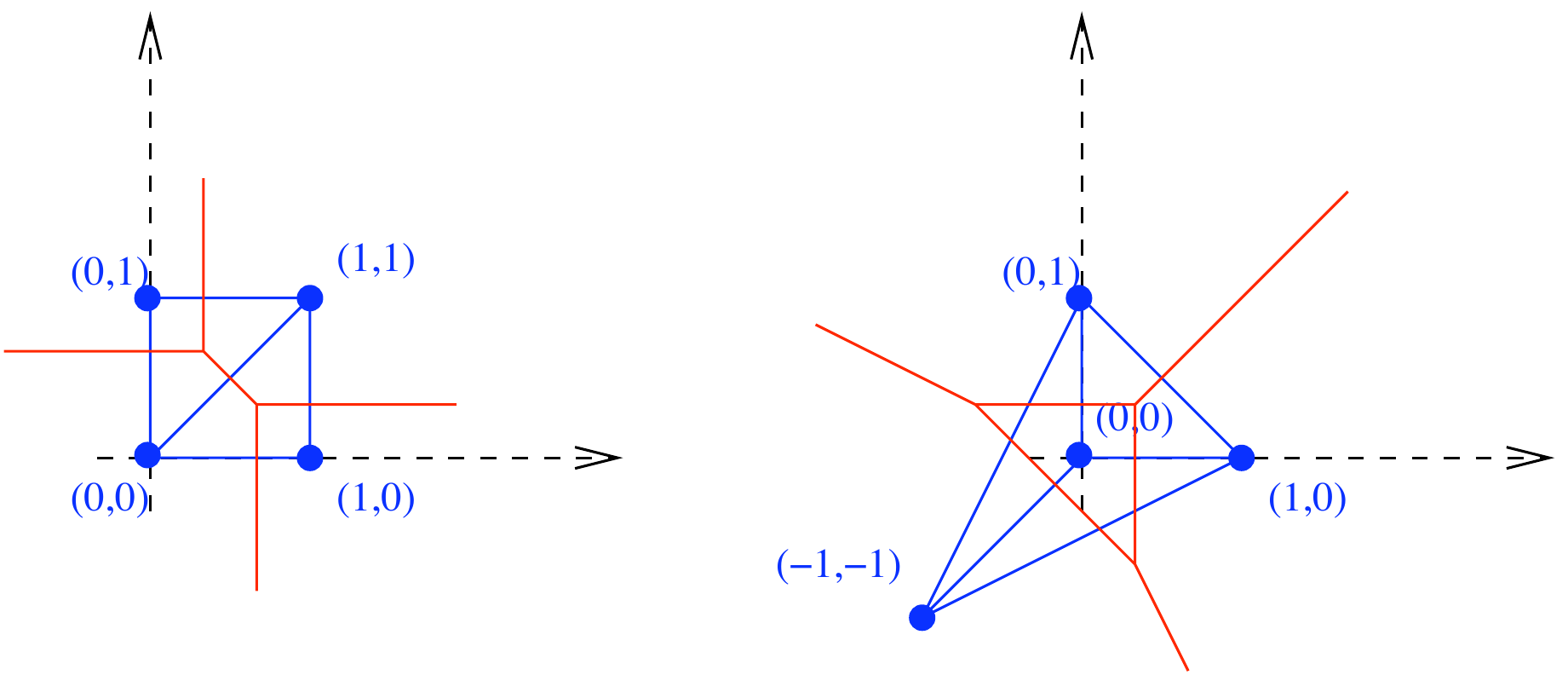}  $$
\caption{The toric graphs (red) and their duals (blue) for the resolved connifold $q=(1,1,-1,-1)$ and of local $\mathbb P^2$ $q=(1,1,-3,1)$. All the blue triangles have area $1/2$.
\label{figtoricgraphex2}}
\end{figure}

See fig.\ref{figtoricgraphex2} for examples.

\subsubsection{Smoothness condition}\label{secsmoothness}

The polyhedra $\pi(\CYX)$ is not smooth at its vertices. Thus  $\CYX$ might not be smooth at these vertices.
It can be shown that the invertibility of the matrix $[q_{i,j}]_{i=1,\dots,r;\, j=1,\dots,r+3, j\neq i_1,i_2,i_3}$ with integer coefficients, i.e. that:
\beq\label{defsmoothness}
\forall (i_1,i_2,i_3)={\rm vertex}
\virg
\det\left(q_{i,j}\right)_{i=1,\dots,r;\, j=1,\dots,r+3, j\neq i_1,i_2,i_3} = \pm 1
\eeq
ensures the smoothness of $\CYX$ near the vertex $|X_{i_1}|^2=|X_{i_2}|^2=|X_{i_3}|^2=0$. We refer to it as the smoothness condition of $\CYX$ near $(i_1,i_2,i_3)$.

Near $(i_1,i_2,i_3)$, $\CYX$ is locally $\mathbb C^3$, and the condition above is related to the fact that we can define the 3 canonical angles of $\mathbb C^3$, and the angles must have periodicities $2\pi$ (so we need integer coefficients).
Once again, we refer the reader to the literature cited at the beginning of this section for further insights on this topic.

\medskip

In terms of the toric graph, a vertex of the polyhedra is a triangular face of the dual graph $\hat\Upsilon_\CYX$, and the condition \ref{defsmoothness} is equivalent to saying that:
\beq\label{defsmoothness2}
{\rm area\, of\, triangle}\,\,(i_1,i_2,i_3)\,\,\, = \, \frac{1}{2}
\eeq
i.e.
\beq
\alpha_{i_1}\beta_{i_2}-\alpha_{i_2}\beta_{i_1}
+\alpha_{i_2}\beta_{i_3}-\alpha_{i_3}\beta_{i_2}
+\alpha_{i_3}\beta_{i_1}-\alpha_{i_1}\beta_{i_3}
=\pm 1.
\eeq
If we relabel the points $i_1,i_2,i_3$ so that the triangle $(i_1,i_2,i_3)$ has trigonometric orientation in $\mathbb R^2$, then the expression above is $+1$:
\beq
\alpha_{i_1}\beta_{i_2}-\alpha_{i_2}\beta_{i_1}
+\alpha_{i_2}\beta_{i_3}-\alpha_{i_3}\beta_{i_2}
+\alpha_{i_3}\beta_{i_1}-\alpha_{i_1}\beta_{i_3}
=+ 1.
\eeq

\bp\label{propsmoothness}
The dual toric graph $\hat\Upsilon_\CYX$ is a triangulated polygon with vertices in $\mathbb Z^2$, made of triangles of area $1/2$. 
\ep

\subsubsection{Local framings}

The previous smoothness condition can be readily rewritten as a determinant, by introducing the following matrix:

\bd\label{defframsigma}
To every vertex $\sigma$ of $\Upsilon_\CYX$, ( i.e. to a positively oriented triangle $(i_1,i_2,i_3)$ of the dual $\hat\Upsilon_\CYX$) we associate the $2\times 2$ matrix $\fram_\sigma$ (called local framing matrix at vertex $\sigma$):
\beq
\fram_\sigma = \left(\begin{array}{cc} \fram_{a,\sigma}  & \fram_{b,\sigma} \cr \fram_{c,\sigma} & \fram_{d,\sigma} \end{array}\right)
= \left(\begin{array}{cc} \beta_{i_3}-\beta_{i_1}  & \beta_{i_1}-\beta_{i_2} \cr \alpha_{i_1}-\alpha_{i_3} & \alpha_{i_2}-\alpha_{i_1} \end{array}\right)
\eeq
The smoothness condition is that
$\det \fram_\sigma = 1$,
i.e.
$
\fram_\sigma \in Sl_2(\mathbb Z).
$
\ed

Observe that a rotation of the triangle $(i_1,i_2,i_3)\to (i_2,i_3,i_1)$ amounts to
\beq
\fram_\sigma \to \fram_\sigma\,\times\, \left(\begin{array}{cc} 0 & -1 \cr 1 & -1 \end{array}\right),
\eeq
and a change of orientation of the triangle $(i_1,i_2,i_3)\to (i_2,i_1,i_3)$ amounts to
\beq
\fram_\sigma \to \fram_\sigma\,\times\, \left(\begin{array}{cc} 1 & 0 \cr 1 & -1 \end{array}\right).
\eeq

\bd\label{defframedges}

Let $e=(\sigma,\sigma')$ be an edge of $\Upsilon_\CYX$, we denote $\epsilon=(\sigma,e)$ (resp. $\epsilon'=(\sigma',e)$) the half edge of $e$ starting from $\sigma$ (resp. $\sigma'$), and let $\sigma$ be dual to the positively oriented triangle $(i_1,i_2,i_3)$ in $\hat\Upsilon_\CYX$ and $\sigma'$ be dual to the positively oriented triangle $(i_2,i_1,i_4)$.
We define the framing of the half-edge $(\sigma,e)$ as:
\beq
\fram_{\epsilon} = - \fram_{\epsilon'} = \beta_{i_1}-\beta_{i_2} = \fram_{b,\sigma} = -\fram_{b,\sigma'}.
\eeq

\ed

Notice that the framings of the 3 half-edges emanating from a vertex $\sigma$ are respectively:
\beq
\fram_{b,\sigma}\quad , \quad -\fram_{a,\sigma}-\fram_{b,\sigma} \quad , \quad \fram_{a,\sigma},
\eeq
and their sum is zero:
\beq
\forall\,\sigma={\rm vertex\,of\,}\Upsilon_\CYX\, ,  \qquad \sum_{\epsilon\,{\rm adjacent\,to\,}\sigma} f_\epsilon=0.
\eeq

\subsubsection{Lagrangian submanifolds}

For a toric Calabi-Yau 3-fold $\CYX$, we define a set of special Lagrangian submanifolds as follows:

\bd 
Consider a 1-dimensional affine subspace $V$ of $\mathbb R^3$, given by relations
\beq
V:\qquad \forall\,i=1,2 \,\, , \qquad \sum_{j=1}^{r+3} r_{i,j} |X_j|^2 = c_i
\eeq 
where $\{r_{i,j}\}_{i=1,\dots,2,\, j=1,\dots,r+3}$ are integers such that $\sum_j r_{i,j}=0$, and $c_1,c_2$ are two real numbers chosen such that $V$ intersects an edge of the polyhedra of $\CYX$.
Then define a special Lagrangian submanifold $L$, as the 3 dimensional submanifold of $\CYX$ given by the following relationships:
\beq
\forall\,i=1,2 \,\, , \qquad \sum_{j=1}^{r+3} r_{i,j} |X_j|^2 = c_i
\eeq
and the realtionships between $\theta_1,\dots,\theta_{r+3}$:
\beq
0 = \det\left(\begin{array}{ccc}
q_{1,1} & \dots & q_{1,r+3} \cr
\vdots &  & \vdots \cr
q_{r,1} & \dots & q_{r,r+3} \cr
r_{1,1} & \dots & r_{1,r+3} \cr
r_{2,1} & \dots & r_{2,r+3} \cr
\theta_1 & \dots & \theta_{r+3}
\end{array}\right) .
\eeq
This condition implies that $L$ is Lagrangian, i.e. the symplectic form $\om=\frac{1}{2}{\displaystyle \sum_{j=1}^{r+3}} d|X_j|^2\wedge d\theta_j=0$ vanishes on $L$.

\smallskip

In addition, we shall require that the plane orthogonal to $V$ in $\mathbb R^3$ be not parallel to any edge of the polyhedra of $\CYX$.
\ed

\begin{figure}[t]
\centering
$$\includegraphics[height=5.5cm]{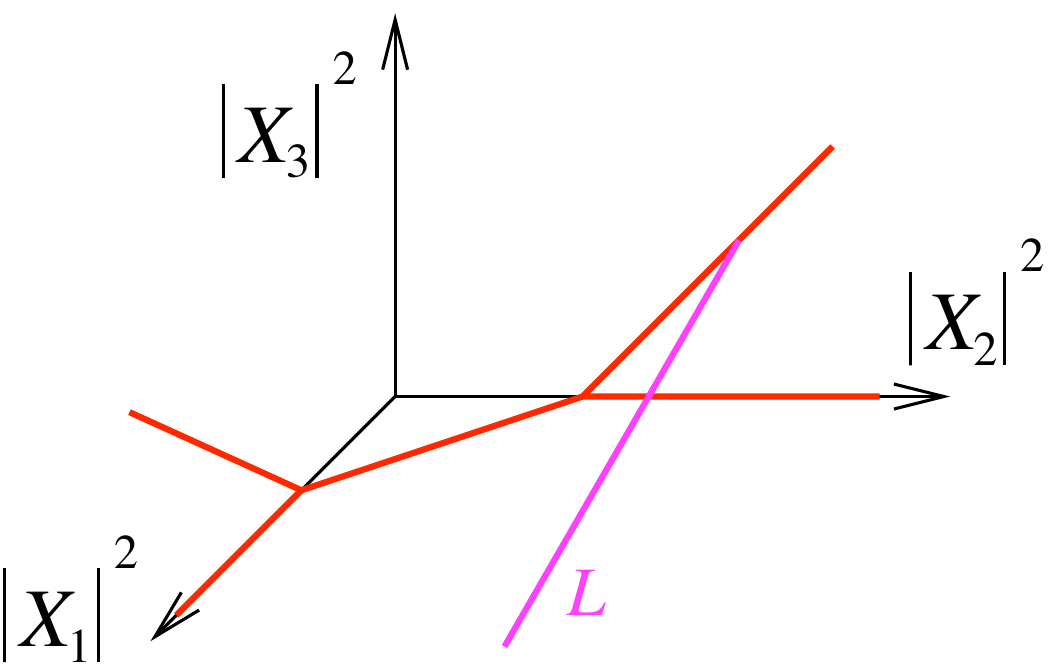}  $$
\caption{Example of a brane $L$ for $\CYX=$resolved connifold. $\pi(L)$ is a line $V\subset\mathbb R^3$, ending on an edge of the polyhedra of $\CYX$. 
\label{figdiagconnifoldbrane}}
\end{figure}

$L$ (in fact $V$) can be pictorially represented as a half--line on the polyhedra of $\CYX$, ending on an edge, or also $L$ can be represented as a line attached to an edge of the toric graph, see fig.\ref{figdiagconnifoldbrane}.

The fact that $L$ ends on an edge of the toric graph, implies that $L$ is topologically
\beq
L \sim \mathbb C\times S^1,
\eeq
and it has Betti number
\beq
b_1(L)=\dim H_1(L,\mathbb Z)=1
\virg
H_1(L,\mathbb Z)\sim \mathbb Z.
\eeq

From now on, we shall always assume that $L$ ends on a non--compact edge.

\subsection{The mirror\label{secmirror}}

For a set of $r$ complex parameters $\{\td t_i\}_{i=1}^r$, and a set of charges $\{\vec q_i\}_{i=1}^r$ as before, one defines the projective curve embedded in $ \mathbb P^2 = (Y_1,Y_2,Y_3)$ by the homogeneous degree 1 polynomial:
\beq
H(Y_1,Y_2,Y_3;\td t_1,\dots,\td t_r)=\sum_{i=1}^{r+3} Y_i =0
\eeq
where $Y_i=\ee{-x_i}$, and  
\beq\label{eqqxtdt}
\forall\,i=1,\dots,r \quad , \quad \sum_{j=1}^{r+3} q_{i,j} x_{j}=\td t_i.
\eeq
Thanks to the Calabi-Yau condition $\sum_j q_{i,j}=0$, it is homogeneous of degree 1 .
If we choose the patch $Y_3=1$, this defines an algebraic curve embedded into $\left(\mathbb {C^*}\right)^2$:
\beq
H(Y_1,Y_2,1;\td t_1,\dots,\td t_r)=0.
\eeq
\br
The polynomial $H(Y_1,Y_2,Y_3)$ depends on the complex parameters $\td t_i$ and charges $\vec q_i$ but, as long as there is no ambiguity, we omit to write down this dependance explicitly in the following.
\er

\medskip
Let us consider our two examples:

$\bullet$ For the resolved conifold $q=(1,1,-1,-1)$, we have $x_4=x_1+x_2-x_3-\td t$, i.e.
\beq
Y_4=\ee{\td t}\,\,\frac{Y_1\,Y_2}{Y_3}
\eeq
and thus
\beq
H(Y_1,Y_2,Y_3)= Y_1+Y_2+Y_3+\ee{\td t}\,\,\frac{Y_1\,Y_2}{Y_3}.
\eeq
The corresponding algebraic curve is:
\beq
Y_1+Y_2+1+\ee{\td t}\,Y_1 Y_2 = 0
\eeq
which is parameterized by a unique parameter $\td t$.

\medskip

$\bullet$ For local $\mathbb P^2$ $q=(1,1,-3,1)$, we have $x_4=3x_3-x_1-x_2-\td t$, i.e.
\beq
Y_4=\ee{\td t}\,\,\frac{Y_3^3}{Y_1\,Y_2}
\eeq
and thus
\beq
H(Y_1,Y_2,Y_3)= Y_1+Y_2+Y_3+\ee{\td t}\,\,\frac{Y_3^3}{Y_1\,Y_2}.
\eeq
The algebraic curve is:
\beq
Y_1+Y_2+1+\,\,\frac{\ee{\td t}}{Y_1\,Y_2} = 0.
\eeq

\medskip

\subsubsection{Newton's polygon}

Notice that the equation of this projective curve is always of the form:
\beq
H(Y_1,Y_2,Y_3) = \sum_{(\alpha,\beta)={\rm vertex\,of}\, \hat\Upsilon_\CYX}\, H_{\alpha,\beta}\,Y_1^{\alpha}\,Y_2^{\beta}\,Y_3^{1-\alpha-\beta}.
\eeq
In the patch $Y_3=1$, it gives the plane curve
\beq
H(Y_1,Y_2,1) = \sum_{(\alpha,\beta)={\rm vertex\,of}\, \hat\Upsilon_\CYX}\, H_{\alpha,\beta}\,Y_1^{\alpha}\,Y_2^{\beta}=0.
\eeq
Newton's polygon is defined as the set of points $(\alpha,\beta)\in\mathbb Z^2$ such that $H_{\alpha,\beta}\neq 0$.
Therefore:
\bp
The Newton's polygon of the plane curve $H(Y_1,Y_2,1)=0$ is the dual toric graph.
\ep

\subsubsection{Topology of the mirror curve\label{sectopologymirror}}

Let $\curve$ be the Riemann surface of polynomial equation $0=H(X,Y)=\sum_{i,j} H_{i,j} X^i Y^j$ with generic coefficients $H_{i,j}$.

It is a classical result in algebraic geometry, that the genus $\genus$ of $\curve$ is the number of integer points strictly contained in the convex envelope of the Newton's polygon.

Since the Newton's polygon is the dual toric graph $\hat\Upsilon_\CYX$ and since  $\hat\Upsilon_\CYX$ is triangulated with triangles of area $1/2$, each triangle contains no integer point in its interior. Integer points are only at vertices, and thus the number $\genus$ of integer points is the number of vertices of $\hat\Upsilon_\CYX$ which are strictly inside the polygon, i.e. in terms of the dual, this is the number of compact faces of $\Upsilon_\CYX$, that is to say the number of "loops" of $\Upsilon_\CYX$.
Since the number of faces of $\Upsilon_\CYX$ is $r+3$\footnote{Remember that each face corresponds to some $|X_i|^2=0$, $i=1,\dots,r+3$.}, the number of non--compact faces is $r+3-\genus$.
This means that:

\smallskip

\bp
The genus $\genus$ of the algebraic curve $\curve$ is the number of loops in the toric graph $\Upsilon_\CYX$. The number of punctures of $H(X,Y)=0$ is $r+3-\genus$.

$\hat\Upsilon_\CYX$ is a triangulated polygon in $\mathbb Z^2$, with $r+3$ vertices, $\genus+2r+3$ edges, $\genus+r+1$ triangular faces (each triangle having area $1/2$).

\smallskip

The toric graph $\Upsilon_\CYX$ is a planar trivalent graph with $2\genus+r$ compact edges, $r+3-\genus$ non--compact edges, $\genus+r+1$ vertices and $\genus$ compact faces.

\ep

\smallskip

For our two examples, this gives:

$\bullet$ The mirror of the resolved connifold has genus $\genus=0$. The toric graph has $r=1$, it has 5 edges, 1 compact and 4 non--compact, and it has 2 vertices, and no internal face. See fig.\ref{figtoricgraphex1}.

$\bullet$ The mirror of local $\mathbb P^2$ has genus $\genus=1$. Its toric graph has $r=1$, it has 3 vertices, 6 edges, 3 compact and 3 non--compact, and one internal face. See fig.\ref{figtoricgraphex1}.

\subsubsection{Branchpoints}

The branchpoints are the zeroes of the meromorphic differential form $dx=-dX/X$ on $\curve$.
The number of branchpoints is given by the Hurwitz formula:
\beq
  \#{\rm zeroes\,\,of\,\,}dx = 2\genus-2-\deg\,(dx) 
\eeq
where $-\deg\,(dx)$ is the number of poles of $dx$, i.e. the number of punctures, which is equal to $r+3-\genus$. This shows that the number of branchpoints is equal to the number of vertices of $\Upsilon_\CYX$:
\beq
  \#{\rm zeroes\,\,of\,\,}dx = 2\genus-2-\deg\,(dx) = \genus+r+1.
\eeq

To each vertex $\sigma$ of  $\Upsilon_\CYX$, we can associate a branchpoint $a_\sigma$. This labeling of branchpoints by vertices of the $\Upsilon_\CYX$ is made explicit in the next section.

\subsubsection{Framing}

For $f \in \mathbb{Z}$, we shall consider the plane curve $H_f(X,Y)=0$ defined by
\beq
H_f(X,Y) = H(X\,Y^{f}\,,Y,1)
\eeq
i.e. we have replaced $Y_1=X\,Y^f$ and $Y_2=Y$ and $Y_3=1$.
Such an integer $f\in \mathbb Z$ is called the "framing".

The framed curve has an equation of the form:
\beq
H_f(X,Y) = \sum_{(\alpha,\beta)={\rm vertex\,of}\, \hat\Upsilon_\CYX}\, H_{\alpha,\beta}\,X^{\alpha}\,Y^{\beta+f \alpha}.
\eeq

\begin{figure}[t]
\centering
$$\includegraphics[height=4.5cm]{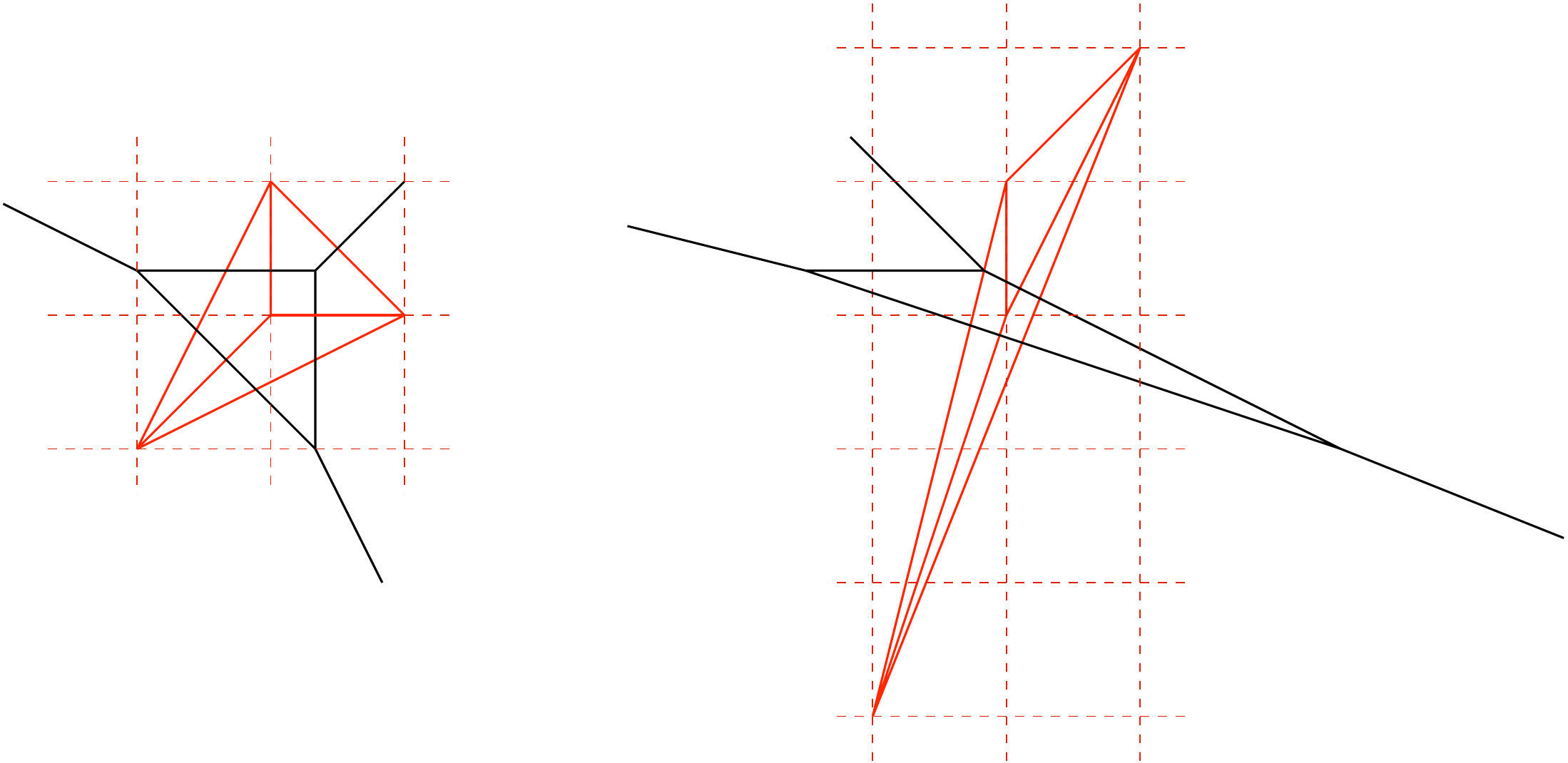}  $$
\caption{The toric graph of local $\mathbb P^2$ and its dual, and the same graph after a framing transformation $X\to X Y^f$ such that there is no more vertical edge (no horizontal edge in the dual). Here the framing is $f=2$.
\label{figdiaglocP2framed}}
\end{figure}

Its Newton's polygon is an affine transformation of the dual toric graph $\hat\Upsilon_\CYX$.
See fig.\ref{figdiaglocP2framed} for the example of local $\mathbb P^2$.

In all what follows we shall always assume that we have chosen a framing such that the framed toric graph has no vertical edge.

\subsection{Atlas of the mirror curve with cylinders and pants}

\subsubsection{Amoeba and tropical limit}

For describing the geometry of the mirror curve, it is very useful to introduce some basic results of tropical geometry.

First of all, given a sub-manifold of $\mathbb{C}^n$, one defines its Amoeba as follows:
\bd
For $n \in \mathbb{N}$ and a polynomial $P(Y_1,\dots,Y_n)$, we define the Amoeba of the submanifold ${\cal V} := \left\{(Y_1,\dots,Y_n) \in \mathbb{C}^n| P(Y_1,\dots,Y_n) = 0\right\}$ by
\beq
{\cal A}(P) := \left\{(\Re\,x_1,\dots,\Re\,x_n)  |\, P(e^{-x_1},\dots,e^{-x_n}) = 0\right\}.
\eeq
\ed

Another way to define this Amoeba is  the image of ${\cal V}$ under the so-called $Log$ map which we shall now define.

\bd
For $\lambda \in \mathbb{R}^+$, let us define the "Log" map
\beq
Log_\lambda \, : \, 
\begin{array}{rcl}
\left(\mathbb{C^*}\right)^n &\to& \mathbb{R}^n \cr
(Y_1,\dots,Y_n) &\to& \left(-\frac{\log \left|Y_1\right|}{ \log \lambda}, \dots,- \frac{\log \left|Y_n\right|}{\log \lambda}\right).
\end{array}
\eeq

\ed

With this definition, one sees that
\beq
{\cal A}(P) = Log_e\left({\cal V}\right),
\eeq
while changing the value of $\lambda$ amounts to applying a rescaling to the Amoeba.

In particular, the limit $\lambda \to \infty$ is known as the tropical limit, following a result of Mikhalkin \cite{Mikhalkin} and Rullg\aa rd \cite{Rullgard}:
\bt
When $\lambda\to +\infty$, the $\lambda$ rescaled Amoeba of $P(X,Y) = {\displaystyle \sum_{i,j}} \alpha_{i,j} \lambda^{-a_{i,j}} X^i Y^j $ converges to a tropical curve:
\beq
Log_\lambda\left({\cal V}\right) \to_{\lambda \to \infty} P_{\infty}(x,y)
\eeq
where the tropical curve 
\beq
P_{\infty}(X,Y):= "  \sum_{i,j} a_{i,j} X^i Y^j "
\eeq
is  defined by
\bea
"\sum_{i,j} a_{i,j} \, X^i \, Y^j"  &=& \left\{ (x_0,y_0) \in \mathbb{R}^2 | \exists (i,j) \neq (k,l) \, , \; \forall (m,n) \notin  \left\{(i,j),(k,l)\right\} \, , \;   \right. \cr
&&  \quad \left.   a_{i,j}+ i x_0 + j y_o = a_{k,l}+ k x_0 + l y_0 \leq a_{m,n}+ m x_0 + n y_0
\right\}. \cr
\eea

\et

A tropical curve is thus a union of straight segments in $\mathbb R^2$, forming a graph with trivalent vertices drawn in $\mathbb{R}^2$ whose faces are associated to the monomials defining polynomial, edges to pairs of such monomials and vertices to triple of them.
In particular, the $\lambda$ rescaled Amoeba of $P(x,y) = \sum_{i,j} \alpha_{i,j} \lambda^{-a_{i,j}} X^i Y^j$ converges in the tropical limit to a graph whose faces correspond to sectors where the log of one of the monomials $\alpha_{i,j} \lambda^{-a_{i,j}} X^i Y^j$ has a modulus larger than the other monomials. The edges and vertices of the limiting Amoeba are thus the locus where two, respectively three, of these monomial are of equal magnitude, and larger than the other ones. Let us remark that the pairs $(i,j) \in \mathbb{Z}^2$ for which $\alpha_{i,j}\neq0$ fix the slope of the possible edges of the tropical curve whereas the position of the vertices as well as the connectivity of the graph representing this tropical curve depend on the exponents $a_{i,j}$'s.

Let us apply the rescaling technic to the study of the geometry of the mirror curve.
 \bp
The tropical limit of the plane curve $H_f(X,Y)=0$ with complex parameters $\td t_j=\td T_j\,\log\lambda$, is the framed toric graph rescaled by $\log\lambda$ of $\CYX$ with K\"ahler parameters $t_j=\td t_j+O(1)$.
\ep
\proof{

We are interested in the large complex parameter limit  $\tilde{t}_k \to \infty$. For reaching this limit, let us define
\beq
\tilde{T}_k := \frac{\tilde{t}_k}{\log \lambda}
\eeq
where we assume $\td T_k=O(1)$ when $\lambda\to +\infty$.

Let us first remark that the coefficient of $X^\alpha Y^\beta$ in the polynomial $H_f(X,Y)$ reads
\beq
H_{\alpha,\beta-f \alpha} = e^{\tilde{t}_{\alpha,\beta}}
\eeq
where the times $\tilde{t}_{\alpha,\beta}$ are linear combinations of the complex parameters:
\beq
\tilde{t}_{\alpha,\beta} = \sum_{k} C_{\alpha,\beta;k} \tilde{t}_k.
\eeq
With these notations, the framed mirror curve can be written
\beq
H_f(X,Y) = \sum_{(\alpha,\beta-f\alpha)={\rm vertex\,of}\, \hat\Upsilon_\CYX}\, \lambda^{{\displaystyle \sum_k} C_{\alpha,\beta;k} \tilde T_k} \,X^{\alpha}\,Y^{\beta}.
\eeq

We can now study the tropical limit of the plane curve $H_f(X,Y)=0$. When $\lambda \to \infty$, its rescaled Amoeba converges to
\beq
H_f^{[\infty]}(X,Y) = " \sum_{(\alpha,\beta-f\alpha)={\rm vertex\,of}\, \hat\Upsilon_\CYX}\,  \left(-{\displaystyle \sum_k} C_{\alpha,\beta ;k} \tilde T_k\right)  \,X^{\alpha}\,Y^{\beta}  \, " .
\eeq
It is supported by the lines of equation:
\beq
-{\displaystyle \sum_k} C_{\alpha,\beta ;k} \tilde T_k + \alpha x + \beta  y =
-{\displaystyle \sum_k} C_{\alpha',\beta' ;k} \tilde T_k + \alpha' x + \beta'  y 
\eeq
for $(\alpha,\beta-f\alpha)$ and $(\alpha',\beta'-f\alpha')$ two vertices of  $\hat\Upsilon_\CYX$. It is easy to see that these lines are parallel to the edges of the framed toric graph of $\CYX$\footnote{Remark that this property does not depend on the value of the parameters $C_{\alpha,\beta ;k}$ and $\td T_k$.} .

More precisely, the tropical curve $H_f^{[\infty]}(X,Y)$ is a graph whose faces are sectors of $\mathbb{R}^2$ where the logarithm of the modulus of one of the monomials in $H_f(X,Y)$ is larger than all the other ones. 
It is now  important to remember that such a monomial is just the expression of one of the $Y_i/Y_3$ in terms of $X$ and $Y$. 
If $i=3$ we have that $|Y_i/Y_3|=1$ and $|Y_j/Y_3|\ll 1$ for $j\neq 3$.
If $i\neq 3$, we have that $|Y_j/Y_i|\ll 1$ for $j\neq i$.
Let us write
$$
|X_j|^2 = - \log |Y_j/Y_i|.
$$
Hence, a face of the graph defining the tropical limit, corresponds to
$$
|X_i|^2 = 0 
\quad , \,\, \forall \,j\neq i\,\,\, |X_j|^2>0
$$
and by definition $\sum_j q_{k,j} |X_j|^2 = \td t_k$.
In other words a face of the tropical curve corresponds to 
 the vanishing of one of the $|X_i|^2$, i.e. to a face of the toric graph. 
 
 In this way, we can associate a face of the Toric graph to each face of the graph defined by the tropical curve. The adjacency of these latter faces is thus given by the Toric graph. Combining this result with the fact that the edges of the tropical curve are parallel to the ones of the framed toric graph proves the proposition (up to a translation of the edges which does not change the graph and is of no interest in the following).

}

\subsubsection{Large radius}

\begin{figure}[t]
\centering
$$\includegraphics[height=4.5cm]{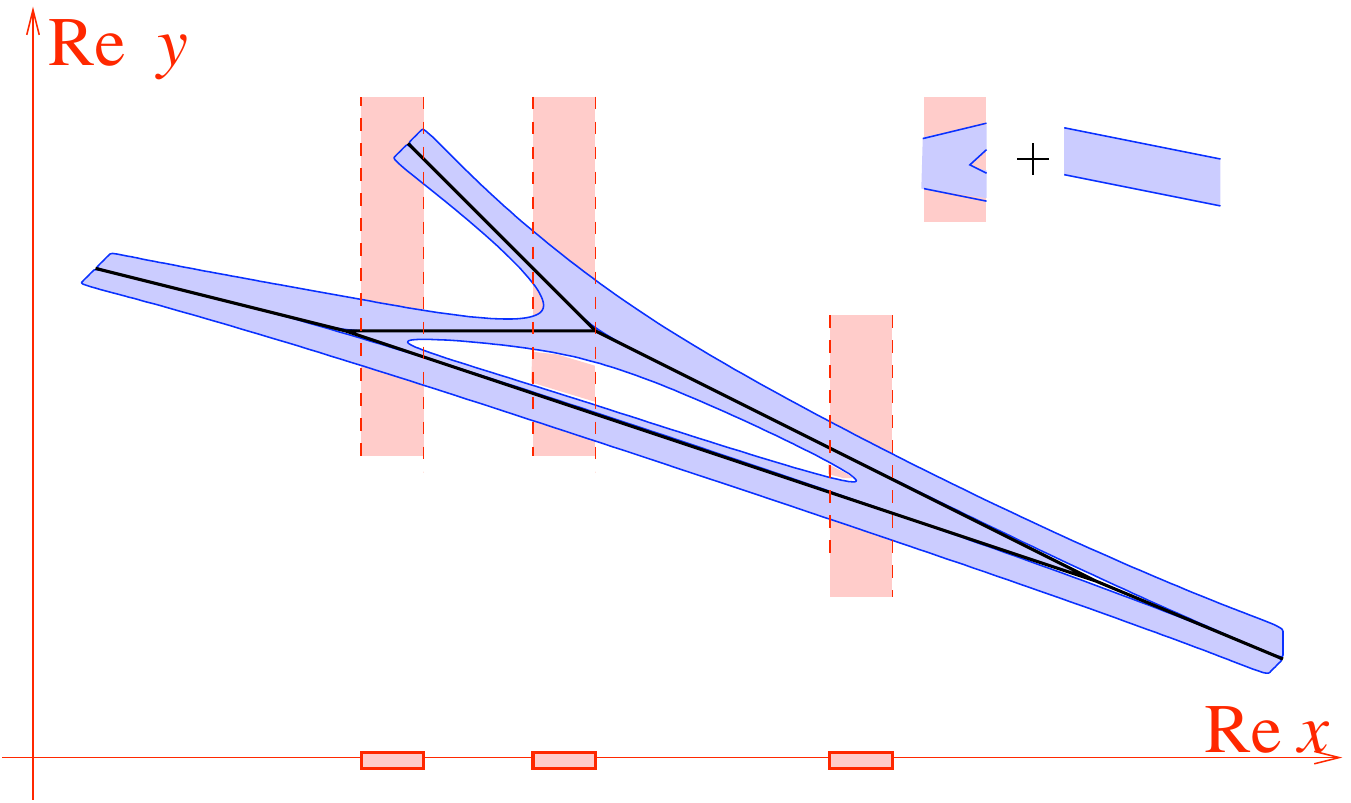}  $$
\caption{For $\td t_i$'s large enough, the Amoeba surrounds the toric graph.
Since we have chosen the framing so that the toric graph has no vertical edge, it is possible to cut the amoeba by vertical strips of width $O(1)$ (when $\td t_i\to \infty$), such that the amoeba is a union of 3-legged amoebas of fixed width $R$, and 2-legged amoebas of arbitrary width.
\label{figamoebalocP2}}
\end{figure}

When the parameters $\td{t}_j$ are large enough, the amoeba of the mirror curve is a small region which surrounds its tropical limit (cf. fig. \ref{figamoebalocP2}), that is to say, it is a fattening of the framed toric graph.

Thus, when the $\td t_j$'s are large enough, it is possible to cut the amoeba by vertical strips of fixed width $R=O(1)$, such that the amoeba is a union of 3-legged amoeba pieces of fixed width, and 2-legged amoeba pieces of arbitrary width. See fig \ref{figamoebalocP2}.

\smallskip

$\bullet$ {\bf Cylinders}
 
Regions of the amoeba which are close to edges (i.e. 2-legged amoeba pieces) correspond to two of the $|Y_i|$'s of the same magnitude and all the others much smaller, i.e. they are approximated by:
\beq\label{cylinceredges1}
(\ee{\td t_{\alpha,\beta}} X^{\alpha}Y^{\beta} + \ee{\td t_{\alpha',\beta'}} X^{\alpha'}Y^{\beta'} )(1+  o(1))=0
\eeq
with $(\alpha,\beta)$ and $(\alpha',\beta')$ two adjacent vertices of the dual toric graph.
Let us consider an half-edge $\epsilon$, so that $(\alpha,\beta)$ is to the right of $\epsilon$ and $(\alpha',\beta')$ is to the left of $\epsilon$, so that $\fram_\epsilon=\beta-\beta'$.

The curve $\ee{\td t_{\alpha,\beta}} X^{\alpha}Y^{\beta} + \ee{\td t_{\alpha',\beta'}} X^{\alpha'}Y^{\beta'} =0$, is a rational curve with 2 punctures, it is topologically a cylinder.
It can be parametrized by a complex parameter $z\in \mathbb C^*$:
\beq\label{eqparamcylinder}
\left\{\begin{array}{l}
X=\ee{-x} = z^{\beta-\beta'} \,(1+o(1))\cr
Y = \ee{-y}=\ee{\frac{t_{\alpha',\beta'}-t_{\alpha,\beta}+\ii\pi}{\beta-\beta'}}\,\,z^{\alpha'-\alpha} \,(1+o(1))
\end{array}\right.
\eeq
The differential $dx = -\,\fram_\epsilon\, dz/z$ never vanishes, so this curve contains no branchpoint.

\smallskip

$\bullet$ {\bf Pairs of pants}
 Regions of the amoeba which are close to vertices (i.e. 3-legged amoeba pieces) correspond to three of the $|Y_i|$'s of the same magnitude and all the others much smaller, i.e. they are approximated by:
\beq
(\ee{\td t_{\alpha,\beta}} X^{\alpha}Y^{\beta} + \ee{\td t_{\alpha',\beta'}} X^{\alpha'}Y^{\beta'} +\ee{\td t_{\alpha'',\beta''}} X^{\alpha''}Y^{\beta''})(1+  o(1))=0
\eeq
where $(\alpha,\beta),(\alpha',\beta'),(\alpha'',\beta'')$ is an oriented triangle of $\hat\Upsilon_{\CYX}$, i.e. a vertex $\sigma$ of $\Upsilon_{\CYX}$.
This curve can be parametrized by a complex parameter $z\in \mathbb C\setminus\{0,1,\infty\}$:
\beq
\left\{\begin{array}{l}
x =-\ln X =   -\fram_{b}\,\ln z\,\,-\fram_{a}\ln{(1-z)} \quad +\fram_{b}\,\ln{\frac{\fram_{b}}{\fram_{a}+\fram_{b}}}\,\,+\fram_{a}\ln{\frac{\fram_{a}}{\fram_{a}+\fram_{b}}} +a_\sigma +o(1)\cr
y=-\ln Y = -\fram_{d}\,\ln z\,\,-\fram_{c}\ln{(1-z)} \quad +\fram_{d}\,\ln{\frac{\fram_{b}}{\fram_{a}+\fram_{b}}}\,\,+\fram_{c}\ln{\frac{\fram_{a}}{\fram_{a}+\fram_{b}}}  +b_\sigma +o(1)\cr
\end{array}\right.
\eeq
where $\fram$ is the framing matrix defined in def. \ref{defframsigma}.

\begin{figure}[t]
\centering
$$\includegraphics[height=4.5cm]{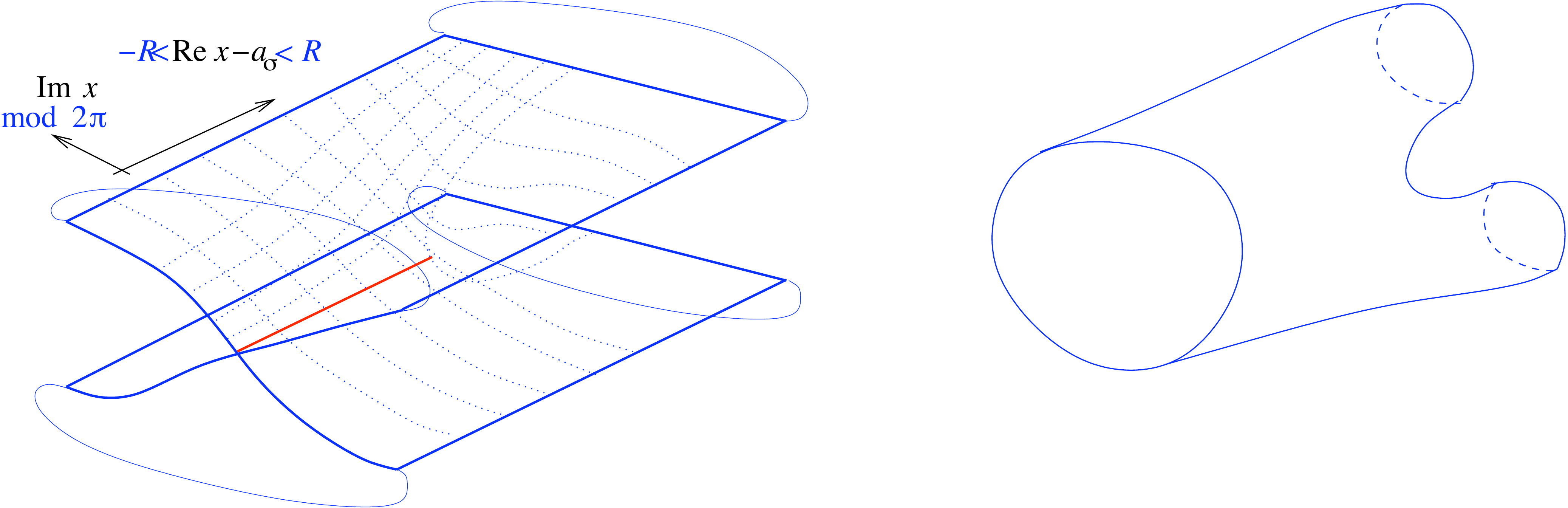}   $$
\caption{Each $\curve_\sigma$ is topologically a "pair of pants" i.e. a Riemann sphere with 3 holes, $\d\curve_\sigma$ is the union of 3 circles.
$C_\sigma$ is realized by gluing 2 copies of the $x$-complex plane, with a cut $]\infty, a_\sigma]$, glued together along the cut.
Moreover, since the curve is algebraic in the variable $X=\ee{-x}$, this means that $x$ has to be identified with $x\equiv x+2\pi i\,\fram$. Also, we have defined $C_\sigma$ by restricting $-\frac {R}{2}<\Re\, (x-a_\sigma)<\frac {R}{2}$ for large enough $\td t_j$'s.
\label{figCsigmaxplane}}
\end{figure}

It is easy to see that this curve is topologically a pair of pants
see fig.\ref{figCsigmaxplane}.

The differential 
\beq
dx =  \left(\frac{-\fram_b}{z}+\frac{\fram_a}{1-z}\right)\,dz
\eeq
vanishes at exactly one point $z=\fram_b/(\fram_a+\fram_b)$, so this pair of pants contains exactly one branchpoint (here we use the fact that we have chosen the framing so that the framed toric graph has no vertical edge, i.e. $\fram_a$, $\fram_b$ and $\fram_a+\fram_b$ are all non-vanishing).

%
%
%



\subsubsection{Atlas of the mirror curve\label{secatlas}}

\begin{figure}[t]
\centering
$$\includegraphics[height=5.5cm]{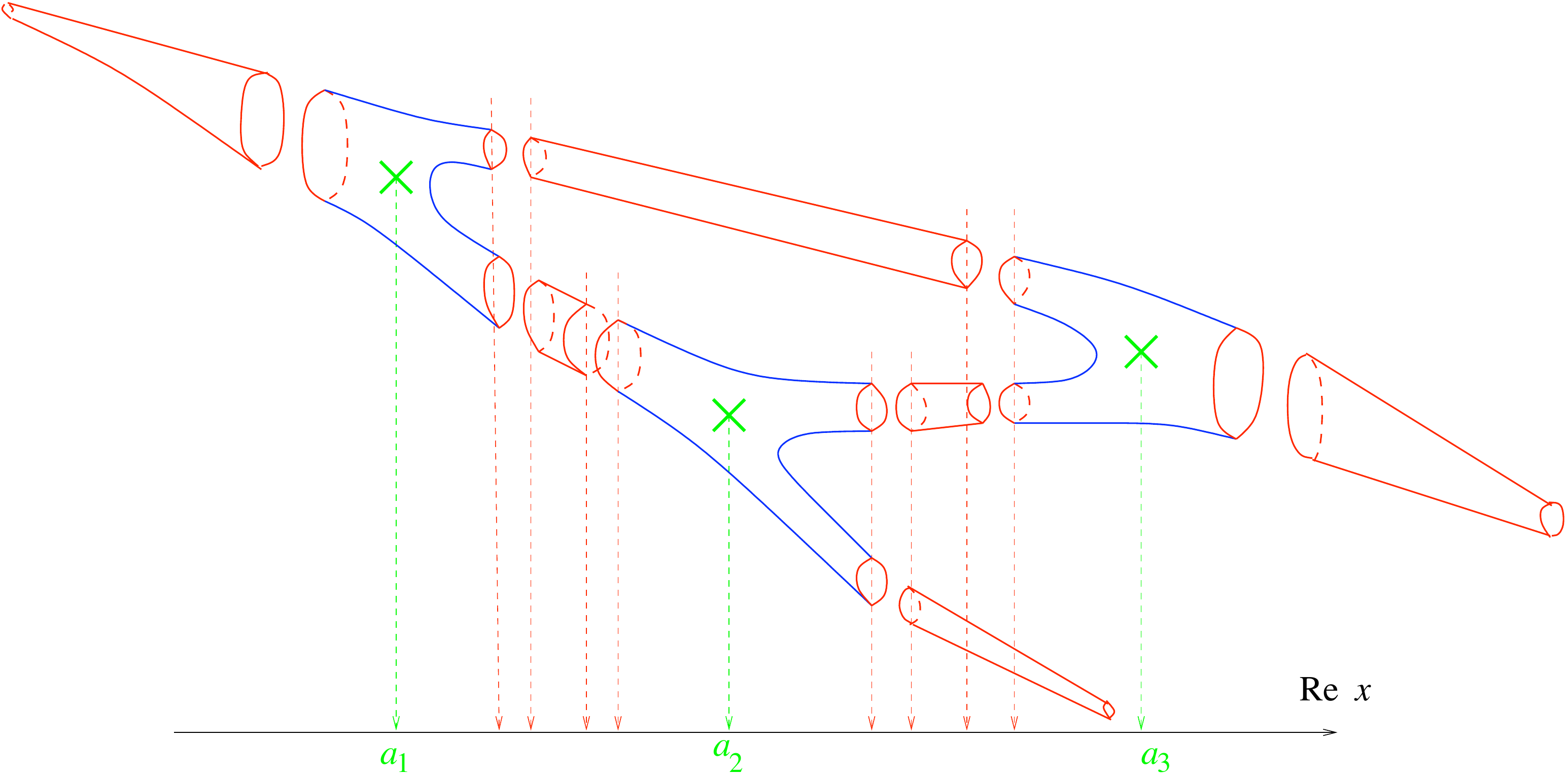}  $$
\caption{For $\td t_j$'s large enough, and if we have chosen an appropriate framing (such that the toric graph has no vertical edge), then we can cut the curve by planes $\Re\, x=$constant, such that the curve is a union of pairs of pants and cylinders.
Each pair of pant contains exactly one branchpoint $a_\sigma$, and we chose the cutting planes such that the domain $\{\Re\,(x-a_\sigma)\}$ remains finite in the limit $\td t_j\to +\infty$.
The cylinders become of infinite lenght in the limit $\td t_j\to +\infty$, their length is of the order $|\Re (a_\sigma-a_{\sigma'})|$ i.e. of the order of $\td t_j$. The example here is local $\mathbb P^2$, of equation $1+XY^2 +Y+\frac{\ee{\td t}}{X Y^3}=0$, i.e. the curve $1+X+Y+\frac{\ee{\td t}}{XY} = 0$ framed by $X\to XY^2$. 
\label{figspcurvelocP2framed}}
\end{figure}

%
%
%

\medskip
This gives an explicit atlas of charts to describe the plane curve $H_f(X,Y)=0$.

In other words, an atlas of $\curve$ is obtained as follows:

\bp\label{propatlascurve}
The curve $\curve$ is covered by a union of cylinders $\curve_e$ (with $e\in$edges) and spheres with 3 holes $\curve_{\sigma}$ (with $\sigma\in$vertices).
\beq
\curve = \cup_{\sigma={\rm vertices\,of}\,\Upsilon_\CYX} \, C_\sigma \quad \cup_{e={\rm edges\,of}\,\Upsilon_\CYX} \, C_e.
\eeq
The transition maps are obtained by identifying the coordinate $\ee{-x}$ in each patch. Indeed, if $\sigma$ is a vertex and $e$ is an adjacent edge, then the map $\curve\to\mathbb C^*$, $\mapsto \ee{-x}$ is analytical and invertible on $\curve_{\sigma} \cap \curve_e$ (this intersection has the topology of a cylinder).

Each pair of pants $C_\sigma$ contains exactly one branchpoint, and thus we label the branchpoints by vertices $\sigma\in \Upsilon_\CYX$.

\ep


\subsubsection{The Harnack property}

Notice that the map $\curve\to {\cal A}:\,\, (\ee{-x},\ee{-y})\mapsto (\Re\,x,\Re\,y)$ is $2\to 1$ in each cylinder and each pair of pants, and our assumption that $\td t_j$'s are large enough implies that the amoeaba pieces of distinct cylinders and pairs of pants don't overlap, so the map  $\curve\to {\cal A}:\,\, (\ee{-x},\ee{-y})\mapsto (\Re\,x,\Re\,y)$ is globally $2\to 1$.
This is called the "Harnack property", and this means that $H(\ee{-x},\ee{-y})=0$ is a Harnack curve.

\smallskip

It was shown by \cite{Kenyon2003, hanany-2006} that a Harnack curve can always be realized as the limit shape of a crystal model, which makes the link with combinatorics, but we shall not use that here.

\medskip

\subsubsection{Torelli marking of the curve}\label{secTorelli}

\begin{figure}[t]
\centering
$$\includegraphics[height=4.5cm]{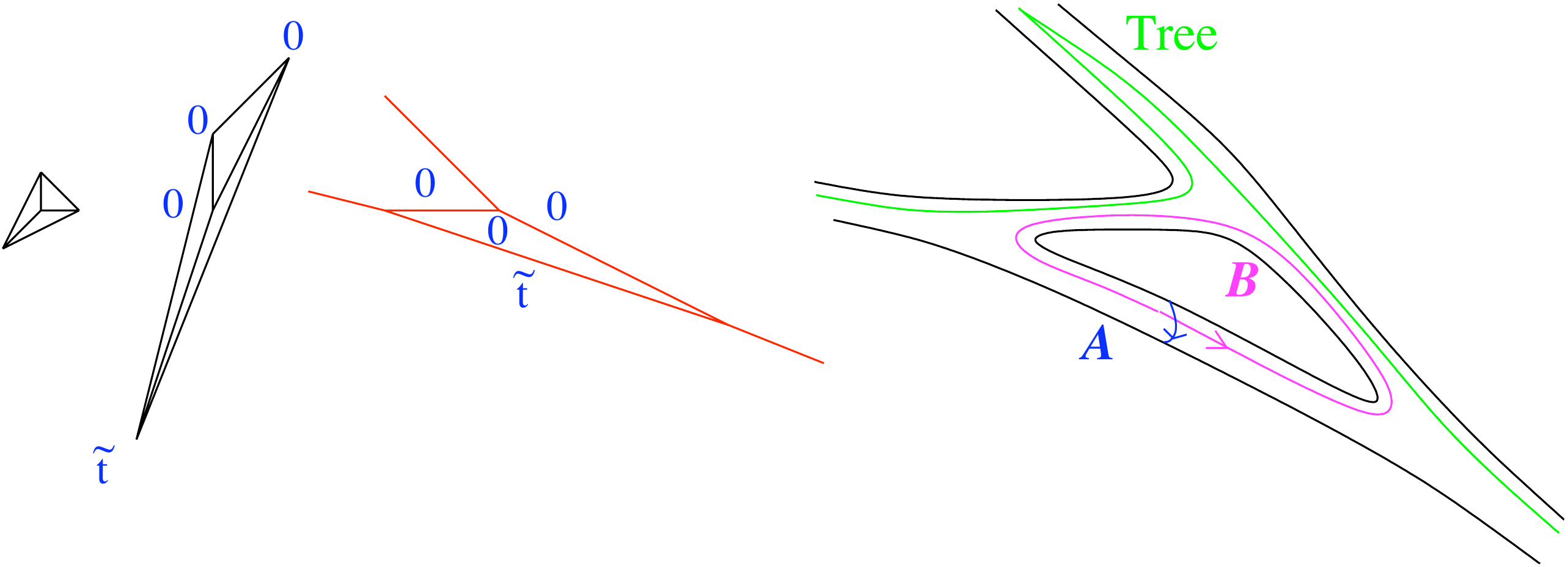}  $$
\caption{First consider a tree $T$ going through all vertices and all non-compact edges of $\Upsilon_\CYX$, choose a root and orient $T$ from root to leaves.
For each compact edge $I$ of $\Upsilon_\CYX\setminus T$ (there are $\genus$ of them), choose an oriented cycle $\acycle_I$ wrapping the corresponding cylinder of $\curve$.
Choose the cycle $\bcycle_I$ to be the (only) loop of $T\cup I$, pulled back on $\curve$ by the amoeba map, and orient it so that $\acycle_I \cap \bcycle_I=+1$.
That provides a symplectic basis of cycles on $\curve$: $\acycle_I \cap \bcycle_J=\delta_{I,J}$. 
\label{figcyclesmirrormap}}
\end{figure}

We shall need to define a symplectic basis of cycles on the curve. See fig. \ref{figcyclesmirrormap}.

\smallskip

$\bullet$ First, let us choose a tree $T$ covering all vertices and all non-compact edges of the toric graph $\Upsilon_\CYX$. Choose a root and orient the edges of $T$ from root to leaves.

\smallskip

$\bullet$ there are exactly $\genus$ compact edges of $\Upsilon_\CYX$ which are not covered by $T$. All such edges $e$ are such that $T\cup e$ has  exactly one loop, and label that loop, and that edge, by the vertex of $\hat\Upsilon_\CYX$ adjacent to it inside the loop.
The set of labels is the set of interior points of $\hat\Upsilon_\CYX$
\beq
\{I\,|\,\,I={\rm compact\,face\,of}\,\Upsilon_\CYX = {\rm interior\,vertex\,of}\,\hat\Upsilon_\CYX\,\}.
\eeq

For each such edge, choose arbitrarily an orientation, i.e. an half edge $\epsilon_I$.
Define the cycle $\acycle_{\epsilon_I}$ to be a circle of constant $\Re\,x$ wrapping the cylinder $\curve_{\epsilon_I}$ leaving the other half-cylinder on its left side.

\smallskip

$\bullet$ For each such half-edge $\epsilon_I$, define the cycle $\bcycle_{\epsilon_I}$ on $\curve$, to be a pullback by the amoeba map, of the loop of $T\cup \epsilon_I$, and orient it such that $\acycle_{\epsilon_I}\cap\bcycle_{\epsilon_I}=+1$.

\smallskip

All this defines a symplectic basis of cycles on $\curve$, satisfying:
\beq
\acycle_I\cap\bcycle_J=\delta_{I,J}
\quad , \quad
\acycle_I\cap\acycle_J=0
\quad , \quad
\bcycle_I\cap\bcycle_J=0.
\eeq
A curve $\curve$ with a symplectic basis of cycles is said to have a "Torelli marking".

\medskip

{\bf Factor $s_\epsilon$:}

For later purposes (for defining the mirror map in section \ref{secmirrormap} below), we need to associate a weight $s_\epsilon$ to each half-edge $\epsilon$ of $\Upsilon_\CYX$, such that:
\beq\label{eqdefsepsilon}
\left\{
\begin{array}{l}
(\epsilon,\epsilon')={\rm compact\, edge\,of}\,\Upsilon_\CYX \qquad \quad \to \quad s_\epsilon=-s_{\epsilon'} \cr
(\epsilon_1,\epsilon_2,\epsilon_3)={\rm vertex\,of}\,\Upsilon_\CYX \qquad \quad \to \quad 
s_{\epsilon_1}+s_{\epsilon_2}+s_{\epsilon_3}=1.
\end{array}\right.
\eeq

\medskip
Remark: a choice of $s_\epsilon$ satisfying \eqref{eqdefsepsilon} is not unique, for instance one can change it to $s_\epsilon+K\,\fram_\epsilon$ where $K$ is an arbitrary constant, since $\fram_\epsilon$ satisfies the homogeneous part of \eqref{eqdefsepsilon}.

Let us show that some choice of $s_\epsilon$ do exist.
\medskip

\bd\label{defsepsilon}

The factor $s_\epsilon$ can be constructed as follows:

$\bullet$ to each half-edge $\epsilon$ of $\Upsilon_\CYX$ which doesn't belong to the tree $T$, associate $s_\epsilon=0$.

\smallskip

$\bullet$ to each half-edge $\epsilon$ which is a leaf of $T$, associate $s_\epsilon=0$.

\smallskip

$\bullet$ Recursively, starting from leaves to root, to each vertex of $T$ whose children half-edges have already been computed, define for the parent half-edge $\epsilon$:
\beq
s_\epsilon = 1-\sum_{{\rm childdren}\,\epsilon'} s_{\epsilon'}
\eeq
and proceed until all half edges have been computed.
\ed

\subsection{The mirror map\label{secmirrormap}}

Our mirror curve is of the form:
\beq
H(X,Y) =  \sum_{(\alpha,\beta)\in \hat\Upsilon_\CYX} \ee{\td t_{\alpha,\beta}}\,\,X^\alpha\,Y^\beta
\eeq
where each coefficient $\td t_{\alpha,\beta}$ is a linear combination of the $\td t_j$'s:
\beq\label{tdtijCijktk}
\td t_{\alpha,\beta} =  \sum_{j=1}^r\,\,C_{(\alpha,\beta);j}\,\, \td t_j.
\eeq
Notice that each $I=(\alpha,\beta)$ is a vertex of the dual toric graph $\hat\Upsilon_\CYX$, and thus corresponds to a face of the toric graph $\Upsilon_\CYX$.

\bl
The  $r\times r$ matrix $C$ such that
\beq
\forall\,(\alpha,\beta)\in \hat\Upsilon_\CYX,\,(\alpha,\beta)\notin\{(0,0),(0,1),(1,f)\}\, ,
\qquad
\quad
\td t_{\alpha,\beta} = \sum_{j=1}^r C_{(\alpha,\beta);j}\,\td t_j
\eeq
is invertible (we have set $\td t_{0,0}=\td t_{0,1}=\td t_{1,f}=0$).

\el

\proof{
The vertices of $\hat\Upsilon_\CYX$ are labeled $I=1,2,\dots,r+3$ in \eqref{eqqxtdt}, and are also labeled $I=(\alpha,\beta)$ the coordinates of $I$ in $\mathbb Z^2$, so that $C$ is a square $r\times r$ matrix.
If we write
\beq
x_{0,0} = 0,\,\, x_{1,f}=x+fy , \,\,x_{0,1}=y,\,\,\,
x_{\alpha,\beta} = \td t_{\alpha,\beta} + \alpha x + \beta y
\eeq
the definition of the $x_{\alpha,\beta}$ given by  \eqref{eqqxtdt}, reads
\beq
\forall\,i=1,\dots,r \, , \qquad q_{i,1} (x+fy)+q_{i,2} y+\sum_{I=4}^{r+3} q_{i,I} x_I = \td t_i.
\eeq
i.e.
\beq
(C^{-1})_{i,I} = q_{i,I} \quad ,\, i=1,\dots,r \, ,\,\,\, I = 4,\dots,r+3,
\eeq
which is invertible due to the smoothness condition \eqref{defsmoothness}.
}

\begin{figure}[t]
\centering
$$\includegraphics[height=5.5cm]{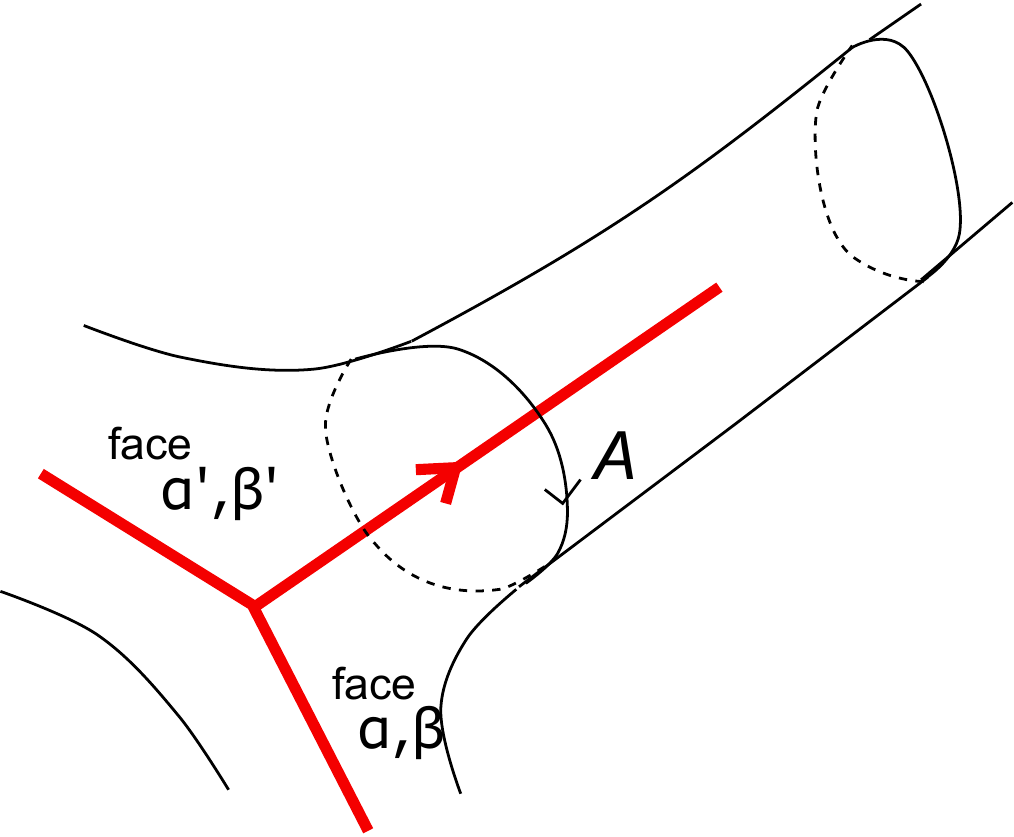}  $$
\caption{Consider an half-edge $\epsilon$, it corresponds to a cylinder on the mirror curve, and we choose a cycle $\acycle_\epsilon$ wrapping the cylinder, oriented so that the vertex from which $\epsilon$ comes, is to the right of $\acycle_\epsilon$.
\label{figcylmirror1}}
\end{figure}

\bl\label{lemmamonodromieAe}
Let $\epsilon$ be an half-edge of the toric graph, so that $(\alpha,\beta)$ is the vertex of $\hat\Upsilon_\CYX$ to the right of $\epsilon$, and $(\alpha',\beta')$ is the vertex of $\hat\Upsilon_\CYX$ to the left of $\epsilon$, and thus $\fram_\epsilon=\beta-\beta'$.
Let $\curve_\epsilon$ be the corresponding cylinder of the curve, and $\acycle_\epsilon$, a cycle wrapping the cylinder $\curve_\epsilon$ positively (see fig.\ref{figcylmirror1}).
In the large radius limit we have that
$\ln{(Y\,\,X^{-\,\frac{\fram_{d,\sigma(\epsilon)}}{\fram_\epsilon}})}$ has no monodromy around $\acycle_\epsilon$, and
\beq
\frac{1}{2\ii\pi}\oint_{\acycle_\epsilon} \ln{(Y\,\,X^{\frac{\alpha-\alpha'}{\fram_\epsilon}})}\,\,\frac{dX}{X} = \td t_{\alpha',\beta'}-\td t_{\alpha,\beta} + O(1)
\eeq

\el

\proof{
Let $\epsilon$ be the half edge from the vertex $(\alpha,\beta)$ to $(\alpha',\beta')$.
According to \eqref{cylinceredges1}, the equation of the curve in the cylinder is in the large radius regime:
\beq
(X^\alpha\,Y^\beta\,\ee{\td t_{\alpha,\beta}} + X^{\alpha'}\,Y^{\beta'}\,\ee{\td t_{\alpha',\beta'}} )(1+o(1)) = 0
\eeq
i.e.
\beq
Y^{\beta-\beta'}\,X^{\alpha-\alpha'} = - \ee{\td t_{\alpha',\beta'}-\td t_{\alpha,\beta}}\,(1+o(1))
\eeq
where the $o(1)$ term is analytical in $\curve_\epsilon$,
and thus
\beq
\ln{(Y\,\,X^{\frac{\alpha-\alpha'}{\beta-\beta'}})}
= \frac{\td t_{\alpha',\beta'}-\td t_{\alpha,\beta}\pm \ii\pi}{\beta-\beta'}  + o(1),
\eeq
where the right hand side is analytical on the cylinder $\curve_\epsilon$.
This guarantees that the contour integral around $\acycle_\epsilon$ makes sense.
Moreover, using the parametrization \eqref{eqparamcylinder}, the cycle $\acycle_\epsilon$ is a trigonometricaly  oriented circle around $0$ in the variable $z$, and thus we get:
\beq
\frac{1}{2\ii\pi}\oint_{\acycle_\epsilon} \ln{(Y\,\,X^{-\,\frac{\fram_{d,\sigma(\epsilon)}}{\fram_\epsilon}})}\,\,\frac{dX}{X} = (\td t_{\alpha',\beta'}-\td t_{\alpha,\beta} \pm \ii\pi) \Res_{z\to 0}\,\frac{dz}{z} + o(1).
\eeq

}

%

\bl It is possible to define (uniquely) $t_{\alpha,\beta}$ for all vertices $(\alpha,\beta)$ of $\hat\Upsilon_\CYX$ such that:
\beq
t_{0,0}=t_{1,0}=t_{0,1}=0, 
\eeq
and for every half-edge $\epsilon$ (we call $(\alpha,\beta)$ the vertex of $\hat\Upsilon_\CYX$ to the right of $\epsilon$ and $(\alpha',\beta')$ the one to its left):
\beq
t_{\alpha',\beta'}- t_{\alpha,\beta}  - s_\epsilon\,\,\ii\pi = \frac{1}{2\ii\pi}\oint_{\acycle_\epsilon} \ln{(Y\,\,X^{-\,\frac{\fram_{d,\sigma(\epsilon)}}{\fram_\epsilon}})}\,\,\frac{dX}{X} 
\eeq
where $s_\epsilon$ was defined in def.\ref{defsepsilon}.
\el

\begin{figure}[t]
\centering
$$\includegraphics[height=5.5cm]{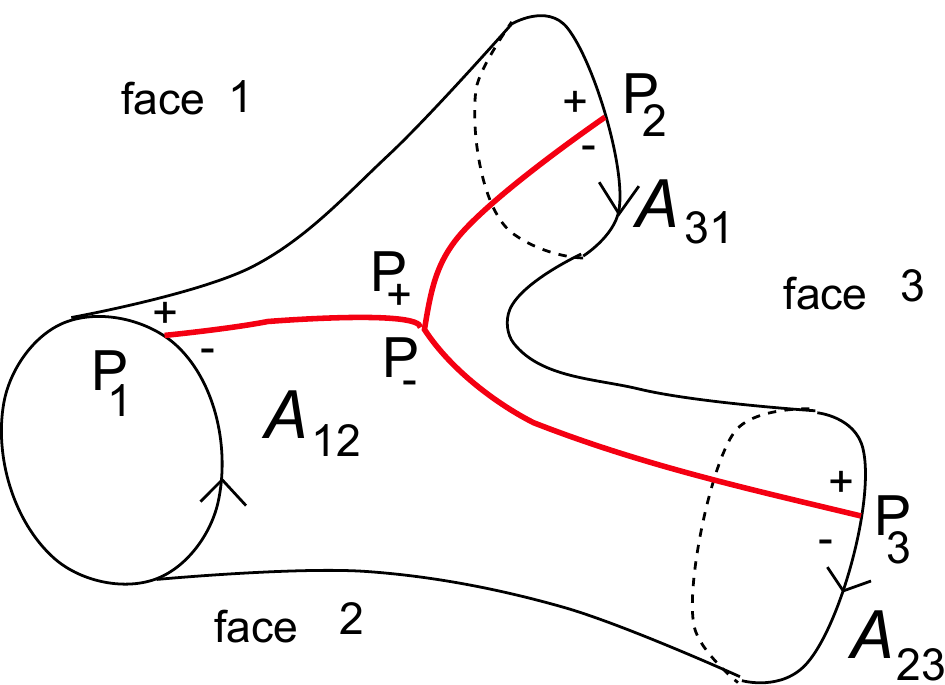}  $$
\caption{In order to compute integrals with logs, we need to introduce cuts.
\label{figpantmirror1}}
\end{figure}

\proof{
Since the graph is connected, we can relate any vertex $(\alpha,\beta)$ to $(0,0)$ by a sequence of edges.
We have to check that the result is independent of which sequence of edges.

First, if $\epsilon$ and $\epsilon'$ are the 2 half-edges of the same compact egde, one has $s_{\epsilon'}=-s_\epsilon$ and $\acycle_{\epsilon'}=-\acycle_\epsilon$ and the integrand is unchanged, so that $\epsilon$ and $\epsilon'$ give the same value for $t_{\alpha',\beta'}-t_{\alpha,\beta}$.

Then, we have to check that for any face $(\alpha_1,\beta_1),(\alpha_2,\beta_2),(\alpha_3,\beta_3)$ of $\hat\Upsilon_\CYX$, we indeed have:
$(t_{\alpha_1,\beta_1}-t_{\alpha_2,\beta_2}) + (t_{\alpha_2,\beta_2}-t_{\alpha_3,\beta_3}) + (t_{\alpha_3,\beta_3}-t_{\alpha_1,\beta_1}) =0$.

\smallskip

Consider an oriented face $\sigma = (\alpha_1,\beta_1),(\alpha_2,\beta_2),(\alpha_3,\beta_3)$ of $\hat\Upsilon_\CYX$, and its 3 half-edges $\epsilon_{1,2},\epsilon_{2,3},\epsilon_{3,1}$. $\sigma$ is also a vertex of $\Upsilon_\CYX$, and labels a pair of pants of $\curve$, whose 3 oriented boundaries are $\acycle_{\epsilon_{1,2}}$, $\acycle_{\epsilon_{2,3}}$, $\acycle_{\epsilon_{3,1}}$ (they are oriented so that the pair of pants lies to the right of its boundaries).

Notice that $\ln{Y}$ and $\ln{X}$ are not analytical on the whole pair of pants.
We need to introduce  cuts, so that the pair of pants minus the cuts is simply connected. We choose 3 cuts from some point $P_1\in \acycle_{\epsilon_{1,2}}$, $P_2\in \acycle_{\epsilon_{3,1}}$,  $P_3\in \acycle_{\epsilon_{2,3}}$, to a point $P$ in the middle, as in fig.\ref{figpantmirror1}.

According  to lemma \ref{lemmamonodromieAe}, on a cycle $\acycle_{\epsilon_{1,2}}$, since the integrand of the following integral is analytical, we have (we denote $\alpha_{i,j}=\alpha_i-\alpha_j$ and $\beta_{i,j}=\beta_i-\beta_j$):
\beq
\oint_{\acycle_{\epsilon_{1,2}}} \left(\ln{Y} + \frac{\alpha_{1,2}}{\beta_{1,2}}\,\ln X \right)\,\frac{dX}{X}
= -\int_{P_{1-}}^{P_{1+}} \left(\ln{Y} + \frac{\alpha_{1,2}}{\beta_{1,2}}\,\ln X \right)\,\frac{dX}{X}
\eeq
\beq
\oint_{\acycle_{\epsilon_{2,3}}} \left(\ln{Y} + \frac{\alpha_{2,3}}{\beta_{2,3}}\,\ln X \right)\,\frac{dX}{X}
= - \int_{P_{3+}}^{P_{3-}} \left(\ln{Y} + \frac{\alpha_{2,3}}{\beta_{2,3}}\,\ln X \right)\,\frac{dX}{X}
\eeq
\beq
\oint_{\acycle_{\epsilon_{3,1}}} \left(\ln{Y} + \frac{\alpha_{3,1}}{\beta_{3,1}}\,\ln X \right)\,\frac{dX}{X}
= -\int_{P_{2+}}^{P_{2-}} \left(\ln{Y} + \frac{\alpha_{3,1}}{\beta_{3,1}}\,\ln X \right)\,\frac{dX}{X}
\eeq

For the integrals of $\ln Y$, by deforming integration contours across the pair of pants, we have:
\bea
&& \int_{P_{1-}}^{P_{1+}} \ln{Y} \,\frac{dX}{X}
+\int_{P_{2+}}^{P_{2-}} \ln{Y} \,\frac{dX}{X}
+\int_{P_{3+}}^{P_{3-}} \ln{Y} \,\frac{dX}{X} \cr
&=& \int_{P_{1-}}^{P_{-}} \left(\ln{Y_-}-\ln{Y_+}\right) \,\frac{dX}{X}
+\int_{P_{2+}}^{P_{+}} \left(\ln{Y_+}-\ln{Y_-}\right) \,\frac{dX}{X}
+\int_{P_{3-}}^{P_{-}} \left(\ln{Y_+}-\ln{Y_-}\right) \,\frac{dX}{X} . \cr
\eea
Again, according  to lemma \ref{lemmamonodromieAe}, we have that the discontinuities of $\ln Y$ are proportional to discontinuities of $\ln X$, and the discontinuity of $\ln X$ around $\acycle_{\epsilon}$ is $2\ii\pi\fram_\epsilon$, that gives:
\bea
&& -\int_{P_{1-}}^{P_{1+}} \ln{Y} \,\frac{dX}{X}
-\int_{P_{2+}}^{P_{2-}} \ln{Y} \,\frac{dX}{X}
-\int_{P_{3+}}^{P_{3-}} \ln{Y} \,\frac{dX}{X} \cr
&=& \frac{\alpha_{12}}{\beta_{12}}\,\int_{P_{1-}}^{P_{-}} \left(\ln{X_-}-\ln{X_+}\right) \,\frac{dX}{X}
+\frac{\alpha_{31}}{\beta_{31}}\,\int_{P_{2+}}^{P_{+}} \left(\ln{X_+}-\ln{X_-}\right) \,\frac{dX}{X} \cr
&&  +\frac{\alpha_{23}}{\beta_{23}}\,\int_{P_{3-}}^{P_{-}} \left(\ln{X_+}-\ln{X_-}\right) \,\frac{dX}{X} \cr
&=& 2\ii\pi\alpha_{12}\,\int_{P_{1-}}^{P_{-}}  \,\frac{dX}{X}
+2\ii\pi\alpha_{31}\,\int_{P_{2+}}^{P_{+}}  \,\frac{dX}{X}
+2\ii\pi\alpha_{23}\,\int_{P_{3-}}^{P_{-}}  \,\frac{dX}{X} \cr
&=& 2\ii\pi \left(\alpha_{12}\,\ln{X(P_-)}+\alpha_{23}\,\ln{X(P_-)}+\alpha_{31}\,\ln{X(P_+)}\right)\cr
&& -2\ii\pi \left(\alpha_{12}\ln{X(P_{1-})}+\alpha_{31}\ln{X(P_{2+})}
+\alpha_{23}\ln{X(P_{3-})} \right) \cr
&=& 2\ii\pi \alpha_{31} \left(\,\ln{X(P_+)}-\ln{X(P_-)}\right) -2\ii\pi \left(\alpha_{12}\ln{X(P_{1-})}+\alpha_{31}\ln{X(P_{2+})}
+\alpha_{23}\ln{X(P_{3-})} \right) \cr
&=& -\,(2\ii\pi)^2 \alpha_{31}\beta_{12} -2\ii\pi \left(\alpha_{12}\ln{X(P_{1-})}+\alpha_{31}\ln{X(P_{2+})}
+\alpha_{23}\ln{X(P_{3-})} \right) .
\eea

We now compute the integrals with $\ln X$:
\bea
\frac{\alpha_{12}}{\beta_{12}}\,\int_{P_{1-}}^{P_{1+}} \ln X\,\,\frac{dX}{X}
&=& \frac{1}{2}\,\frac{\alpha_{12}}{\beta_{12}}\,\left(\ln{X(P_{1+})}^2-\ln{X(P_{1-})}^2\right) \cr
&=& -\,\frac{1}{2}\,\frac{\alpha_{12}}{\beta_{12}}\,\left(\ln{X(P_{1+})}-\ln{X(P_{1-})}\right)\,\left(\ln{X(P_{1+})}+\ln{X(P_{1-})}\right) \cr
&=& -2\ii\pi\alpha_{12}\,\left(\ln{X(P_{1-})}-\ii\pi\beta_{12}\right) .\cr
\eea
Similarly, we get
\bea
\frac{\alpha_{23}}{\beta_{23}}\,\int_{P_{3+}}^{P_{3-}} \ln X\,\,\frac{dX}{X}
&=& \frac{1}{2}\,\frac{\alpha_{23}}{\beta_{23}}\,\left(\ln{X(P_{3-})}^2-\ln{X(P_{3+})}^2\right) \cr
&=& -\,\frac{1}{2}\,\frac{\alpha_{23}}{\beta_{23}}\,\left(\ln{X(P_{3-})}-\ln{X(P_{3+})}\right)\,\left(\ln{X(P_{3-})}+\ln{X(P_{3+})}\right) \cr
&=& -\,2\ii\pi\alpha_{23}\,\left(\ln{X(P_{3-})}+\ii\pi\beta_{23}\right) \cr
\eea
and
\bea
\frac{\alpha_{31}}{\beta_{31}}\,\int_{P_{2+}}^{P_{2-}} \ln X\,\,\frac{dX}{X}
&=& \frac{1}{2}\,\frac{\alpha_{31}}{\beta_{31}}\,\left(\ln{X(P_{2-})}^2-\ln{X(P_{2+})}^2\right) \cr
&=& -\,\frac{1}{2}\,\frac{\alpha_{31}}{\beta_{31}}\,\left(\ln{X(P_{2-})}-\ln{X(P_{2+})}\right)\,\left(\ln{X(P_{2-})}+\ln{X(P_{2+})}\right) \cr
&=& -\,2\ii\pi\alpha_{31}\,\left(\ln{X(P_{2+})}-\ii\pi\beta_{31}\right) . \cr
\eea

Finally we have:
\bea
&& \oint_{\acycle_{\epsilon_{1,2}}} \left(\ln{Y} + \frac{\alpha_1-\alpha_2}{\beta_1-\beta_2}\,\ln{X}\right)\,\frac{dX}{X}
+
\oint_{\acycle_{\epsilon_{2,3}}} \left(\ln{Y} + \frac{\alpha_2-\alpha_3}{\beta_2-\beta_3}\,\ln{X}\right)\,\frac{dX}{X} \cr
&& +
\oint_{\acycle_{\epsilon_{3,1}}} \left(\ln{Y} + \frac{\alpha_3-\alpha_1}{\beta_3-\beta_1}\,\ln{X}\right)\,\frac{dX}{X} \cr
&=& -(2\ii\pi)^2 \alpha_{31}\beta_{12} -2\ii\pi \left(\alpha_{12}\ln{X(P_{1-})}+\alpha_{31}\ln{X(P_{2+})}
+\alpha_{23}\ln{X(P_{3-})} \right) \cr
&& +2\ii\pi\alpha_{12}\,\left(\ln{X(P_{1-})}-\ii\pi\beta_{12}\right) \cr
&& +2\ii\pi\alpha_{23}\,\left(\ln{X(P_{3-})}+\ii\pi\beta_{23}\right) \cr
&& +2\ii\pi\alpha_{31}\,\left(\ln{X(P_{2+})}-\ii\pi\beta_{31}\right) \cr
&=& 2\,\pi^2 \left(2\alpha_{31}\beta_{12}  +\alpha_{12}\,\beta_{12}-\alpha_{23}\,\beta_{23}+\alpha_{31}\,\beta_{31}\right) \cr
&=& 2\,\pi^2 \left(\alpha_{23}\beta_{31} -\alpha_{31}\,\beta_{23}\right) \cr
&=& 2\,\pi^2  
\eea
i.e.
\bea
-\ii\pi&=& \frac{1}{2\ii\pi}\,\oint_{\acycle_{\epsilon_{1,2}}} \left(\ln{Y} + \frac{\alpha_1-\alpha_2}{\beta_1-\beta_2}\,\ln{X}\right)\,\frac{dX}{X}
+
\frac{1}{2\ii\pi}\,\oint_{\acycle_{\epsilon_{2,3}}} \left(\ln{Y} + \frac{\alpha_2-\alpha_3}{\beta_2-\beta_3}\,\ln{X}\right)\,\frac{dX}{X} \cr
&& +
\frac{1}{2\ii\pi}\,\oint_{\acycle_{\epsilon_{3,1}}} \left(\ln{Y} + \frac{\alpha_3-\alpha_1}{\beta_3-\beta_1}\,\ln{X}\right)\,\frac{dX}{X} .
\eea

Then, notice that 
\beq
s_{\epsilon_{1,2}}+s_{\epsilon_{2,3}}+s_{\epsilon_{3,1}} = 1.
\eeq
This proves that $t_{\alpha,\beta}$ is well defined for all vertices $(\alpha,\beta)$ of $\hat\Upsilon_\CYX$.
}

\bd[Mirror map, Mirror curve]\label{defmirrormap} The map $\{\td t_k\}_{k=1,\dots,r} \mapsto \{t_k\}_{k=1,\dots,r}$  defined by:
\beq
t_j = \sum_{(\alpha,\beta)\in \hat\Upsilon_\CYX}\,C^{-1}_{j,(\alpha,\beta)}\,t_{\alpha,\beta}
\eeq
where
$t_{0,0}=t_{0,1}=t_{1,0}=0$, and for every half-edge $\epsilon$ (with faces $I_{\epsilon+}$ to the left and $I_{\epsilon-}$ to the right) of the toric graph,
\beq
t_{I_{\epsilon+}}-t_{I_{\epsilon-}}  = \ii\pi s_\epsilon+\frac{1}{2\ii\pi}\oint_{\acycle_\epsilon} \ln{(Y\,\,X^{-\,\frac{\fram_{d,\sigma(\epsilon)}}{\fram_\epsilon}})}\,\,\frac{dX}{X} 
\eeq
is well defined and invertible (for large enough $\td t_k$'s). It is called "the {\bf mirror map}".

In the large radius limit, it satisfies
\beq
t_j = \td t_j+O(1).
\eeq

If the K\"ahler parameters $t_j$'s defining the manifold $\CYX$, and the complex parameters $\td t_j$ defining the plane curve $H(X,Y;\{\td t_j\})$ are related by those relations, then we say that $H(X,Y;\{\td t_j\})=0$ is the mirror curve of $\CYX$\footnote{
The mirror curve defined in this way is not the image of $\CYX$ under mirror symmetry. This image is indeed a 3-fold defined by the equation $ H(X,Y;\{\td t_j\})= u v$.
}.
\ed

\br
Since the $t_j$'s are (up to addition of regularizing terms) periods $\frac{1}{2\ii\pi}\oint \ln Y \,\frac{dX}{X}$, they automatically satisfy some Picard-Fuchs equations, and they are solutions of Picard-Fuchs equations which behave as $t_j = \td t_j+O(1)$ in the large radius limit.
This is how the mirror map is usually defined.

\er



\section{A-model side}

\subsection{Closed Gromov-Witten invariants}

\subsubsection{Definition}

Consider $\CYX$ a toric Calabi--Yau manifold of dimension $3$, with some toric symmetry $T^3$.

\bd
Let $\beta\in H_2(\CYX,\mathbb Z)$. We define the moduli space of "stable maps"
\beq
\overline\modsp_{g,0}(\CYX,\beta)=\{(\Sigma,f)\}
\eeq
where $\Sigma$ is a (possibly nodal) connected oriented Riemann surface of genus $g$, and $f:\Sigma\to \CYX$ is  holomorphic in the interior of $\Sigma$, $f(\Sigma)\in\beta$, and  $f$ is a stable map.
Stability means that if $\Sigma$ is a nodal surface, any sphere component with at most 2 nodal points cannot be collapsed to a point by $f$, and any torus component with no nodal point cannot be collapsed to a point by $f$.
$\overline\modsp_{g,0}(\CYX,\beta)$ is the set of equivalence classes of stable maps modulo isomorphisms.
It is an orbifold, meaning that stable maps with symmetries are quotiented by their automorphism group.
\ed

\smallskip

It is a classic result of algebraic geometry that $\overline\modsp_{g,0}(\CYX,\beta)$ is a compact moduli space (see for example \cite{FultonPandha} for a review on the subject).

It has been shown that it has a fundamental class $1$ and a virtual cycle $[\overline\modsp_{g,0}(\CYX,\beta)]^{\rm vir}$ \cite{LiTian1}, and thus we can define the Gromov--Witten invariants as:
\bd
We define the Gromov--Witten invariants as a formal power series (the formal parameter being $\mathbf Q=\ee{-\mathbf t}$):
\bea
{\cal W}_{g,0}(\CYX,\mathbf t) 
&=& \sum_{\beta\in H_2(\CYX,\mathbb Z)}\,\,\ee{-\mathbf t.\beta}\,\,\, \int_{[\overline\modsp_{g,0}(\CYX,\beta)]^{\rm vir}}\,\, 1.
\eea
where $\mathbf t=(t_1,\dots,t_{b_2(\CYX)})$ is a vector of dimension $b_2(\CYX)= \dim H_2(\CYX,\mathbb Z)$, of complex formal parameters $t_i$.
\ed
We emphasize that this defines a formal series in the parameters $\mathbf Q=\ee{-\mathbf t}$, and every equality we are going to consider, will be an equality of formal series, meaning equality of the coefficients of the series.

\smallskip
We will not study the construction of the virtual cycles in this article, as we will not need it explicitly. We refer the reader interested in more details to the literature \cite{mirrorbook}.

\subsubsection{Localization}

The toric symmetry of $\CYX$ allows to use Attiya-Bott localization and thus
\bea
{\cal W}_{g,0}(\CYX,\mathbf t) 
&=& \sum_{\beta\in H_2(\CYX,\mathbb Z)}\,\,\ee{-\mathbf t.\beta}\,\sum_{\Xi\in \overline\modsp_{g,0}(\CYX,\beta)^{\rm fixed}} \,\,\int_{[\Xi]^{\rm vir}} \frac{1}{e_T(N_\Xi^{\rm vir})}
\eea
where $\overline\modsp_{g,0}(\CYX,\beta)^{\rm fixed}$ is the subset of $\overline\modsp_{g,0}(\CYX,\beta)$ of maps  invariant under the toric action of $\CYX$. For $\Xi\in\overline\modsp_{g,0}(\CYX,\beta)^{\rm fixed}$, the virtual cycle of $\overline\modsp_{g,0}(\CYX,\beta)$ descends to a virtual cycle $[\Xi]^{\rm vir}$, and to a normal bundle $N_\Xi^{\rm vir}$, and $e_T$ is the equivariant Euler class.

This construction is well studied in the litterature, and we shall now briefly describe the fixed locus to motivate the definition of the relative Gromov-Witten invariants given in section \ref{secopenGW}.

\subsubsection{Torus orbits in $\CYX$}

Before describing the contribution to the localization formula, it is useful to recall the torus action on $\CYX$.
Remember that we have defined the moment map
\beq
\pi:\,\,\CYX \to \mathbb R^3 \quad , \quad \mapsto (|X_1|^2,|X_2|^2,|X_3|^2).
\eeq
It sends $\CYX$ to the polyhedra.

\smallskip
The torus action consists in shifting the 3 independent angles, it doesn't change the radius. Hence, it keeps the polyhedra fixed. Let us now study the orbits of the points of $\CYX$ under this action:

- For a generic point $p\in\CYX$, $\pi(p)$ is in the bulk of the polyhedra, so the 3 independent radius are non--vanishing, and the orbit of $p$ under the torus action is a 3-dimensional torus.

- For a point $p$ such that $\pi(p)$ is on a face of the polyhedra, there are only 2 non--vanishing radius, and thus an orbit of the torus action is a 2-dimensional torus.

- For a point $p$ such that $\pi(p)$ is  on an edge of the polyhedra, there is only 1 non--vanishing radius, and thus an orbit of the torus action is a circle.

- For a point $p$ such that $\pi(p)$ is  a vertex of the polyhedra, there is no non--vanishing radius, and thus an orbit of the torus action is a point, i.e. vertices of the torus graph correspond to fixed points of the torus action.

\medskip

Notice that:

- a fixed point i.e. $\pi^{-1}({\rm vetex})$  is a 0-dimensional manifold. For a vertex $\sigma$ of $\Upsilon_\CYX$, we denote $\sigma  = \pi^{-1}(\sigma)$ the corresponding fixed point of $\CYX$;

- $\pi^{-1}({\rm edge})$ is a 2-dimensional manifold (a 1-dimensional family of circles). If the edge is a closed edge, this is a sphere with 2 punctures (the 2 punctures being the 2 fixed points), and if the edge is open, this is a half-sphere with one puncture. For a closed edge $e = (\sigma,\sigma')$ (resp. an open edge $\epsilon= (\sigma,e)$), we denote $\tau_{(\sigma,\sigma')} = \pi^{-1}(e)$ (resp. $\tau_{(\sigma,e)} = \pi^{-1}(\epsilon)$) the corresponding sphere (resp. half-sphere).

- $\pi^{-1}({\rm face})$ is a 4-dimensional manifold (a 2-dimensional family of 2-dimensional tori).

- $\pi^{-1}({\rm bulk})$ is a 6-dimensional manifold (a 3-dimensional family of 3-dimensional tori).

\smallskip

In other words, the only fixed locus of $\CYX$ which are manifolds of dimension at most 2, correspond to either edges or vertices of the toric graph.

\medskip

\bd
Let $\CYX^{\rm fixed}$ be the subset of $\CYX$:
\beq
\CYX^{\rm fixed}=\pi^{-1}({\rm edges\,\cup\,vertices}).
\eeq
 $\CYX^{\rm fixed}$ is a circle bundle over the toric graph.

\begin{itemize} 
\item To each vertex $\sigma$ of the toric graph corresponds a circle $\sigma$ of vanishing radius (a point);

\item to each compact edge $e = (\sigma,\sigma')$ corresponds a sphere $\tau_{(\sigma,\sigma')}$;

\item to each non--compact edge $\epsilon = (\sigma,e)$ corresponds a half--sphere $\tau_{(\sigma,e)}$.
\end{itemize}

\begin{figure}[t]
\centering
$$\includegraphics[height=5.5cm]{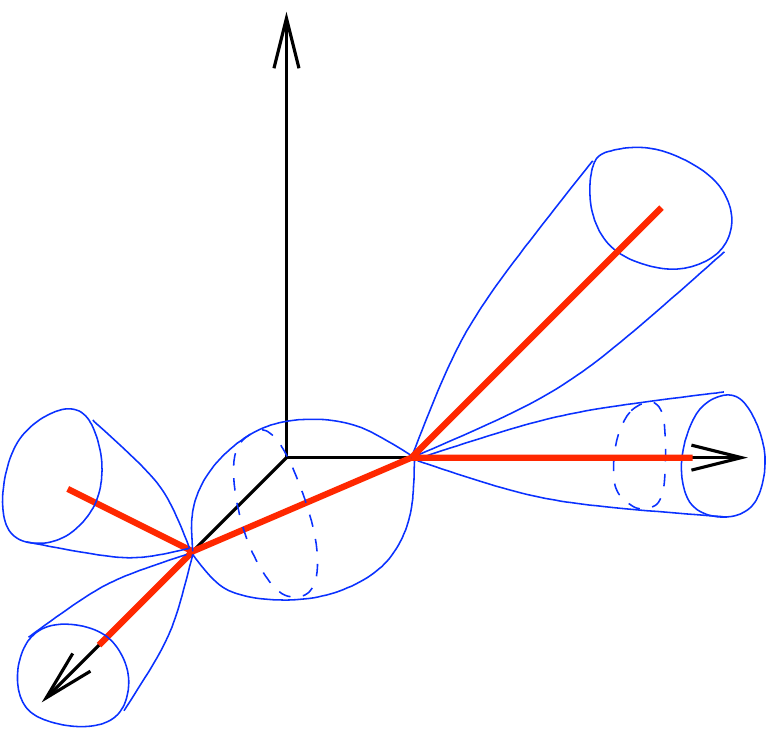}  $$
\caption{The fixed locus $\CYX^{\rm fixed}=\pi^{-1}({\rm edges\,\cup\,vertices})$ is a circle bundle over the toric graph. Vertices correpond to points, compact edges  correspond to spheres, and non compact edges correspond to half-spheres.
\label{figtoricgraphfixed}}
\end{figure}

\smallskip
See fig.\ref{figtoricgraphfixed} for the resolved conifold example.
\ed

\subsubsection{Framing\label{secframingglobal}}

There are several possible torus actions in $\CYX$, obtained by changing the cannonical basis of $\mathbb C^3$ by an $U(3)$ change of coordinate. 
So far we have choosen the basis $X_1,X_2,X_3$, but we can take $(X'_1,X'_2,X'_3)=U\,(X_1,X_2,X_3)$ where $U\in U(3)$, so that the symplectic form is conserved:
\beq
\om = \sum_{i=1}^{3} dX_i\wedge d\bar X_i = \sum_{i=1}^{3} dX'_i\wedge d\bar X'_i.
\eeq
The torus action shifts the angles $\theta_1,\theta_2,\theta_3$ where $X_i=|X_i|\,\ee{i\theta_i}$, but we could also shift the $\theta'_i$'s where $X'_i=|X'_i|\,\ee{i\theta'_i}$.

\medskip

Here, we shall choose a basis of $X_i$'s such that  the torus action leaves the $S^1$ circle  $L\cap\{$toric graph$\}$ invariant. It depends on our choice of $L$.
Therefore, up to a change of the variables $\{X_i\}_{i=1,\dots,r+3}$ we shall assume that $L$ corresponds to $|X_i|^2=0$ for $i\neq 1$. In other words $L$ is the line along the coordinate $X_1$.
This can always be achieved by a linear transformation of the type:
\beq
|X_1|^2\to |X_1|^2 + f |X_2|^2 + f' |X_3|^2.
\eeq
Notice that for the mirror, it corresponds to
\beq
Y_1\to Y_1\,Y_2^f\,Y_3^{f'}
\eeq
and if we choose the patch $Y_3=1$ that means
\beq
X\to X\, Y^f,
\eeq
i.e. it is a framing transformation.

\subsubsection{Fixed loci}\label{secfixedlociclose}

For a map $[f:\Sigma \to \CYX]$ to belong to  $\overline\modsp_{g,0}(\CYX,\beta)^{\rm fixed}$, it has to map $\Sigma$ to 
$f(\Sigma) \subset \CYX^{\rm fixed}$.

Let ${\cal O}_i$ be an irreducible component of $\Sigma$. It's image under $f$ enters one of these three cases\footnote{The interested reader can find proofs and references in \cite{LiuReview}. We only want to motivate some of the forthcoming definition in this part.}:
\begin{itemize}

\item Either it is collapsed, i.e. it is mapped to a 0-dimensional component of $\CYX^{\rm fixed}$. This means that $\left. f\right|_{{\cal O}_i}$ is constant and maps all the points of ${\cal O}_i$ to a vertex $\sigma({\cal O}_i)$ of the polyhedra. This component can have an arbitrary topology as long as it satisfies the stability condition: its genus $g({\cal O}_i)$ and the number of nodal points $n({\cal O}_i)$ that it contains must satisfy
\beq
2 -2g({\cal O}_i) - n({\cal O}_i) <0.
\eeq

\item Either it is a sphere with two nodal points and is mapped to a compact 2-dimensional fixed submanifold of $\CYX$, that is to say, to an inner edge $e$ of the polyhedra. The map  $\left. f\right|_{{\cal O}_i}$ is then a map between two spheres mapping the nodal points to the two vertices to which $e({\cal O}_i)$ is incident. This map can be of arbitrary degree $d({\cal O}_i)\in \mathbb N^*$ and is totaly ramified at the two nodal points.

\item Either it is a sphere with only one nodal point and is mapped to a compact 2-dimensional fixed submanifold of $\CYX$, that is to say, to an inner edge $e$ of the polyhedra. The nodal point is mapped to one of the two vertices adjacent to the inner edge, therefore there exists a smooth point on ${\cal O}_i$ which is mapped to the other vertex.
 The map  $\left. f\right|_{{\cal O}_i}$ is then a map between two spheres mapping the nodal point to a vertex, and the other smooth point to the other vertex. This map can be of arbitrary degree $d({\cal O}_i)\in \mathbb N^*$ and is totaly ramified at the two fixed points.

\end{itemize}
A map $(\Sigma,f) \in \overline\modsp_{g,0}(\CYX,\beta)^{\rm fixed}$ is thus the union of such irreducible components $\Sigma = \bigcup_i {\cal O}_i$ of respective genus $g\left({\cal O}_i\right)$ with respectively $n\left({\cal O}_i\right)$ nodal points and a map $f$ such that its restrictions $f_i = f_{{\cal O}_i}$ of degree $d_i$ satisfy the constraints:
\begin{itemize}

\item If ${\cal O}_i$ is stable, i.e. if $2-2g\left({\cal O}_i\right)-n\left({\cal O}_i\right)<0$, then $d_i=0$ and $f_i$ is the constant map mapping all the points of  ${\cal O}_i$ to a vertex $\sigma\left({\cal O}_i\right) = \sigma_{l(i)}$ of the Toric graph. The label $l(i)$ tells which fixed point of $\CYX$,  ${\cal O}_i$ is mapped to.

\item If ${\cal O}_i$ is unstable, i.e. if $g\left({\cal O}_i\right) = 0$ and $n\left({\cal O}_i\right) \in \left\{1,2\right\}$, then $f_i$ has an arbitrary degree $d_i = d\left({\cal O}_i\right)>0$ mapping the sphere ${\cal O}_i \simeq \mathbb{P}^1$ to one of the 1-dimensional fixed submanifolds of $\CYX$, $f_i\left({\cal O}_i\right) = \tau_{e_{l(i)}}$. The label $l(i)$ tells to which 1-dimensional fixed locus of $\CYX$,  ${\cal O}_i$ is mapped to.

\item The intersection of two irreducible components is a nodal point which is mapped to a fixed point of $\CYX$, i.e. to a vertex of the toric graph;

\item The genus of $\Sigma$ is equal to $g$;

\item The image of $\Sigma$ belongs to the class $\beta$, which translates into
\beq\label{eqbetadiedges}
\sum_{{\cal O}_i \,  unstable} d_i \, \left[\tau_{e_{l(i)}}\right]  =  \beta.
\eeq

\end{itemize}

\subsection{Graphs for the fixed locus}

A good way to encode such a fixed map is through a map between $\overline\modsp_{g,0}(\CYX,\beta)^{\rm fixed}$ and a set of graphs. For later convenience, we now introduce a set of graphs which is slightly larger than the one requested for describing $\overline\modsp_{g,0}(\CYX,\beta)^{\rm fixed}$.

\subsubsection{Graphs}

This leads us to define the following set of graphs ${\cal G}_{g,n}$ 
as:
\bd\label{defraphs}
Let $g,n$ be non-negative integers.
${\cal G}_{g,n}$ is the set of graphs defined as follows:
$G\in {\cal G}_{g,n}$ if $G$ is a connected graph, made of vertices and half--edges, each closed edge is a pair of half-edges, and:

- each vertex $v$ has a "color" $\sigma_v$ which is a vertex of $\Upsilon_\CYX$, a "genus" $g_v\in \mathbb N $, and a "valence" $n_v=\#$ of half--edges incident to $v$.
We denote 
\beq
E_v=\{h\,|\,h\,=\,{\rm half-edge\,incident\,to}\, v\}
\virg
\# E_v=n_v.
\eeq

- each half-edge $h\in E_v$ carries a "degree" $d_{h}\in\mathbb N$ and a "color" $\epsilon_h$ which is an half-edge of $\Upsilon_\CYX$, incident to $\sigma_v$ :
\beq
\epsilon_h \in \{{\rm half-edges\,of}\,\Upsilon_\CYX\,\,{\rm incident\,to}\,\sigma_{v(h)}\}
\qquad {\rm where}\, v(h)\, {\rm is\, the\, vertex\, adjacent\, to}\, h,
\eeq
which implies that for a given $h$, $\epsilon_h$ can take only 3 values (there are 3 half-edges incident to a vertex in $\Upsilon_\CYX$).

- t2here are exactly $n$ open half--edges, they are labeled
$h_1,\dots,h_n$.

- each closed edge $e=(h_+,h_-)$ is made of two half--edges.

- we impose to have
\beq
2-2g-n = \sum_v (2-2g_v-n_v) .
\eeq

\medskip

$\bullet$ We define ${\cal G}^{\rm stable}_{g,n}\subset {\cal G}_{g,n}$ the subset of ${\cal G}_{g,n}$
with the additional condition that $G\in {\cal G}^{\rm stable}_{g,n}$ iff
\beq
\forall \, v\, , \quad 2-2g_v-n_v<0,
\eeq
in other words $(g_v,n_v)\neq (0,1)$ and $(g_v,n_v)\neq (0,2)$.

\medskip

$\bullet$ We also define $\td {\cal G}_{g,n}$ (resp. $\td {\cal G}^{\rm stable}_{g,n}$) as the same set of graphs, but without degree labels $d_{h}$ attached to half-edges.

\ed

\subsubsection{Fixed map and graphs}


Let $\Xi=(\Sigma,f)$ be a fixed stable map. 
Notice that only the irreducible components of $\Sigma$ which are spheres with 1 or 2 nodal points, are not mapped to a fixed point of $\CYX^{\rm fixed}$, and they are mapped to spheres of $\CYX^{\rm fixed}$.

Therefore we define:
\bd
Let $\Xi=(\Sigma,f)\in \overline\modsp_{g,0}(X,\beta)^{{\rm fixed}}$ be a fixed stable map.
Let us define:
\beq
\Sigma_{{\rm vertices}} = f^{-1}({\rm fixed\, points\,in\,}\CYX^{\rm fixed})
\quad , \qquad
\Sigma_{{\rm edges}}  = \Sigma\setminus \Sigma_{{\rm vertices}}. 
\eeq
Let us write $\Sigma_{{\rm vertices}}$ and $\Sigma_{{\rm edges}}$ as the disjoint union of their connected components:
\beq
\Sigma_{{\rm vertices}}  = \uplus_i\,\, \hat{\cal O}_i 
\quad , \qquad
\Sigma_{{\rm edges}}  = \uplus_i\,\, {\cal O}_i .
\eeq
Each ${\cal O}_i$ is a sphere with 2 point removed, i.e. it is a cylinder homeomorphic to $\mathbb C^*$, and each $\hat{{\cal O}_i }$ is either a nodal surface, or it is an isolated nodal point, or an isolated smooth point on a sphere with only one other nodal point.
\smallskip

To $\Xi=(\Sigma,f)$, we associate a graph of ${\cal G}_{g,0}$ as follows:

- to each ${{\cal O}_i }\subset \Sigma_{\rm edges}$ we associate an edge $e_i$. The edge $e_i$  is made of two half--edges $h_{i+}$ and $h_{i-}$ corresponding to the two nodal points of ${\cal O}_i$. The edge $e_i$, and thus the two half--edges $h_{i\pm}$  carry the degree $d_{e_i}$ of the map $f:{\cal O}_i\to \CYX^{\rm fixed}$. Each half--edge $h_{i\pm}$ carries a label $\epsilon_{h_{i\pm}}$ equal to the label of the corresponding half--edge in the toric graph $\Upsilon_\CYX$.

- to each $\hat{{\cal O}_i }\subset \Sigma_{\rm vertices}$ we associate a vertex $v_i$. The vertex carries a label $\sigma_{v_i}$ which is the vertex of the toric graph corresponding to the fixed point $f(\hat{{\cal O}_i })$. It carries a genus $g_{v_i}=$genus of $\hat{{\cal O}_i }$ (and we set $g_{v_i}=0$ if $\hat{{\cal O}_i }$ is a point).
It carries a valence $n_{v_i}=\#(\overline\Sigma_{\rm edges}\cap\hat{{\cal O}_i })$.

- the incidence relations are determined as follows: an edge $e_i$ is adjacent to a vertex $v_j$ iff $\overline{\cal O}_i\cap\hat{{\cal O}_j}\neq \emptyset$.
For each vertex $v$ we define $E_v=\{{\rm half-edges\, adjacent\,to\,}v\}$.

\ed

\begin{figure}[t]
\centering
$$\includegraphics[width=12.5cm]{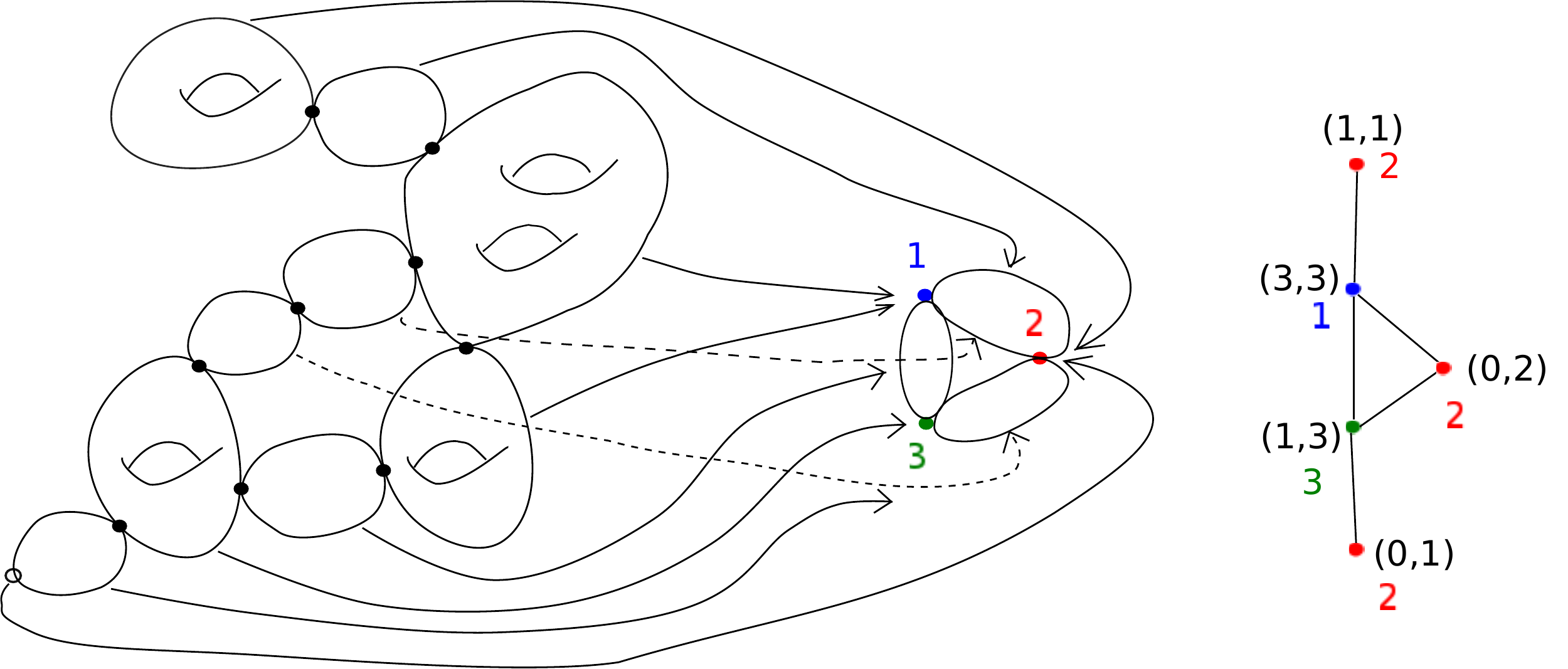}  $$
\caption{In this example, $\CYX$ is local $\mathbb P^2$, whose toric graph has 3 vertices, labeled $(1), (2), (3)$. $(\Sigma,f)\in \overline\modsp_{g,0}(X,\beta)^{{\rm fixed}}$ is a stable nodal map, invariant under the torus action. Each sphere component ${\cal O}_i$ with 1 or 2 nodal point is mapped by $f$ to an invariant sphere of $\CYX^{{\rm fixed}}$.
Each higher genus component $\hat{\cal O}_i$ or each sphere with at least 3 nodal points  is mapped to a fixed point of $\CYX$, with label $1,2$ or $3$.
Each nodal point is also mapped to a fixed point of $\CYX$.
For sphere components with only one nodal point (see the bottom left sphere in this example) there is also a smooth point mapped to a fixed point.
To each sphere component ${\cal O}_i$ of $\Sigma$ with 1 or 2 nodal point we associate an edge.
To each connected component $\hat{\cal O}_i$ of the preimage of a fixed point (this can be either a stable nodal surface, or an isolated nodal point, or an isolated smooth point) we associate a vertex, to which we associate the label of the fixed point and the pair $(g,n)$ corresponding to the total genus $g$ of $\hat{\cal O}_i$ (and we set $g=0$ if $\hat{\cal O}_i$ is a point), and where $n$ is the number of edges adjacent to it.
\label{figfixedmap1}}
\end{figure}

See fig.\ref{figfixedmap1} for an example.

We have defined those graphs so that it defines 
 an injective  orbifold morphism\footnote{For a precise definition as well as the description of the image of this morphism see \cite{LiuReview}.}:
\bea
\overline\modsp_{g,0}(X,\beta)^{\rm fixed} 
&\to & \mathop{{\oplus}}_{G\in{\cal G}_{g,0}}\,\,\prod_{v={\rm vertices}(G)}\,\,{\overline\modsp}_{g_v,n_v} \quad \times \mathbb N^{\#\,{\rm half-edges}(G)} \cr
(\Sigma,f) &\mapsto &
\mathop{{\oplus}}_{i={\rm vertices}} \hat{\cal O}_i
\quad \mathop{{\oplus}}_{h={\rm half-edges}} d_h.
\eea

\br
Notice that this application  is not surjective. Indeed not all graphs $G\in {\cal G}_{g,0}$ satisfy that the degrees of closed half-edges $d_{(v,e)}$ be such that they combine to be the homology class $\beta$.
Moreover, the image $f(\Sigma)\subset \CYX^{\rm fixed}$ must be such that neighboring vertices $v,v'$ can only have labels $\sigma_v,\sigma_{v'}$ which are adjacent in the toric graph, and if $h$ and $h'$ are the two half edges of an edge $e=(h,h')$ we should have that $(\epsilon_h,\epsilon_{h'})$ be an edge of $\Upsilon_\CYX$, and we should have $d_h=d_{h'}$.
As we have defined them, not all graphs in ${\cal G}_{g,n}$ satisfy those conditions.
In \cite{LiJun2004} the authors prefer to define a smaller set of graphs containing only the graphs which can be images of fixed stable maps.

Here, we prefer to define a larger set of graphs ${\cal G}_{g,n}$
 in order to make the link with the topological recursion formalism in the B-model side, later in section \ref{omgngraphs}.

Our strategy will be to assign weights to graphs, in order that unwanted graphs receive a vanishing weight.
\er

\subsubsection{Gromov-Witten invariants and weighted sum over graphs}

Since this morphism is injective, the sum over fixed stable maps can be translated into a sum over $\mathop{{\oplus}}_{G\in{\cal G}_{g,0}}\,\,\prod_{v={\rm vertices}(G)}\,\,{\overline\modsp}_{g_v,n_v} \quad \times \mathbb N^{\#\,{\rm half-edges}(G)}$, i.e. over decorated graphs.

It can be shown that the measure on this set of graphs factorises, up to a symmetry factor, into a product of measures on the vertices and edges forming the graph \cite{LiuReview}:

\bt [Localization formula \cite{LiJun2004,Diaco}]\label{thlocclosedGW}

The Gromov--Witten invariants of a calabi-Yau $\CYX$ can be written as a sum over graphs, weighted by products of weights $H_{g,n;\sigma}(\{k_1,\dots,k_n\})$ associated to vertices and weights $F_{\epsilon,\epsilon'}(k,k')$ associated to edges:
\bea\label{eqWg0localizationgraphs}
{\cal W}_{g,0}(\CYX,{\bf t})
&=& \sum_{G\in {\cal G}_{g,0}} \frac{(-1)^n}{\#{\rm Aut}(G)}\, 
\prod_{v}\, H_{g_v,n_v;\sigma_v}(\{d_{h}/\fram_{\epsilon_h}\}_{h\in E_v})  \cr
&& \prod_{{\rm edges}\,\,e=(h_+(e),h_-(e))} \, F_{\epsilon_{h_+(e)},\epsilon_{h_-(e)}}(d_{h_+(e)},d_{h_-(e)}) . \,\,\,\cr
\eea
where

$\bullet$ if $2-2g-n<0$:
\bea\label{defHgn}
H_{g,n,\sigma}(k_1,\dots,k_n)\,
&=& 2^{g-1}\,\left(\frac{1}{2}\,\prod_{\epsilon={\rm half-edges\,of}\,\sigma} \fram_\epsilon\right)^{g-1+\frac{n}{2}}\, \cr
&& \int_{\bar\modsp_{g,n}}\, \prod_{\epsilon={\rm half-edges\,of}\,\sigma} \CL_{\rm Hodge}(\fram_{\epsilon}) \prod_{i=1}^n \frac{\gamma_{\fram_\sigma}(k_i)}{1-k_i\,\psi_i}
\eea
\beq\label{eqdefgamma}
{\rm where}\qquad \quad \gamma_{\fram}(k) = \frac{1}{\sqrt{\pi\,k}}\,\,\frac{\hat\Gamma(k\,(\fram_a+\fram_b))}{\hat\Gamma(k\,\fram_a)\hat\Gamma(k\,\fram_b)}\,
\eeq
and where $\hat \Gamma$ is the "regularized" $\Gamma$ function defined as
\beq\label{defhatGamma}
\hat\Gamma(u) = \frac{e^u\,\sqrt{u}}{u^u\,\sqrt{2\pi}}\Gamma(u)
= \ee{{\displaystyle \sum_{k=1}^\infty}  \frac{\Ber_{2k}}{2k(2k-1)}\,u^{1-2k} }
\eeq
($\Gamma(u)$ is the Gamma function and $\Ber_k$ is the $k^{\rm th}$ Bernoulli number),
 and $\CL_{\rm Hodge}(f)$ is the Hodge class in $\overline\modsp_{g,n}$ and $\psi_i$ is the first Chern class of the cotangent bundle at the $i^{\rm th}$ marked point.

The same formula applies to $(g,n)=(0,1)$ and $(0,2)$ if we define

$\bullet$ if $(g,n)=(0,1)$:
\beq
\int_{\bar\modsp_{0,1}}\, \prod_{\epsilon={\rm half-edges\,of}\,\sigma} \CL_{\rm Hodge}(\fram_{\epsilon}) \frac{1}{1-k\,\psi} \stackrel{{\rm def}}{=} \frac{1}{k^2}
\eeq

$\bullet$ if $(g,n)=(0,2)$:
\beq
\int_{\bar\modsp_{0,2}}\, \prod_{\epsilon={\rm half-edges\,of}\,\sigma} \CL_{\rm Hodge}(\fram_{\epsilon}) \frac{1}{1-k_1\,\psi_1}\,\frac{1}{1-k_2\,\psi_2} \stackrel{{\rm def}}{=}  \frac{1}{k_1+k_2}.
\eeq

$\bullet$ and
\beq\label{defFeTm}
F_{\epsilon_+,\epsilon_-}(d,d') = A_{\epsilon_+,\epsilon_-}\,\delta_{d,d'}\,\,\frac{d}{f_{\epsilon_+}^2}\,\ee{-d\,\frac{\arond_{\sigma(\epsilon_+)}-\arond_{\sigma(\epsilon_-)}}{\fram_{\epsilon_+}}}\,
\eeq
where $A_{\epsilon_+,\epsilon_-}=1$ if there exists an edge $e=(\epsilon_+,\epsilon_-)$ in the toric graph and zero otherwise, and $\arond_\sigma$ is the projection along the axis $|X_1|^2$ of the vertex $\sigma$ of the toric graph, defined in def. \ref{deftoricgraph} (it is a linear combination of the $t_i$'s).
Notice that if $A_{\epsilon_+,\epsilon_-}=1$, we have $\fram_{\epsilon_+}= - \fram_{\epsilon_-}$, and thus
\beq
F_{\epsilon_+,\epsilon_-}(d,d') = F_{\epsilon_-,\epsilon_+}(d',d).
\eeq

\et

{\bf Sketch of a proof:}

Intuitively, this decomposition comes from the fact that, once a graph is fixed, the enumeration of corresponding fixed maps can be performed independently for each irreducible component of $\Sigma$. 

The functions $H_{g,n;\sigma}(\{k_1,\dots,k_n\})$ correspond to vertices, i.e. to constant maps $f:\hat{\cal O}_i \to {\rm fixed\,point\,in\,}\CYX$, and can thus be computed only with the knowledge of a vicinity of a fixed point of $\CYX$, and in the vicinity of a fixed point, $\CYX$ can be replaced by $\mathbb C^3$, thus $H_{g,n;\sigma}(\{k_1,\dots,k_n\})$ are related to the Gromov-Witten invariants of $\mathbb C^3$, i.e. to the topological vertex, and are computed by the Mari\~no--Vafa formula \cite{MV01,Liu2003a} as triple Hodge integrals, and result in \eq{defHgn}.

For edges, we already mentioned that graphs in ${\cal G}_{g,n}$ which are not images of fixed stable maps, should receive a vanishing weight, so in particular $F_{\epsilon_-,\epsilon_+}(d',d)$ must vanish if $\epsilon_\pm$ are not the two half--edges forming an edge of $\Upsilon_\CYX$, and also it must vanish if $d\neq d'$, so it must be proportional to $A_{\epsilon_+,\epsilon_-}\,\delta_{d,d'}$.
The weight $\ee{-d\,\frac{\arond_{\sigma(\epsilon_+)}-\arond_{\sigma(\epsilon_-)}}{\fram_{\epsilon_+}}}$
is such that thanks to \eqref{eqbetadiedges}
\beq
\prod_{{\rm edges}\,e=(h_+,h_-)}
\ee{-d_{e}\,\frac{\arond_{\sigma(\epsilon_{h_+})}-\arond_{\sigma(\epsilon_{h_-})}}{\fram_{\epsilon_{h_+}}}}
= \ee{-t.\beta}.
\eeq
The factor $\prod_{{\rm edges}\,e} d_e/\fram_e^2$ is a symmetry factor.

All those factors are encoded by $F_{\epsilon,\epsilon'}(k,k')$ defined  in  \eq{defFeTm}.
$\square$

\subsection{Open Gromov-Witten invariants}
\label{secopenGW}

We wish to generalize this definition of Gromov-Witten invariants to the enumeration of open surfaces whose boundaries are mapped to Lagrangian sub-manifolds. For this purpose, we define:

\bd
For $\beta\in H_2(\CYX,L,\mathbb Z)$ (relative homology class of 2-chains in $\CYX$ whose boundaries lie on $L$) and $\vec w=(w_1,\dots,w_n)$ with $w_i\in H_1(L,\mathbb Z)$, and such that ${\displaystyle \sum_i} w_i =\d \beta$, we define the moduli space
\beq
\overline\modsp_{g,n}(\CYX,L,\beta,\vec  w) = \{(\Sigma,f)\}
\eeq
where $\Sigma$ is a (possibly nodal) connected oriented Riemann surface of genus $g$ with $n$ circle boundaries labeled  $\d_1\Sigma,\dots,\d_n\Sigma$, and $f:\Sigma\to \CYX$ is a holomorphic stable map, such that $f(\d_i\Sigma)\subset L$, and $f$ is a stable map.
Stability means that if $\Sigma$ is a nodal surface, any sphere component with at most 2 nodal or marked points or boundaries cannot be collapsed to a point by $f$, and any torus component with no nodal point or boundary cannot be collapsed to a point by $f$.
And $f(\Sigma)\in\beta$ and $f(\d_i\Sigma)\in w_i$.
Again, $\overline\modsp_{g,n}(\CYX,L,\beta,\vec w)$ is the set of equivalence classes of stable maps modulo isomorphisms.
It is an orbifold, meaning that stable maps with symmetries are quotiented by their automorphism group.
\ed

\medskip

When $n>0$, Katz and Liu \cite{KatzSheldon2001} have constructed a virtual class and virtual cycle in $\overline\modsp_{g,n}(\CYX,L,\beta,\vec  w)$.
Their method is based on the fact that $L$ is the fixed locus of an antiholomorphic involution in $\CYX$, and thus by "doubling" $\Sigma$ (i.e. extending $\Sigma$ to a larger closed Riemann surface by Schwarz principle across the boundaries), they embed $\overline\modsp_{g,n}(\CYX,L,\beta,\vec  w)$ in  a closed moduli space $\overline\modsp_{g',0}(\CYX,\beta')$, where the virtual cycle and class are well known, and they take the restriction to the part invariant under the antiholomorphic involution.  
This allows them to show that there is a localization formula, which we use below.
%

\smallskip

For our purpose here, we shall start directly from the localization formula of \cite{Liu2003a}, and which is the straightforward generalization of theorem \ref{thlocclosedGW}:

\bd\label{thlocalizationgraphs}
For $n\geq 0$ we define the open Gromov--Witten invariants as
\bea\label{eqGWlocalizationdef}
{\cal W}_{g,n}(\CYX,L,\mathbf t;x_1,\dots,x_n)
&=& \sum_{G\in {\cal G}_{g,n}} \frac{(-1)^n}{\#{\rm Aut}(G)}\, 
\prod_{v}\, H_{g_v,n_v;\sigma_v}(\{d_{h}/\fram_{\epsilon_h}\}_{h\in E_v})  \cr
&& \prod_{{\rm closed\,edges}\,\,e=(h_+(e),h_-(e))} \, F_{\epsilon_{h_+(e)},\epsilon_{h_-(e)}}(d_{h_+(e)},d_{h_-(e)}) \,\,\,\cr
&& 
\qquad  \prod_{{\rm open\,half\,edges}\,\,h_i,\,\, i=1,\dots,n} \,\frac{\delta_{\epsilon_{h_i},\epsilon_i}\,d_{h_i}}{\fram_{\epsilon_{h_i}}^2}\,\ee{-\frac{d_{h_i}}{\fram_{\epsilon_i}}\, (x_i-\arond_{\sigma(\epsilon_i)})} ,\cr
\eea
where we recall that $\epsilon_i$ is the half--edge of $\Upsilon_\CYX$ on which $L$ ends, $\fram_{\epsilon_i}$ is its framing, and $\arond_{\sigma(\epsilon_i)}$ is the position of the vertex of $\Upsilon_\CYX$ adjacent to the half--edge $\epsilon_i$ on which $L$ ends. 
The factors $H_{g,n,\sigma}(k_1,\dots,k_n)$ and $F_{\epsilon_+,\epsilon_-}(d,d')$ are defined in \eq{defHgn} and \eq{defFeTm}.
\ed

\subsubsection{Heuristic origin of this definition}

One would like to define the Open Gromov--Witten invariants
as formal power series (the formal parameters being $\mathbf Q=\ee{-\mathbf t}$ and $X_i=\ee{-x_i}$)  computing the integral of the fundamental class $1$ over the virtual fundamental cycle of $[\modsp_{g,n}(\CYX,L,\beta,\vec w)]^{\rm vir}$:
\bea\label{defopenGW}
{\cal W}_{g,n}(\CYX,L,\mathbf t;x_1,\dots,x_n) 
&"="& \sum_{\beta,w_1,\dots,w_n}\,\,\ee{-\mathbf t.\beta}\,\prod_{i=1}^n \ee{-\,\frac{w_i}{\fram_{\epsilon_i}}\,x_i}\,\, \int_{[\overline\modsp_{g,n}(\CYX,L,\beta,\vec w)]^{\rm vir}}\,\, 1  \cr
\eea
where $\fram_{\epsilon_i}$ is the framing of the half-edge of $\Upsilon_\CYX$ on which $L$ ends (see def. \ref{defframedges}).

\smallskip
Since this virtual cycle is not so well understood, we prefer to attempt to define  the Open Gromov--Witten invariants through the localization formula:
\bea\label{defopenGWloc}
{\cal W}_{g,n}(\CYX,L,\mathbf t;x_1,\dots,x_n) 
&"="& \sum_{\beta\in H_2(\CYX,L,\mathbb Z)}\,
\sum_{w_i\in H_1(L,\mathbb Z)}\,\ee{-\mathbf t.\beta}\,\ee{-{\displaystyle \sum_{i=1}^n} \frac{w_i}{\fram_{\epsilon_i}} x_i}\sum_{\Xi\in \overline\modsp_{g,n}(\CYX,L,\beta,w)^{\rm fixed}} \,\cr
&& \qquad \quad \,\int_{[\Xi]^{\rm vir}} \frac{1}{e_T(N_\Xi^{\rm vir})} \cr 
\eea
where $\overline\modsp_{g,n}(\CYX,L,\beta,w)^{\rm fixed}$ is the fixed locus of the moduli space of stable maps under the torus action under study.
In order to give a meaning to that definition, we need to describe the fixed locus in more details.

\subsubsection{Fixed maps}

The fixed locus $\overline\modsp^{\rm fixed}_{g,n}(\CYX,L,\beta,\om)$ is well studied and well known \cite{LiuReview,Aganagic2004,mirrorbook}. Let us  describe it briefly. It is just a generalization of the $n=0$ case studied in the preceding section \ref{secfixedlociclose} obtained by including maps from half-spheres components of $\Sigma$ to fixed half-spheres $\tau_{(\sigma,e)}$ of $\CYX$.

\medskip

%
%
%
%
%
%
%
%
%
%
%
%
%
%

\medskip

Let $\Xi=(\Sigma,f) \in  \overline\modsp_{g,n}(\CYX,L,\beta,w)^{\rm fixed}$ be a fixed stable map.

Since the boundaries of $\Sigma$ have to be sent to $L$, this means that they must be sent to $L\cap \CYX^{\rm fixed}$ which is a circle of fixed radius, and
\beq
\pi(f(\d_i \Sigma)) = {\rm point\,on\,the\,toric\,graph}\,=\,L\cap{\rm toric\,graph}.
\eeq

Also, we see that the image of any irreducible component of $\Sigma$ must be either a point or a sphere with 2 fixed points, or a half-sphere (a disc bounded by the circle $L\cap \CYX^{\rm fixed}$)  with 1-fixed point (this last case only if the component contains a boundary).

In particular this implies that a component of $\Sigma$ containing a boundary is never collapsed to a point, and therefore it can be only a disc with only 1 nodal point and one boundary.
Therefore, each boundary is on a disc component.

Appart from those disc components, which are necessarily sent to the half--edge of the toric graph on which $L$ ends, all the other components are mapped by $f$ in the same way as described in section \ref{secfixedlociclose}.

%
%
%
%

\subsubsection{Fixed maps and graphs}

\begin{figure}[t]
\centering
\label{figfixedcurve1}
$$
\includegraphics[height=4cm]{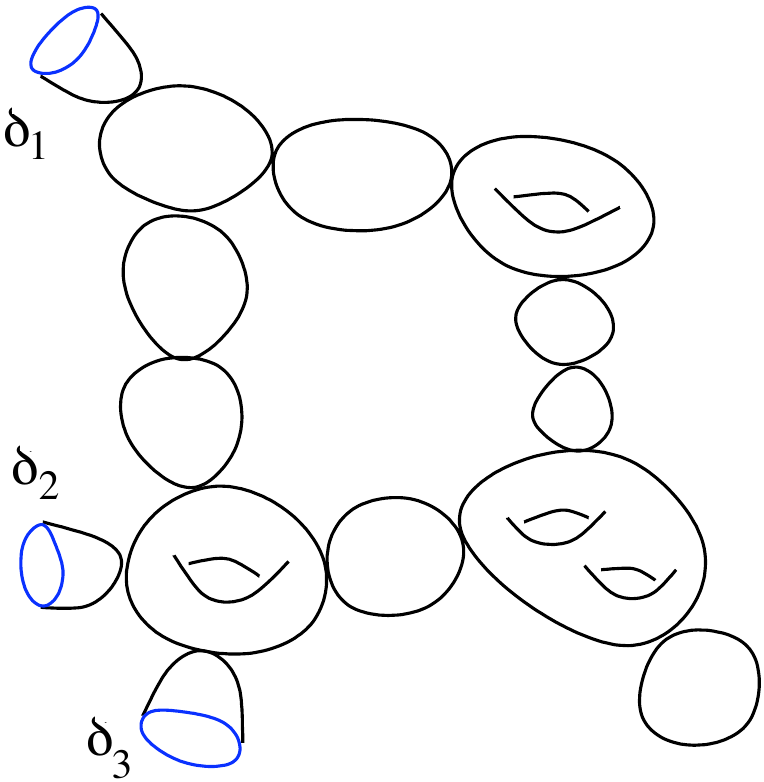} 
\qquad 
\includegraphics[height=4cm]{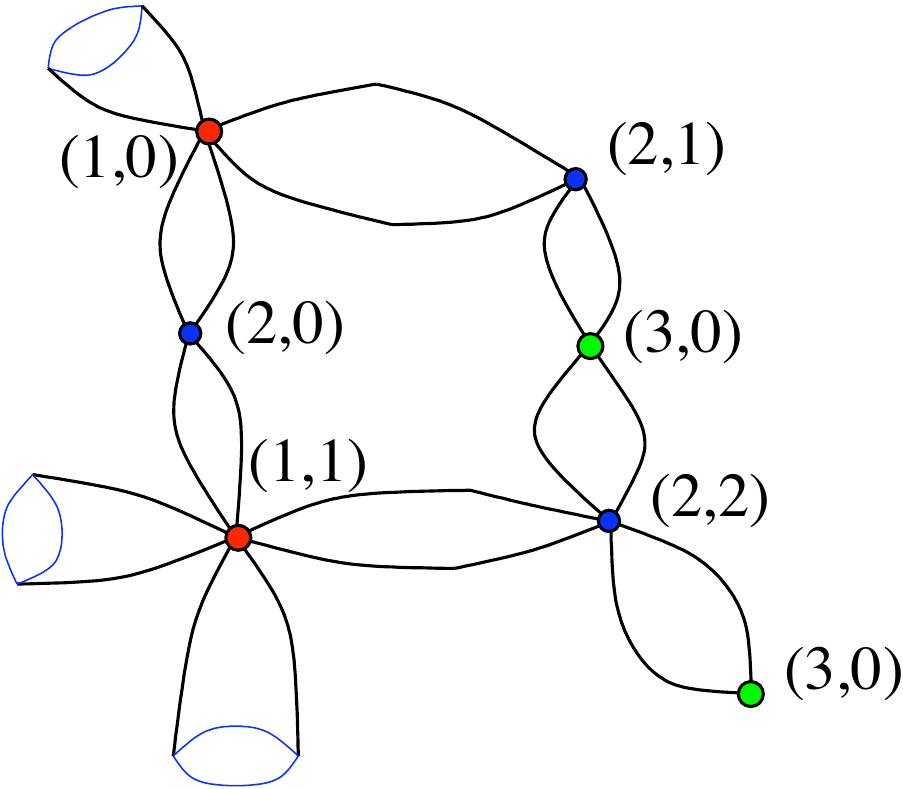}
\qquad
\includegraphics[height=4cm]{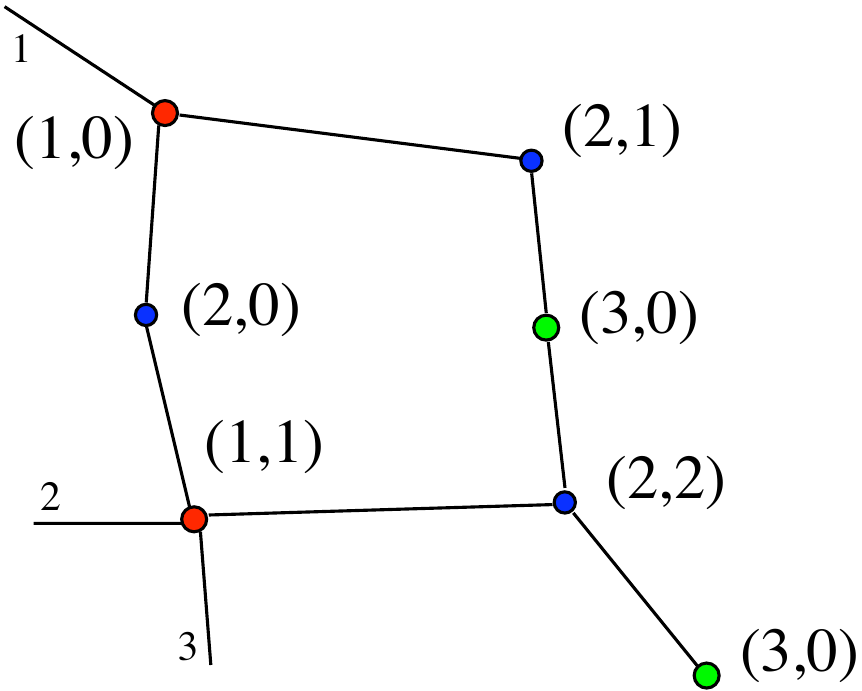} 
$$
\caption{Example of a fixed map of a (nodal) curve $\Sigma$ of genus $5$ and $3$ boundaries into $\CYX$. The image $f(\Sigma)\subset \CYX$ is made of fixed points in $\CYX$, and of spheres and half spheres invariant under the torus action. 
All components of a topology with $\chi<0$ are necessarily collapsed to fixed points. Only components which are spheres with 1 or 2 nodal points  can be mapped to invariant curves. Nodal points are mapped to fixed points, and it may also happen that some smooth point gets mapped to a fixed point.
The fixed map can be represented by a diagram whose vertices are labeled by (fixed points, genus of the component), and edges correspond to invariant curves. Each vertex carries 2 indices: the label of the fixed point and the genus of the component which was mapped to it.} 
\end{figure}

Using the sets of graphs defined in definition \ref{defraphs}, the study of the fixed locus of the torus action allows to introduce the following morphism.
\bd
We define an injective  orbifold morphism:
\beq
\overline\modsp_{g,n}(X,L,\beta,w)^{\rm fixed} \to \mathop{{\oplus}}_{G\in{\cal G}_{g,n}}\,\,\prod_{v={\rm vertices}(G)}\,\,{\overline\modsp}_{g_v,n_v} \quad \times \mathbb N^{\#\,{\rm half-edges}(G)}
\eeq
where each fixed map $\Xi=(\Sigma,f)$ is mapped to a graph $G$ as follows:

1) to each sphere component of $\Sigma$ with 1 or 2 nodal points is associated a closed edge $e$ of the graph, i.e. two half--edges.

2) to each half-sphere component whose boundary is the circle $\d_i\Sigma$, is associated the open half--edge $h_i$ with label $i$, and we have $\epsilon_{h_i}=\epsilon_i$ the open half--edge of $\Upsilon_\CYX$ on which the brane $L$ ends, and the degree $d_{h_i}=w_i$ the degree of the map $f:\d_i\Sigma\to S^1=L\cap \CYX^{\rm fixed}$.
\beq
\epsilon_{h_i}=\epsilon_i
\quad , \qquad
d_{h_i}=w_i.
\eeq

3) to each connected component of $\Sigma_{\rm vertices}=f^{-1}({\rm fixed\, points})$, we associate a vertex $v$ whose labels $(g_v,n_v,\sigma_v)$ are such that $\sigma_v$ is the label of the fixed point to which that component is sent by $f$, $g_v$ is the genus of that component (if the component is an isolated point we set $g_v=0$), and $n_v$ is the number of adjacent spheres or half--spheres.

5) the incidence relations between vertices and half--edges are obviously the incidence relations of components of $\Sigma$.

6) to each half-edge $h=(v,e)$ is associated the degree $d_{h}\in \mathbb N$ of the map $f$ from the sphere of $\Sigma$ corresponding to edge $e$ to the sphere of $\CYX^{\rm fixed}$ at the nodal or possibly smooth point of that sphere sent to the vertex $v$.


7) to each vertex $v$ corresponding to a connected component of $\Sigma$ as described in 3), we associate the corresponding nodal Riemann surface in $\overline\modsp_{g_v,n_v}$.
If $(g_v,n_v)=(0,1)$ we define $\overline\modsp_{0,1}"="$point, and if $(g_v,n_v)=(0,2)$ we define $\overline\modsp_{0,2}"="$point.

\ed

From that definition, it is easy to see how the localization formula \eq{defopenGWloc} should give \eq{eqGWlocalizationdef} in def.\ref{thlocalizationgraphs}.
The only difference between \eq{eqGWlocalizationdef} and \eq{eqWg0localizationgraphs}, is that we now have a factor counting the half--spheres.

\section{B-model side\label{secBmodel}}

The B-model side is also a "counting" of embeddings of Riemann surfaces into  a Calabi--Yau 3-fold $\hat\CYX$, but with a weight different from the Gromov--Witten side. 
Mirror symmetry, and here more precisely the BKMP conjecture, claims that the "amplitudes" computed in the B-model with $\hat\CYX=$mirror of $\CYX$ equal the (open or closed) Gromov-Witten invariants of $\CYX$ of the A-model side.

We refer the reader to the literature \cite{mirrorbook} for a precise definition of the B-model. Here we shall use the "remodeled" B-model (as named in \cite{BKMP}), which defines the B-model amplitudes as some "topological recursion invariants" which we explain below.

\subsection{Mirror}

Mirror symmetry assigns another Calabi--Yau manifold to $\CYX$, namely:
\beq
\hat\CYX = \{w_+\,w_- = H(X,Y)\}\subset \mathbb C^2\times  \left( \mathbb C^*\right)^2
\eeq
i.e. $\hat\CYX$ is a 3-dimensional complex submanifold of $\mathbb C^2\times  \left(\mathbb C^*\right)^2$, defined in coordinates $(w_+,w_-)\in \mathbb C^2$ and $(X,Y)\in  \left( \mathbb C^*\right)^2$ by the relationship:
\beq
w_+\,w_- = H(X,Y)
\eeq
where $H$ is the mirror curve defined in section \ref{secmirror}.

It has the Calabi-Yau property, and posses a nowhere vanishing holomorphic 3-form defined by:
\beq
\Omega = \frac{dw_+\wedge dX\wedge dY}{w_+\,X\,Y}= -\,\frac{dw_-\wedge dX\wedge dY}{w_-\,X\,Y}.
\eeq

\medskip

\bd
The plane curve
\beq
H(\ee{-x}\,\ee{-fy},\ee{-y})=0
\eeq
is called the framed "spectral curve" of $\hat\CYX$, by abuse of notation, we shall also call it $\hat\CYX$.
It is also often called the "mirror curve" of $\CYX$.
\ed

It is the singular locus in $\hat\CYX$ at which $w_+=0$ or $w_-=0$.
In general, the Calabi--Yau 3-fold $\hat\CYX$ is an hyperbolic bundle over $\left(\mathbb C^*\right)^2$, whose fiber degenerates on the spectral curve.

\subsection{Spectral curves}\label{secdefspcurve}

A spectral curve is in fact the data of a plane curve with some additional structure. Here for our purposes we shall define:

\bd[Spectral curve]
A spectral curve $\spcurve=(\curve,x,y,B)$, is the data of:

$\bullet$ a Torelli marked Riemann surface $\curve$, with a symplectic basis of cycles $\acycle_I\cap \bcycle_j=\delta_{I,J}$,

$\bullet$ two analytical functions $x:\curve\to\mathbb C$, $y:\curve\to\mathbb C$,

$\bullet$ a Bergman\footnote{\label{footnoteB}$B$ is the fundamental 2nd kind form as in Mumford's Tata lectures series \cite{MumTata}. We call it the Bergman kernel after the work of Korotkin and Kokotov \cite{KoKo}, and also because Bergman together with Schiffer are the main contributors to its study \cite{BergSchif}.} kernel $B: \curve\times \curve\to T^*(\curve)\otimes T^*(\curve)$, i.e. a symmetric 2nd kind bilinear meromorphic differential, having a double pole on the diagonal and no other pole, and normalized (in any local coordinate $z$) as:
\beq
B(z_1,z_2) \mathop{{\sim}}_{z_2\to z_1} \frac{dz_1\otimes dz_2}{(z_1-z_2)^2} + {\rm analytical} \,\, ,
\eeq
\beq
{\rm and}\,\,\,\forall\,z_2\in \curve,\, \forall\,I=1,\dots,\genus,\,\,\, \oint_{z_1\in\acycle_I} B(z_1,z_2)=0.
\eeq

\bigskip

Moreover, the spectral curve $\spcurve$ is called regular if the meromorphic form $dx$ has a finite number of zeroes on $\curve$, denoted $\{\alpha_1, \dots,\alpha_\bpt\}$ which are all simple zeroes, and $dy$ doesn't vanish at the zeroes of $dx$. In other words, locally near a branchpoint $\alpha_i$, $y$ behaves like a square root of $x$:
\beq\label{bpysqrtx}
y(z) \mathop{{\sim}}_{z\to \alpha_i} y(\alpha_i) + y'(a_i)\,\sqrt{x(z)-a_i} + O(x(z)-a_i) \qquad , \,\, y'(a_i)\neq 0
\eeq
and where $a_i=x(\alpha_i)$ is the $x$--projection of the branchpoint $\alpha_i$:
\beq
x(\alpha_i)=a_i.
\eeq

\ed

\medskip

From now on, all spectral curves considered shall be mirrors of toric CY 3folds:
\beq
\spcurve=(\curve, x,y,B)
\eeq
where:

$\bullet$ $\curve$ is the Riemann surface described in section \ref{sectopologymirror}, and whose atlas of charts (obtained by gluing pairs of pants and cylinders) $\curve = \cup_\sigma \curve_\sigma \cup_{(\sigma,\sigma')} \curve_{\sigma,\sigma'}$ was described in section \ref{secatlas}, cf fig.\ref{figspcurvelocP2framed}.
 $\curve$ is a compact Riemann surface of some genus $\genus$ equal to the number of loops of the toric graph $\Upsilon_\CYX$.
 Its Torelli marking is given in section \ref{secTorelli}. It is such that the $\acycle$-cycles $\acycle_{\epsilon_I}$ wrap cylinders $\curve_{\epsilon_I}$ corresponding to half edges of $\Upsilon_\CYX$.

$\bullet$ the analytical functions $x$ and $y$ are:
\beq
x=-\ln X \virg y=-\ln Y.
\eeq
Because of the logarithm, they are not globally defined on $\curve$, they can be defined on $\curve\setminus T$ where $T$ is the tree introduced in section \ref{secTorelli}.
Notice that $X:\curve\to \mathbb C,\,\,z\mapsto X(z)$ (resp. $Y$) is a meromorphic function on $\curve$, its number of poles = its number of zeroes = the degree in $Y$ (resp, in $X$) of the polynomial $H(X,Y)$.

This shows that $x$ and $y$ have logarithmic singularities, but their differentials $dx=-dX/X$ and $dy=-dY/Y$ are meromorphic forms on $\curve$, having only simple poles, and their residues are rational numbers (related to the degrees of $X$ and $Y$ at their poles or zeroes).
The poles of $dx$ and $dy$ are the punctures, i.e. the zeroes or poles of $X$ and/or of $Y$.

$\bullet$ The Bergman${}^{\ref{footnoteB}}$ kernel $B$ is the unique fundamental form of the 2nd kind on $\curve$ (i.e. having a double pole on the diagonal and no other pole, see \cite{fay}), normalized on $\acycle_I$-cycles defined in section \ref{secTorelli}:
\beq
\forall\,\,I=1\dots,\genus\, , \,\,\,\forall\,z_1\in\curve\, , \qquad \quad
\oint_{z_2\in\acycle_j} B(z_1,z_2)=0.
\eeq
Notice that this implies that for any half-edge $\epsilon$ of $\Upsilon_\CYX$, we have:
\beq
\forall\,\,\epsilon={\rm half-edge}\, , \,\,\,\forall\,z_1\in\curve\, , \qquad \quad\oint_{z_2\in\acycle_\epsilon} B(z_1,z_2)=0.
\eeq

Alternatively, $B$ can be obtained as the second logarithmic derivative of the prime form on $\curve$ (cf \cite{fay}):
\beq
B(z_1,z_2) = d_1\otimes \,d_2\,\,\ln{E(z_1,z_2)}.
\eeq
$B$ is also related to the "heat kernel" on $\curve$, or to the "Green function" on $\curve$, see \cite{farkas, fay}.
Its "physical meaning" is that it gives the electric field measured at $z_2$ created by a unit dipole located at $z_1$ (the log of the prime form $\ln E(z_1,z_2)$ would be the electric potential measured at $z_2$ created by a unit charge located at $z_1$).

\subsection{Invariants of spectral curves and the BKMP conjecture\label{secBKMPconj}}

To any spectral curve $\spcurve$ is associated a set of "invariants" $\om_{g,n}(\spcurve;z_1,\dots,z_n)$ first defined in \cite{EOFg}.
We emphasize that those invariants are defined for any spectral curve: the latter does not need to be related to the mirror of a CY manifold.

\smallskip
For completeness, we recall the definition of invariants $\om_{g,n}(\spcurve)$ of a spectral curve $\spcurve$ in appendix \ref{secappdefWgn}, however we emphasize that it shall not be needed in this article.
Instead we shall need only a few of their properties, and in particular the fact that they can be written in terms of graphs, in section \ref{omgngraphs} below, and can be written in terms of intersection numbers in the moduli space of curves $\overline\modsp_{g,n}$ in theorem \ref{theoremWngclasses} below.

\smallskip
For our purpose we just need:

\bd To a spectral curve $\spcurve$, we associate the family of its invariants $\om_{g,n}$ ($g,n\in \mathbb N$) $:\curve^n \to T^*(\curve)^{\otimes n}$, defined by the topological recursion of \cite{EOFg} (see full definition in appendix \ref{secdefomng}).
\beq
\om_{g,n}(\spcurve;z_1,\dots,z_n) \in \overbrace{T^*(\curve)\otimes\dots\otimes T^*(\curve)}^{n\,{\rm times}}.
\eeq
It is a symmetric multilinear meromorphic differential. 
If $2-2g-n<0$, it has poles only at $z_i=$branchpoints of the spectral curve (zeroes of $dx$), of order at most $6g+2n-4$, and have no residues.
In particular, for $n=0$, we denote $F_g=\om_{g,0}$:
\beq
F_g(\spcurve)=\om_{g,0}(\spcurve) \,\,\, \in \, \mathbb C.
\eeq

\ed

As special cases, for the lowest values of $g$ and $n$, we mention that:
\beq
\om_{0,1}(\spcurve;z) = y(z)\,dx(z)
\virg
\om_{0,2}(\spcurve;z_1,z_2) = B(z_1,z_2),
\eeq
\beq
\om_{0,3}(\spcurve;z_1,z_2,z_3) = \sum_{\sigma={\rm branch\,point}} \Res_{z\to \sigma} \frac{B(z,z_1)\, B(z,z_2)\, B(z,z_3)}{dx(z)\,dy(z)}.
\eeq
$F_0=\omega_{0,0}$ is the "prepotential", and $F_1=\omega_{1,0}$ is related to the log determinant of a Laplacian on $\curve$ \cite{KoKo},
and for other values of $(g,n)$ we refer to appendix \ref{secappdefWgn} or to the literature \cite{EOFg}.


\medskip
Those invariants have many fascinating properties (modularity, integrability, special geometry) \cite{EOFg, EOreview}, and can be expressed in terms of intersection numbers \cite{eynclasses1,eynclasses2}.

\medskip

The invariants $\om_{g,n}(\spcurve;z_1,\dots,z_n)$ depend on $n$ points $(z_1,\dots,z_n)\in\curve^n$. However, it is convenient to use $x_i=x(z_i)\in \mathbb C$ as a local complex coordinate on $\curve$, we thus write:
\bd
For $(g,n)\in \mathbb{N}^2$,
\beq
W_{g,n}(\spcurve;x_1,\dots,x_n) = \om_{g,n}(\spcurve;z_1,\dots,z_n) \virg x_i=x(z_i)\in\mathbb C.
\eeq
\ed
\br
 $\om_{g,n}$ is an analytical (meromorphic) function of each $z_i\in \curve$, but since the map $z_i\mapsto x(z_i)$ might be non--invertible (it is not invertible at the branchpoints), the $W_{g,n}$ are not analytical functions of their variables $x_i\in\mathbb C$. They are typically multivalued, and they have branchcuts starting and ending at the branchpoints and at the punctures.
\er

\bigskip
In \cite{Mar1, BKMP},  Bouchard, Klemm, Mari\~no and Pasquetti (i.e. BKMP),  conjectured that Gromov--Witten invariants of toric Calabi--Yau 3--folds coincide with the invariants of their mirror's spectral curve $\spcurve=\hat\CYX$.

\begin{conjecture}[BKMP conjecture 2006-2008 \cite{Mar1,BKMP}]\label{BKMPconj}
If $\CYX$ is a toric Calabi--Yau 3-fold, we have
\beq
\forall (g,n) \in \mathbb{N}^2 \, , \;
{\cal W}_{g,n}(\CYX,L,\mathbf t;x_1,\dots,x_n)\,dx_1\otimes \dots\otimes dx_n
= W_{g,n}(\hat\CYX;x_1,\dots,x_n)
\eeq
where $\hat\CYX$ is the framed mirror curve of $\CYX$.

\end{conjecture}

This conjecture was checked by \cite{Mar1, BKMP} for many manifolds $\CYX$ to low genus, and it was proved to all genus only for the simplest case, namely $\CYX=\mathbb C^3$ (=framed topological vertex), independently by Chen \cite{ChenLin2009} and Zhou \cite{ZhouJian2009} in 2009, by extending the existing proof for the Hurwitz numbers (which is the infinite framing limit of the BKMP conjecture for the topological vertex, known as Bouchard--Mari\~no conjecture \cite{BM}, proved in \cite{Borot1,Eynardb}).

%
We prove the general case below.

\subsection{Invariants as  graphs and intersection numbers\label{omgngraphs}}

We shall not need the explicit definition of the invariants $W_{g,n}$, 
instead we give a combinatoric algorithm to compute them,
following \cite{eynclasses1}, we write the invariants as the following two theorems (valid for any spectral curves):

\bt[E. 2011 \cite{eynclasses1}]\label{WgnCL1bp}
Let $\spcurve_\sigma=(\curve_\sigma,x,y,B_\sigma)$ be a spectral curve with only one branchpoint located at $x=a_\sigma$.
The invariants $W_{g,n}(\spcurve_\sigma)$ with $2-2g-n<0$ (where one can also have $n=0$), 
can be expressed as integrals of combinations of $\psi$ and Mumford's $\kappa$ classes in $\overline\modsp_{g,n}$ as:
\beq
\label{eqWgn1bptCLbis}
W_{g,n}(\spcurve_\sigma;x_1,\dots,x_n)
= \frac{2^{3g-3+n}}{\ee{-\hat t_{\sigma,0}(2g-2+n)}}\!\!\!\sum_{d_1+\dots+d_n\leq 3g-3+n}
\left(\int_{\overline\modsp_{g,n}} \CL_{\spcurve_\sigma}
\prod_{i=1}^n \tau_{d_i} \right)\quad \prod_{i=1}^n d\xi_{\sigma,d_i}(x_i)
\eeq
where we have defined:

$\bullet$ 
\beq
\ee{-\hat t_{\sigma,0}} = 4 y'(a_\sigma) = 2\,\lim_{z\to \alpha_\sigma} \frac{y(z)-y(\bar z)}{\sqrt{x(z)-a_\sigma}}
\eeq
where $y'(a_\sigma)$ was defined in \eq{bpysqrtx};

$\bullet$ the times $\hat t_{\sigma,k}$, or more precisely their generating function $g_\sigma(u)$
\beq
g_\sigma(u) = \sum_{k=1}^\infty \hat t_{\sigma,k}\,u^{-k}
\eeq
is defined by the Laplace transform of the 1-form $ydx$ along a "steepest descent" path $\gamma_\sigma\subset \curve_\sigma$ such that $x(\gamma_\sigma)-x(a_\sigma)=\mathbb R_+$ and $a_\sigma\in \gamma_\sigma$ (i.e. $\gamma_\sigma$ is the horizontal trajectory of $x$ going through the branchpoint $a_\sigma$):
\beq
 \ee{-\hat t_{\sigma,0}}\,\,\ee{-g_\sigma(u)}
= \frac{2\,u^{3/2}\,\ee{u\, a_\sigma}}{\sqrt\pi}\,\int_{\gamma_\sigma} \ee{-ux}\,\,ydx ;
\eeq

$\bullet$ the coefficients $\hat B_{\sigma,k,l}$, or more precisely their generating function $\hat B_\sigma(u,v)$ is defined by the double Laplace transform of $B_\sigma$ along $\gamma_\sigma$:
\beq
\hat B_\sigma(u,v) = 
\sum_{k,l=0}^\infty \hat B_{\sigma;k,l}u^{-k}v^{-l}
= \frac{uv}{u+v}+\frac{\sqrt{uv}\,\ee{(u+v)\, a_\sigma}}{2\pi}\,\int_{\gamma_\sigma\times \gamma_\sigma} \ee{-ux}\,\ee{-vx'}\,B_\sigma(x,x')  
\eeq
where the double integral is properly regularized so that the result has a large $u,v$ expansion, see \cite{eynclasses1};

$\bullet$ the one forms $d\xi_{\sigma,d}(x)$ are defined by
\bea
\label{defdxironddef1}
d\xi_{\sigma,d}(x) 
&=& -\,\frac{(2d-1)!!}{2^d}\, \Res_{x'\to a_\sigma} B_\sigma(x,x')\,(x'-a_\sigma)^{-d-1/2} ;\cr
\eea

$\bullet$  the tautological  class $\CL_{\spcurve_\sigma}$ is a combination of $\tau_d=\psi^d$ and Mumford's $\kappa$ classes in $\overline\modsp_{g,n}$ (see appendix \ref{apppsikappa}), defined as
\beq
\label{eqdefCLkappa1}
\CL_{\spcurve_\sigma} = \ee{\sum_{k\geq 1} \hat t_{\sigma,k}\kappa_k}\,\,\ee{\frac{1}{2}\sum_{\delta\in\d\modsp_{g,n}} \sum_{k,l} \hat B_{\sigma;k,l} l_{\delta*}\,\tau_k\tau_l}
\eeq
$l_{\delta*}$ is the natural inclusion of $\d\modsp_{g,n}$ into $\modsp_{g-1,n+2}\cup \sum_{h+h'=g,m+m'=n} \modsp_{h,m+1}\times \modsp_{h',m'+1}$.

\et

\proof{
The proof is done in \cite{eynclasses1}. Let us just sketch the main steps.

\noindent The prototype of a spectral curve with only one branchpoint is $y={\displaystyle \sum_{k \in \mathbb{N}}} t_{k+2} (x-a)^{k/2}$. This is  the spectral curve of Kontsevich's matrix Airy integral with times $\{t_k\}$ \cite{Kontsevich1992}, whose invariants $\om_{g,n}$ are (almost by definition of Kontsevich's integral, see \cite{Eynard,Kontsevich1992}),  generating functions for intersection numbers of $\psi$ classes in $\overline\modsp_{g,n}$ (see appendix \ref{apppsikappa}).
Finding a Kontsevich spectral curve (i.e. finding the Kontsevich times $t_k$) which has the same Taylor expansion near the branchpoint as $\spcurve_\sigma$, allows to express the invariants of any spectral curve with one branchpoint in terms of intersection of classes in $\overline\modsp_{g,n}$.
Moreover, the result looks even better if we rewrite, using Arbarello and Cornalba's relations \cite{Arbarello1996}, the combinations of $\psi$ classes in terms of Mumford $\kappa$ classes. All this was done in \cite{eynclasses1} and results into theorem \ref{WgnCL1bp}.

}

\br
Formula \eq{eqWgn1bptCLbis} looks very similar to the ELSV formula, and indeed it reduces to the ELSV formula for the spectral curve $(\mathbb C, x(z)=z-\ln z,y(z)=z,B(z_1,z_2)=dz_1\otimes dz_2/(z_1-z_2)^2)$ which appears in the study of simple Hurwitz numbers \cite{BM, Borot1, Eynardb, eynclasses1}. In that case the combination of $\kappa$ and $\psi$ classes in \eq{eqdefCLkappa1} reduces to the Hodge class through Mumford's formula \cite{Mumford1983}. See \cite{eynclasses1} for the detailed proof.
\er

Once we know how to compute the invariants of spectral curves having only 1 branch-point, the following theorem (proved in \cite{eynclasses2}) gives invariants of spectral curves with an arbitrary number of branchpoints:

\bt[corrolary of the theorem in \cite{eynclasses2, Alexandrov2005, OrantinN.2008, Kostov2010}]\label{theoremWngclasses}

Let $\spcurve=(\curve,x,y,B)$ be a spectral curve with branchpoints $\{a_1,\dots,a_\bpt\}$.
For $\sigma = 1,\dots,\bpt$, let $\spcurve_\sigma=(\curve_\sigma,x,y,B_\sigma)$ be the local spectral curve near the branchpoint $a_\sigma$, with $\curve_\sigma\subset \curve$ containing only the branchpoint $a_\sigma$, and $x$ and $y$ are the restrictions of $x$ and $y$ to $\curve_\sigma$, and $B_\sigma$ a Bergman kernel\footnote{$B_\sigma$ needs not be the restriction of $B$ to $\curve_\sigma$, neither needs to be the normalized Bergman kernel on $\curve_\sigma$. For this theorem it just needs to be any symmetric bilinear differential having a normalized double pole on the diagonal.} on $\curve_\sigma$.

When $2-2g-n<0$, and $n\geq 0$, the invariants of $\spcurve$ can be computed in terms of graphs and invariants of local curves near branchpoints, as
\bea
W_{g,n}(\spcurve;x_1,\dots,x_n) 
&=& \sum_{G\in {\td{\cal G}}^{\rm stable}_{g,n}} \frac{1}{\#{\rm Aut}(G)}\,
\prod_{h={\rm half-edges}} \,\Res_{x_{h}\to a_{\sigma_{v(h)}}}  \cr
&& \prod_{v={\rm vertices}} \,W_{g_v,n_v}(\spcurve_{\sigma_v};\{x_{h}\}_{h\in E_v})  \cr
&& \prod_{(h_+,h_-)={\rm closed\,edges}} 
\Big[ \ln{\left(E(\spcurve;x_{h_+},x_{h_-})\right)} \cr
&& \qquad \qquad - \delta_{\sigma_{v(h_+)},\sigma_{v(h_-)}}\,\ln{\left(E({\spcurve_{\sigma_{v(h_+)}}};x_{h_+},x_{h_-})\right)}\Big] \cr
&& \prod_{h_i={\rm open\,half-edges}} 
dS(\spcurve;x_i,x_{h_i})
\eea
where $E(\spcurve;x_1,x_2)$ denotes the prime form associated to the Bergman kernel on $\spcurve$, i.e.
\beq
d_1\otimes d_2\,\ln{E(\spcurve;x_1,x_2)} = B(\spcurve;x_1,x_2)
\eeq
and $dS(\spcurve;x_1,x_2)$ is the Cauchy kernel associated to the Bergman kernel on $\spcurve$
\beq
d_1\,\ln{E(\spcurve;x_1,x_2)} = dS(\spcurve;x_1,x_2) = \int_{x'_2=o}^{x_2} B(\spcurve;x_1,x'_2).
\eeq
And ${\td{\cal G}}^{\rm stable}_{g,n}$ is the same set of graphs as in ${{\cal G}}^{\rm stable}_{g,n}$ defined in def.\ref{defraphs}, but without degree labels $d_{h}$ on half-edges.

\et

\br
We insist on the fact that this theorem, as well as theorem \ref{WgnCL1bp}, is valid also for $n=0$.
\er

For example $W_{0,4}$: since $\sum_v (2-2g_v-n_v) = -2$ and $2-2g_v-n_v<0$ and $\sum_v n_v\geq 4$, $\td{\cal G}^{\rm stable}_{0,4}$ contains graphs with at most 2 vertices. More precisely, $\td{\cal G}^{\rm stable}_{0,4}$ contains either graphs with one 4-valent vertex, or graphs with two 3-valent vertices:
\bea
&& W_{0,4}(\spcurve;x_1,x_2,x_3,x_4) \cr
&=& \sum_\sigma \Res_{x'_1,x'_2,x'_3,x'_4\to a_\sigma} W_{0,4}(\spcurve_\sigma;x'_1,x'_2,x'_3,x'_4)\,\, \prod_{i=1}^4 dS(\spcurve;x_i,x'_i) \cr
&& + \sum_{\sigma_1}\sum_{\sigma_2} \Res_{x'_1,x'_2,x'_5\to a_{\sigma_1}}\Res_{x'_3,x'_4,x'_6\to a_{\sigma_2}} W_{0,3}(\spcurve_{\sigma_1}; x'_1,x'_2,x'_5)\,\,W_{0,3}(\spcurve_{\sigma_2}; x'_3,x'_4,x'_6) \cr
&& \qquad  (\ln E(\spcurve;x'_5,x'_6)-\delta_{\sigma_1,\sigma_2}\ln E(\spcurve_{\sigma_1};x'_5,x'_6))\,\,\prod_{i=1}^4 dS(\spcurve;x_i,x'_i) \cr
&& + {\rm permutations\,of}\,\,\{x_1,x_2,x_3,x_4\}
\eea
$$\includegraphics[width=14cm]{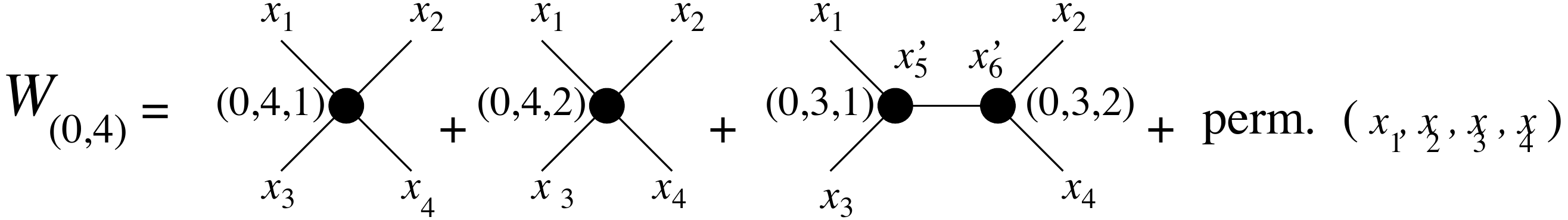} $$

\proof{This is a mere rewriting of \cite{eynclasses2}, and it is a consequence of \cite{OrantinN.2008, Kostov2010}. The graphs are a way of encoding the order in which residues are computed, i.e. it is only a combinatorial way of summing over $\sigma$ the residues at the branchpoints $a_\sigma$. A detailed proof is written in \cite{eynclasses2}.
We emphasize that this theorem applies to any spectral curve $\spcurve$, it doesn't need to be related to any Calabi--Yau mirror geometry.

\smallskip
Also, we mention that this theorem can be seen as a formulation of Givental's relations \cite{givental_sb_2003, givental-2001}, although we shall not pursue in that direction.
}

\subsection{Geometry of the local spectral curve}

In prop \ref{propatlascurve}, we have defined an atlas for the curve $\curve$, in terms of pairs of pants $\curve_\sigma$ and cylinders $\curve_{\sigma,\sigma'}$, labeled by vertices and edges of the toric graph $\Upsilon_\CYX$.
Each $\curve_\sigma$ contains exactly one branchpoint $a_\sigma$.

Since $\curve_\sigma$ is topologically  a pair of pants, it can be realized as the projective complex plane with 3 holes, and we can choose the 3 holes to be connected domains respectively  containing $z=0,1,\infty$, i.e.
\beq
\curve_\sigma \subset \mathbb P^1\setminus\{0,1,\infty\}.
\eeq

The functions $X=\ee{-x}$ and $Y=\ee{-y}$ are holomorphic functions on $\curve_\sigma$, and thus the functions $x$ and $y$ are holomorphic functions on $\curve_\sigma\setminus {\rm T}$ where the tree T was introduced in section \ref{secTorelli}.

\subsubsection{Large radius limit (tropica limit)}

Let $(\arond_\sigma,\brond_\sigma)$ be the vertices of the toric graph (see def. \ref{deftoricgraph}).
$\curve_\sigma$ was defined in section \ref{secatlas} so that $x-\arond_\sigma$ and $y-\brond_\sigma$ have a non-trivial limit when all $t_j\to+\infty$, i.e. we define:

\bd
The functions $\xrond_\sigma: \curve_\sigma\to \mathbb C$ and $\yrond_\sigma: \curve_\sigma\to \mathbb C$ are the large radius limits of $x$ and $y$ in $\curve\cap\curve_\sigma$:
\beq
\xrond_\sigma = \lim_{\td t_j\to +\infty} x-\arond_\sigma
\virg
\yrond_\sigma = \lim_{\td t_j\to +\infty} y-\brond_\sigma.
\eeq
\ed

A vertex $\sigma$ of the toric graph is a triangle $\sigma=(i_1,i_2,i_3)$ of the dual $\hat\Upsilon_\CYX$.
Since we are very close to a vertex of the toric graph, in the tropical limit only  three of the $Y_i$'s don't tend to $0$, i.e. the large radius limit of the mirror curve is:
\beq
Y_{i_1}+Y_{i_2}+Y_{i_3} = 0,
\eeq
or alternatively, only 3 monomials in $H(X,Y)=0=\sum_{i,j} \ee{-\td t_{i,j}}\,X^i\,Y^j$ survive in that limit $H(X,Y) \to \stackrel{\circ}{H}_\sigma(\Xrond_\sigma,\Yrond_\sigma)$:
\beq\label{eqcurvesigmatropicallimit}
H\to \stackrel{\circ}{H}_\sigma \virg
 \stackrel{\circ}{H}_\sigma(\Xrond,\Yrond) = 1+C_\sigma\,\,\,\Xrond^{\fram_{d,\sigma}}\Yrond^{-\fram_{b,\sigma}}+
\td C_\sigma\,\,\,\Xrond^{-\fram_{c,\sigma}}\Yrond^{\fram_{a,\sigma}}.
\eeq 
The exponents $(\fram_{a,\sigma},\fram_{b,\sigma},\fram_{c,\sigma},\fram_{d,\sigma})\in \mathbb Z^4$ are integers corresponding to the vertices of the dual toric graph around the vertex $\sigma$. They form the local framing matrix at vertex $\sigma$ defined in def.\ref{defframsigma}:
\beq
\fram_\sigma=\left(\begin{array}{cc}\fram_{a,\sigma} & \fram_{b,\sigma} \cr \fram_{c,\sigma} & \fram_{d,\sigma}\end{array}\right)
\qquad , \,\,\,
\det\,\fram_\sigma =1.
\eeq

The coefficients $C_\sigma$ and $\td C_\sigma$ are the limits of:
\beq
-\ln{C_\sigma}  = \lim_{\td t_j\to+\infty}\,\,-\fram_{d,\sigma} a_\sigma + \fram_{b,\sigma} y(a_\sigma) + \td t_{i-\fram_{d,\sigma},j+\fram_{b,\sigma}}-\td t_{i,j}
\eeq
\beq
-\ln{\td C_\sigma}  = \lim_{\td t_j\to+\infty}\,\,\fram_{c,\sigma} a_\sigma - \fram_{a,\sigma} y(a_\sigma) + \td t_{i+\fram_{c,\sigma},j-\fram_{a,\sigma}}-\td t_{i,j}
\eeq
where $(i,j)\in\mathbb Z^2$ is any vertex of the dual graph $\hat\Upsilon_\CYX$ adjacent to the vertex $\sigma$. 

\medskip
We can parametrize our curve \eq{eqcurvesigmatropicallimit} by a complex variable $z\in \mathbb C\setminus\{0,1,\infty\}$, and here explicitly:
\bea
\Xrond_\sigma(z) = \ee{-\xrond_\sigma(z)} = (-C_\sigma)^{-\fram_{a,\sigma}}\,(-\td C_\sigma)^{-\fram_{b,\sigma}}\,\,\, z^{\fram_{b,\sigma}}\,\,(1-z)^{\fram_{a,\sigma}} \cr
\Yrond_\sigma(z) = \ee{-\yrond_\sigma(z)}= (-C_\sigma)^{-\fram_{c,\sigma}}\,(-\td C_\sigma)^{-\fram_{d,\sigma}}\,\,\, z^{\fram_{d,\sigma}}\,\,(1-z)^{\fram_{c,\sigma}}
\eea
i.e.
\bea
\xrond_\sigma(z) = \fram_{a,\sigma}\ln{(-C_\sigma)}\,+\fram_{b,\sigma}\,\ln{(-\td C_\sigma)}\,\,\, -\fram_{b,\sigma}\,\ln z\,\,-\fram_{a,\sigma}\ln{(1-z)} \cr
\yrond_\sigma(z) = \fram_{c,\sigma}\ln{(-C_\sigma)}\,+\fram_{d,\sigma}\,\ln{(-\td C_\sigma)}\,\,\, -\fram_{d,\sigma}\,\ln z\,\,-\fram_{c,\sigma}\ln{(1-z)} 
\eea
taking the differentials gives:
\beq
d\xrond_\sigma =  -\,\,\left(\frac{\fram_{b,\sigma}}{z} + \frac{\fram_{a,\sigma}}{z-1}\right)\,dz
\virg
d\yrond_\sigma =  -\,\,\left(\frac{\fram_{d,\sigma}}{z} + \frac{\fram_{c,\sigma}}{z-1}\right)\,dz.
\eeq
Notice that $d\xrond_\sigma$ and $d\yrond_\sigma$ are meromorphic forms on $\mathbb P^1$ having only simple poles at $0,1,\infty$, and the entries of $\fram$ are the residues of those poles.

We have realized $\curve_\sigma$ as an open domain of $\mathbb P^1\setminus \{0,1,\infty\}$.

\medskip

Notice that $d\xrond_\sigma$ vanishes at
\beq
z = \frac{\fram_{b,\sigma}}{\fram_{a,\sigma}+\fram_{b,\sigma}},
\eeq
therefore the branchpoint is located at (by definition it was at $\xrond_\sigma=0, \yrond_\sigma=0$):
\beq
0 
= \fram_{a,\sigma}\ln{(-C_\sigma)}\,+\fram_{b,\sigma}\,\ln{(-\td C_\sigma)}\,\,\, -\fram_{b,\sigma}\,\ln{\frac{\fram_{b,\sigma}}{\fram_{a,\sigma}+\fram_{b,\sigma}}}\,\,-\fram_{a,\sigma}\ln{\frac{\fram_{a,\sigma}}{\fram_{a,\sigma}+\fram_{b,\sigma}}} 
\eeq
and similarly $\yrond_\sigma$ must vanish at the branchpoint. This determines the coefficients $C_\sigma$ and $\td C_\sigma$.

\medskip

We thus define:
\bd\label{defspcurverond}
Let 
\beq
\fram=\left(\begin{array}{cc}\fram_{a} & \fram_{b} \cr \fram_{c} & \fram_{d}\end{array}\right)\in Sl_2(\mathbb Z)
\virg
\fram_{a}\, \fram_{d} - \fram_{b}\,\fram_{c} = 1
\eeq
be a local framing matrix, then we define the "vertex" spectral curve $\spcurverond_\fram$ as:
\beq
\spcurverond_\fram = (\mathbb P^1\setminus\{0,1,\infty\}, \xrond_\fram,\yrond_\fram,\Brond_\fram)
\eeq
where
\beq
\left\{\begin{array}{l}
\xrond_\fram(z) =  -\fram_{b}\,\ln z\,\,-\fram_{a}\ln{(1-z)} \quad +\fram_{b}\,\ln{\frac{\fram_{b}}{\fram_{a}+\fram_{b}}}\,\,+\fram_{a}\ln{\frac{\fram_{a}}{\fram_{a}+\fram_{b}}} \cr
\yrond_\fram(z) = -\fram_{d}\,\ln z\,\,-\fram_{c}\ln{(1-z)} \quad +\fram_{d}\,\ln{\frac{\fram_{b}}{\fram_{a}+\fram_{b}}}\,\,+\fram_{c}\ln{\frac{\fram_{a}}{\fram_{a}+\fram_{b}}} \cr
\Brond_\fram(x_1,x_2) = \frac{dz_1\otimes dz_2}{(z_1-z_2)^2} \qquad \quad {\rm where}\,\, x_1=\xrond_\fram(z_1),\,x_2=\xrond_\fram(z_2).
\end{array}\right.
\eeq

\ed

This curve is the spectral curve of a toric Calabi-Yau whose toric graph has only one vertex, and thus it is $\CYX=\mathbb C^3$, together with a framing matrix $\fram$.
In other words, the large radius limit of the mirror curve near a vertex $\sigma$, is the mirror curve of the toric Calabi--Yau 3-fold $\CYX=\mathbb C^3$ with framing $\fram_\sigma$.

\subsubsection{Local spectral curve at a vertex\label{secdefSpcurvesigma}}

Now, we can consider the spectral curve on its whole without considering the large radius limit.
We define
\bd
Let $\spcurve_\sigma$ be the spectral curve obtained by restriction of the full spectral curve $\spcurve$, to the  vicinity $\curve_\sigma$ of $a_\sigma$:
\beq
\spcurve_\sigma=(\curve_\sigma,x-\arond_\sigma,y-\brond_\sigma,B_\sigma)
\eeq
where $x,y$ are simply the restrictions to $\curve_\sigma$ of $x,y$ on $\curve$, and $B_\sigma$ is the Bergman kernel of the Riemann sphere $\mathbb P^1$ (remember that $\curve_\sigma$ is a sphere with 3 holes, i.e. $\curve_\sigma\subset \mathbb P^1$), 
i.e. it is the same $\Brond_{\fram_\sigma}$ introduced in def.\ref{defspcurverond}, shifted by $a_\sigma-\arond_\sigma$
\beq\label{defBsigma1}
B(\spcurve_\sigma;x_1,x_2) = B_\sigma(x_1,x_2) = \Brond_{\fram_\sigma}(x_1-a_\sigma+\arond_\sigma,x_2-a_\sigma+\arond_\sigma).
\eeq

\ed






By definition, the spectral curve $\spcurve_\sigma$ has only one branchpoint, located at $x=a_\sigma-\arond_\sigma$.
Its invariants are computed 
by theorem \ref{WgnCL1bp}, and can thus be written in terms of integrals of some classes in $\overline\modsp_{g,n}$.

\subsection{Large radius limit: the topological vertex}

Almost by definition def \ref{defspcurverond}, the large radius limit of the spectral curve $\spcurve_\sigma$ is $\spcurverond_{\fram_\sigma}$:
\beq
\spcurverond_\fram = (\mathbb P^1, \xrond_\fram,\yrond_\fram,\Brond_\fram)
\eeq
where
\beq
\left\{\begin{array}{l}
\xrond_\fram(z) =  -\fram_{b}\,\ln z\,\,-\fram_{a}\ln{(1-z)} \quad +\fram_{b}\,\ln{\frac{\fram_{b}}{\fram_{a}+\fram_{b}}}\,\,+\fram_{a}\ln{\frac{\fram_{a}}{\fram_{a}+\fram_{b}}} \cr
\yrond_\fram(z) = -\fram_{d}\,\ln z\,\,-\fram_{c}\ln{(1-z)} \quad +\fram_{d}\,\ln{\frac{\fram_{b}}{\fram_{a}+\fram_{b}}}\,\,+\fram_{c}\ln{\frac{\fram_{a}}{\fram_{a}+\fram_{b}}} \cr
\Brond_\fram(x_1,x_2) = \frac{dz_1\otimes dz_2}{(z_1-z_2)^2} \qquad \quad {\rm where}\,\, x_1=\xrond_\fram(z_1),\,x_2=\xrond_\fram(z_2).
\end{array}\right.
\eeq
It has a unique branchpoint ($d\xrond_\fram(z)=0$) at
\beq
z=\frac{\fram_b}{\fram_a+\fram_b}.
\eeq

Let us then apply theorem \ref{WgnCL1bp} to $\spcurverond_\fram$ (this was done in \cite{eynclasses1}):

\bt["Mari\~no--Vafa formula"]\label{thWngvertex}
 For $2-2g-n<0$, we have:
\bea\label{Wgnvertex1}
&& W_{g,n}(\spcurverond_\fram;x_1,\dots,x_n) \cr
&=& \frac{2^{3g-3+n}}{\ee{\hat t_{\fram,0}(2-2g-n)}}\,\sum_{\{d_i\}}\Big\langle \CL_{\rm Hodge}(\fram_a)\,\CL_{\rm Hodge}(\fram_b)\,\CL_{\rm Hodge}(-\fram_a-\fram_b) \prod_{i=1}^n  \tau_{d_i} \Big\rangle_{g,n}\!\!\prod_{i=1}^n d\td\xirond_{\fram,d_i}(x_i) \cr
\eea
where
\beq
\ee{-\hat t_{\fram,0}} = \frac{2\sqrt{2}}{\sqrt{\fram_a\fram_b(\fram_a+\fram_b)}}
\eeq
and where, if $x$ lies near the puncture of $\mathbb P^1\setminus\{0,1,\infty\}$ (i.e. $z=0$, $1$ or $\infty$) corresponding to the half-edge $\epsilon$, (whose framing is $\fram_{\epsilon}=\fram_b$, $\fram_a$ or $-\fram_a-\fram_b$ respectively):
\beq\label{eqexptdxirond}
\td\xirond_{\fram_\epsilon,d}(x) = (-1)^d\,\left(\frac{{\mathrm d}}{{\mathrm d}x}\right)^d\,\, \xirond_{\fram_\epsilon,0}(x)
= \sum_k \frac{k^{d}}{\fram_{\epsilon_i}^{d+1}}\,\,\gamma_\fram(k/\fram_{\epsilon})\,\,\,\ee{-\,\frac{k}{\fram_{\epsilon}}\,x}.
\eeq

Thus, if $x_i$ lies near the puncture corresponding to the half-edge $\epsilon_i$, (whose framing is $\fram_{\epsilon}$):
\bea
&& W_{g,n}(\spcurverond_\fram;x_1,\dots,x_n) \cr
&=& \frac{2^{3g-3+n}}{\ee{\hat t_{\fram,0}(2-2g-n)}}\,\sum_{\{k_i\}}\Big\langle \CL_{\rm Hodge}(\fram_a)\,\CL_{\rm Hodge}(\fram_b)\,\CL_{\rm Hodge}(-\fram_a-\fram_b) \prod_{i=1}^n  \frac{1}{1-\frac{k_i}{\fram_{\epsilon_i}}\,\psi_i} \Big\rangle_{g,n} \cr
&& \qquad \,\,\prod_{i=1}^n \frac{k_i}{\fram_{\epsilon_i}^2}\,\,\gamma_\fram(k_i/\fram_{\epsilon_i})\,\,\,\ee{-\,\frac{k_i}{\fram_{\epsilon_i}}\,x_i}\,dx_i \cr
\eea
where the sum carries over positive integers $(k_1,\dots,k_n)\in \mathbb Z_+^n$.

\et

\proof{This theorem is a mere application of theorem \ref{WgnCL1bp}, and is fully proved in \cite{eynclasses1}, or alternatively, it can be seen as a consequence of the proof of BKMP for the framed vertex \cite{ChenLin2009, ZhouJian2009}.
For completeness, we redo it in appendix \ref{appvertex}, as an application of theorem \ref{WgnCL1bp}.
}

\br
One can recognize that the right hand side of \eq{Wgnvertex1} is the Mari\~no--Vafa formula for the topological vertex \cite{MV01,Liu2003,Liu2003a}, i.e. the Gromov--Witten invariants of $\CYX=\mathbb C^3$ with framing matrix $\fram$, and thus, theorem \ref{thWngvertex} (proved in \cite{eynclasses1}) can be viewed as another proof of the BKMP conjecture for the topological vertex $\CYX=\mathbb C^3$ with framing matrix $\fram$.
The first proof of the BKMP conjecture for $\CYX=\mathbb C^3$, are those of Chen \cite{ChenLin2009} and Zhou \cite{ZhouJian2009}.
\er


%


\subsection{Invariants of the local spectral curve}


We know that in the tropical limit when all $t_j\to +\infty$, we have  $\spcurve_\sigma\to \spcurverond_{\fram_\sigma}$, and we have expressed the invariants of $\spcurverond_{\fram_\sigma}$ in terms of Hodge classes integrals in $\overline\modsp_{g,n}$. Moreover, it is shown in \cite{EOFg}, that the invariants $\om_{g,n}$ of any spectral curve satisfy "special geometry relations" (similar to Seiberg--Witten for $\om_{0,0}$), which allows to compute the derivatives $\d/\d t_j$, and thus allow to compute the Taylor expansion of the invariants in a vicinity of the tropical limit.
This gives the following lemma: 

\bl\label{lemmarenorm01}
If $2-2g-n< 0$ and $n\geq 0$,
\bea
&& W_{g,n}(\spcurve_\sigma;x_1,\dots,x_n)  \cr
&=& \frac{2^{3g-3+n}}{\ee{\hat t_{\fram_\sigma,0}(2-2g-n)}}\sum_{k=0}^\infty \frac{1}{k!}\,\sum_{d_1,\dots,d_{n+k}} \,\,\prod_{i=1}^k R_{\sigma,d_{n+i}}\cr
&&   \left< \CL_{\rm Hodge}(\fram_{a,\sigma})\,\CL_{\rm Hodge}(\fram_{b,\sigma})\,\CL_{\rm Hodge}(-\fram_{a,\sigma}-\fram_{b,\sigma})\, \prod_{j=1}^{n+k} \tau_{d_j}\right>_{g,n+k} \quad
\prod_{j=1}^n\,d\td \xirond_{\fram_\sigma,d_j}(x_j)
 \cr
&=& \frac{2^{3g-3+n}}{\ee{\hat t_{\fram_\sigma,0}(2-2g-n)}}\, \sum_{d_1,\dots,d_n} \prod_{j=1}^n\,d\td\xirond_{\fram_\sigma,d_j}(x_j) \cr
&&  \left< \CL_{\rm Hodge}(\fram_{a,\sigma})\,\CL_{\rm Hodge}(\fram_{b,\sigma})\,\CL_{\rm Hodge}(-\fram_{a,\sigma}-\fram_{b,\sigma})\,\,\, \ee{l_{1*} \sum_d R_{\sigma,d}\tau_d}\, \prod_{j=1}^{n} \tau_{d_j}\right>_{g,n} \cr
\eea
where 
\beq
R_{\sigma,d} 
= \frac{-\,2\,\ee{\hat t_{\fram_\sigma,0}}}{2\pi i}\,\oint_{\d \curve_\sigma} \td\xirond_{\fram_\sigma,d}(x)\,\,(y(x+\arond_\sigma)-\brond_\sigma-\yrond_{\fram_\sigma}(x)) \,dx,
\eeq
where $\d\curve_\sigma$ is the boundary of $\curve_\sigma$, i.e. the union of three circles, oriented so that $\curve_\sigma$ lies on the left of $\d\curve_\sigma$.
In the second equality,
$l_{1*}$ denotes the natural inclusion of $\modsp_{g,n}\subset \modsp_{g,n+1}$ ,so that $\ee{l_{1*} \sum_d C_{\sigma,d}\psi^d}$ is just a short hand notation for the formula above.

\medskip 

And similarly for $(g,n)=(0,2)$:
\bea\label{defBsigma2}
 W_{0,2}(\spcurve_\sigma;x_1,x_2) 
 &=& B_\sigma(x_1,x_2)  = \Brond_{\fram_\sigma}(x_1-a_\sigma+\arond_\sigma,x_2-a_\sigma+\arond_\sigma)  \cr
 &=& \Brond_{\fram_\sigma}(x_1,x_2)+ \frac{1}{2}\sum_{k=1}^\infty \frac{1}{k!}\,\sum_{d_1,\dots,d_{k+2}}
 \,d\td\xirond_{\fram_\sigma,d_1}(x_1)\,d\td\xirond_{\fram_\sigma,d_2}(x_2)
\prod_{i=3}^{k+2} R_{\sigma,d_{i}}
 \cr
&&  \left< \CL_{\rm Hodge}(\fram_{a,\sigma})\,\CL_{\rm Hodge}(\fram_{b,\sigma})\,\CL_{\rm Hodge}(-\fram_{a,\sigma}-\fram_{b,\sigma})\, \prod_{j=1}^{k+2} \tau_{d_j}\right>_{0,k+2} \cr
\eea
and for $(g,n)=(0,1)$:
\bea
 W_{0,1}(\spcurve_\sigma;x_1) 
 &=& (y(x_1+\arond_\sigma)-\brond_\sigma)\,dx_1 \cr
&=& \yrond_{\fram_\sigma}(x_1)dx_1
+ \frac{1}{2\pi i}\oint_{\d \curve_\sigma} \Brond_{\fram_\sigma}(x_1,x')\,\Phi(x') \cr
&& +  \frac{\ee{-\hat t_{\fram_\sigma,0}}}{4}\sum_{k=2}^\infty \frac{1}{k!}\,\sum_{d_1,\dots,d_{k+1}} 
d\td\xirond_{\fram_\sigma,d_1}(x_1)
\prod_{i=2}^{k+1} R_{\sigma,d_{i}} \cr
&&  \left< \CL_{\rm Hodge}(\fram_{a,\sigma})\,\CL_{\rm Hodge}(\fram_{b,\sigma})\,\CL_{\rm Hodge}(-\fram_{a,\sigma}-\fram_{b,\sigma})\,\tau_{d_1} \prod_{j=2}^{k+1} \tau_{d_j}\right>_{0,k+1} \,\,
\eea


\el

\proof{
Since the proof is quite technical and long, we do it in appendix  \ref{appprooflemmarenorm01}.

Let us just mention that it is proved using the "special geometry" property of the topological recursion.
This property says that the derivative of $W_{g,n}$ with respect to a parameter $t$ on which the spectral curve depends, is the integral of $W_{g,n+1}$ on the dual cycle of $\d ydx/\d t$.
The dual cycle $t^*$ is a cycle such that:
\beq
\frac{\d}{\d t}\,y(x)dx = \int_{x'\in t^*}\,B(x,x').
\eeq

The special geometry property of the topological recursion is  that:
\bt[Special geometry, proved in \cite{EOFg}]\label{thspgeom}
For any spectral curve we have:
\beq
\frac{\d}{\d t}\,W_{g,n}(\spcurve;x_1,\dots,x_n) = \int_{x'\in t^*}\,W_{g,n+1}(\spcurve;x_1,\dots,x_n,x').
\eeq

\et

The proof of lemma \ref{lemmarenorm01} uses that property to show that both sides of lemma  \ref{lemmarenorm01} satisfy the same differential equations with respect to the variables $t_j$'s. 
Moreover thanks to the large radius limit (tropical limit), the two sides obviously coincide when all $t_j=+\infty$, which concludes the proof. 

}

As a corollary of lemma \ref{lemmarenorm01} as well as the expression of $\xirond(x)$ \eq{eqexptdxirond}, we get:
\bc\label{corWgnMgn}
Let $\sigma$ be a vertex of $\Upsilon_\CYX$, and assume that $x_j$, $j=1,\dots,n$ are such that $x_j$ belongs to a cylinder $\curve_{\epsilon_j}$ where $\epsilon_j$ is an half--edge of $\Upsilon_\CYX$ adjacent to $\sigma$, we have for $2-2g-n<0$:
\bea
&& W_{g,n}(\spcurve_\sigma;x_1,\dots,x_n)  \cr
&=& \frac{2^{3g-3+n}}{\ee{\hat t_{\fram_\sigma,0}(2-2g-n)}}\, \sum_{k_1,\dots,k_n}
  \Big< \CL_{\rm Hodge}(\fram_{a,\sigma})\,\CL_{\rm Hodge}(\fram_{b,\sigma})\,\CL_{\rm Hodge}(-\fram_{a,\sigma}-\fram_{b,\sigma})\,\, \ee{l_{1*} \sum_d  R_{\sigma,d}\tau_d}\, \cr
  && \prod_{j=1}^{n} \frac{\gamma_{\fram_\sigma}(k_j/\fram_{\epsilon_j})}{1-\frac{k_j}{\fram_{\epsilon_j}} \,\psi_j}\Big>_{g,n} \quad
\prod_{j=1}^n\, \frac{k_j}{(\fram_{\epsilon_j})^2} \ee{- \,\frac{k_j}{\fram_{\epsilon_j}}\,x_j}\,dx_j  \, , \cr
\eea
and similarly,
\bea
W_{0,2}(\spcurve_\sigma;x_1,x_2) 
&=& B_\sigma(x_1,x_2) = \Brond_{\fram_\sigma}(x_1+\arond_\sigma-a_\sigma,x_2+\arond_\sigma-a_\sigma)  \cr
&=& \Brond_{\fram_\sigma}(x_1,x_2) + \frac{1}{2}\, \sum_{k_1,k_2}\,\, \sum_{k=1}^\infty \frac{1}{k!}  \cr
&&   \Big< \CL_{\rm Hodge}(\fram_{a,\sigma})\,\CL_{\rm Hodge}(\fram_{b,\sigma})\,\CL_{\rm Hodge}(-\fram_{a,\sigma}-\fram_{b,\sigma}) \cr
  && \left(\sum_d  R_{\sigma,d}\tau_d\right)^k\,\,\prod_{j=1}^{2} \frac{\gamma_{\fram_\sigma}(k_j/\fram_{\epsilon_j})}{1-\frac{k_j}{\fram_{\epsilon_j}} \,\psi_j}\,\, \Big>_{0,k+2} \cr
&& \prod_{j=1}^2\, \frac{k_j}{(\fram_{\epsilon_j})^2} \ee{- \,\frac{k_j}{\fram_{\epsilon_j}}\,x_j}\,dx_j \cr
\eea
and
\bea\label{eqW01HodgeMgnR}
W_{0,1}(\spcurve_\sigma;x_1) 
&=& (y(x_1+\arond_\sigma)-\brond_\sigma)\,dx_1  \cr
&=&  \yrond_{\fram_\sigma}(x_1)\,dx_1 + \frac{2}{2\pi i}\oint_{x'\in\d\curve_\sigma} \Brond_{\fram_\sigma}(x_1,x')\,\Phi(x')
+ \frac{\ee{-\hat t_{\fram_\sigma,0}}}{4}\, \sum_{k_1}\,\, \sum_{k=2}^\infty \frac{1}{k!}  \cr
&&   \Big< \CL_{\rm Hodge}(\fram_{a,\sigma})\,\CL_{\rm Hodge}(\fram_{b,\sigma})\,\CL_{\rm Hodge}(-\fram_{a,\sigma}-\fram_{b,\sigma}) \cr
  && \left(\sum_d  R_{\sigma,d}\tau_d\right)^k\,\, \frac{\gamma_{\fram_\sigma}(k_1/\fram_{\epsilon_1})}{1-\frac{k_1}{\fram_{\epsilon_1}} \,\psi_1}\,\, \Big>_{0,k+1} \,\,
 \, \frac{k_1}{(\fram_{\epsilon_1})^2} \ee{- \,\frac{k_1}{\fram_{\epsilon_1}}\,x_1}\,dx_1 \, . \cr
\eea

\ec

\subsection{Invariants of the mirror curve as graphs}

This allows to rewrite theorem \ref{theoremWngclasses} as:

\bc\label{corWnggraphsLhodgerenormtadpoles}
\beq
\begin{array}{l}
W_{g,n}(\spcurve;x_1,\dots,x_n)  \cr
= {\displaystyle \sum_{G\in {\cal G}^{\rm stable}_{g,n}} }\frac{1}{\#{\rm Aut}(G)} \quad
 \prod_{v={\rm vertices}} {2^{3g_v-3+n_v}\,\ee{\hat t_{\sigma_v,0}(2g_v-2+n_v)}} \cr
 \left(\int_{\overline\modsp_{g_v,n_v}} \CL_{\rm Hodge}(\fram_{a,\sigma_v})\CL_{\rm Hodge}(\fram_{b,\sigma_v})\CL_{\rm Hodge}(-\fram_{a,\sigma_v}-\fram_{b,\sigma_v}) \ee{l_{1*}\sum_d  R_{\sigma_v,d}\tau_d} \prod_{h\in E_v} \tau_{d_{h}}\,\right)\, \cr
{\displaystyle  \prod_{(h_+,h_-)={\rm closed\,edges}}} \hat E_{\sigma_{v(h_+)},d_{h_+};\sigma_{v(h_-)},d_{h_-}}  \cr
 \qquad {\displaystyle  \prod_{j=1,\, h_j={\rm open\,half-edges}}^n} \frac{-1}{2\pi i}\oint_{x'_j\in \d\curve_{\sigma_{v(h_j)}}} \td \xirond_{\sigma_{v(h_j)},d_{h_j}}(x'_j-\arond_{\sigma_{v(h_j)}})\,\,B(\spcurve;x'_j,x_j) \cr
 \end{array}
\eeq
where the sum is only over stable graphs (every vertex $v$ is such that $2-2g_v-n_v<0$),
and where
\bea
\hat E_{\sigma,d;\sigma',d'} 
&=& \frac{1}{(2\pi i)^2}\oint_{x\in \d\curve_\sigma}\,\oint_{x'\in \d\curve_{\sigma'}} \td\xirond_{\sigma,d}(x-\arond_\sigma)\,\Big(B(\spcurve;x,x') \cr
&& -\delta_{\sigma,\sigma'} B(\spcurve_\sigma;x-\arond_\sigma,x'-\arond_\sigma) \Big)\,\td\xirond_{\sigma',d'}(x'-\arond_{\sigma'}) . \cr
\eea
\ec
If $x_j$ lies on a cylinder $\curve_{\epsilon_j}$ where $\epsilon_j$ is any half-edge (not necessarily corresponding to the non-compact half-edge of $\Upsilon_\CYX$ where the special Lagrangian brane $L$ is ending), we have the expansion
\beq\label{eqexpxirondgamma}
\td\xirond_{\sigma(\epsilon_j),d}(x_j) = -\,\sum_{k=1}^\infty \frac{\gamma_{\fram_{\sigma(\epsilon_j)}}(k/\fram_{\epsilon_j})}{\fram_{\epsilon_j}}\,\, \ee{-\,\frac{k}{\fram_{\epsilon_j}}\,x_j}\,\,\frac{k^{d}}{(\fram_{\epsilon_j})^{d}}.
\eeq
Using this expansion of $\td \xirond$, we easily arrive at:

\bt\label{thWngtdHtdF}
If $x_j\in \curve_{\epsilon_j}$ where $\epsilon_j$ is an half edge of $\Upsilon_\CYX$ (not necessarily corresponding to a non-compact edge, neither the one on which the brane $L$ ends), we have, for $(g,n) \in \mathbb{N}^2 \backslash \{(0,0),(1,0)\}$,
\bea
&& W_{g,n}(\spcurve;x_1,\dots,x_n)  \cr
&=& \sum_{G\in {\cal G}^{\rm stable}_{g,n}} \frac{1}{\#{\rm Aut}(G)} \quad
 \prod_{v={\rm vertices}} 
\td {\cal H}_{g_v,n_v,\sigma_v}(\{d_{h}/\fram_{\epsilon_h} \}_{h\in E_v})
 \cr
&& \prod_{e=(h_+,h_-)={\rm closed\,edges}} \td {\cal F}_{\epsilon_{h_+},d_{h_+};\epsilon_{h_-},d_{h_-} } 
\qquad \qquad \prod_{j=1,\, h_j={\rm open\, half-edges}}^n  d\td J_{\epsilon_{h_j},d_{h_j}}(x_j) \cr
\eea
where the sum is only over stable graphs (every vertex $v$ is such that $2-2g_v-n_v<0$),
and where
\bea
\td {\cal H}_{g,n,\sigma}(k_1,\dots,k_n) 
&=&  \frac{2^{3g-3+n}}{\ee{-\hat t_{\sigma,0}(2g-2+n)}}\,\int_{\overline\modsp_{g,n}} \CL_{\rm Hodge}(\fram_{a,\sigma})\CL_{\rm Hodge}(\fram_{b,\sigma})\CL_{\rm Hodge}(-\fram_{a,\sigma}-\fram_{b,\sigma}) \cr
&& \ee{l_{1*}\sum_d  R_{\sigma,d}\tau_d} \quad \prod_{i=1}^n \,\,\frac{\gamma_{\fram_\sigma}(k_i)}{1-k_i\,\psi_i} \, ,
\eea
\bea\label{eqFeded}
\td {\cal F}_{\epsilon,d;\epsilon',d'} 
&=& \frac{1}{\,\fram_\epsilon\,\fram_{\epsilon'}\,\,(2\pi i)^2}\oint_{x\in \d_\epsilon \curve_{\sigma(\epsilon)}}\,\oint_{x'\in \d_{\epsilon'} \curve_{\sigma(\epsilon')}} \ee{- \frac{d}{\fram_\epsilon}\,(x-\arond_{\sigma(\epsilon)})}\,\Big(B(\spcurve;x,x') \cr
&& -\delta_{\sigma(\epsilon),\sigma(\epsilon')} B(\spcurve_{\sigma(\epsilon)};x-\arond_{\sigma(\epsilon)},x'-\arond_{\sigma(\epsilon)}) \Big)\,\ee{- \frac{d'}{\fram_{\epsilon'}}\,(x'-\arond_{\sigma(\epsilon')})} 
\eea
and 
\beq\label{defdJsigmad}
d\td J_{\epsilon,d}(x)
=\frac{1}{\fram_\epsilon\,\,2\pi i}\oint_{x'\in \d_\epsilon\curve_{\sigma(\epsilon)}} \ee{-\,\frac{d}{\fram_\epsilon}\,x'}\,\, \,\,B(\spcurve;x'+\arond_{\sigma(\epsilon)},x) .
\eeq

\et

%

\subsection{Weight of edges} \label{secrenormalize}

In the graph sum, the weight of edges is given by formula \eq{eqFeded}, it involves the double Fourrier transform of the Bergman kernel.

\subsubsection{More geometry: the Bergman kernel}

\begin{figure}[t]
\centering
$$\includegraphics[width=14cm]{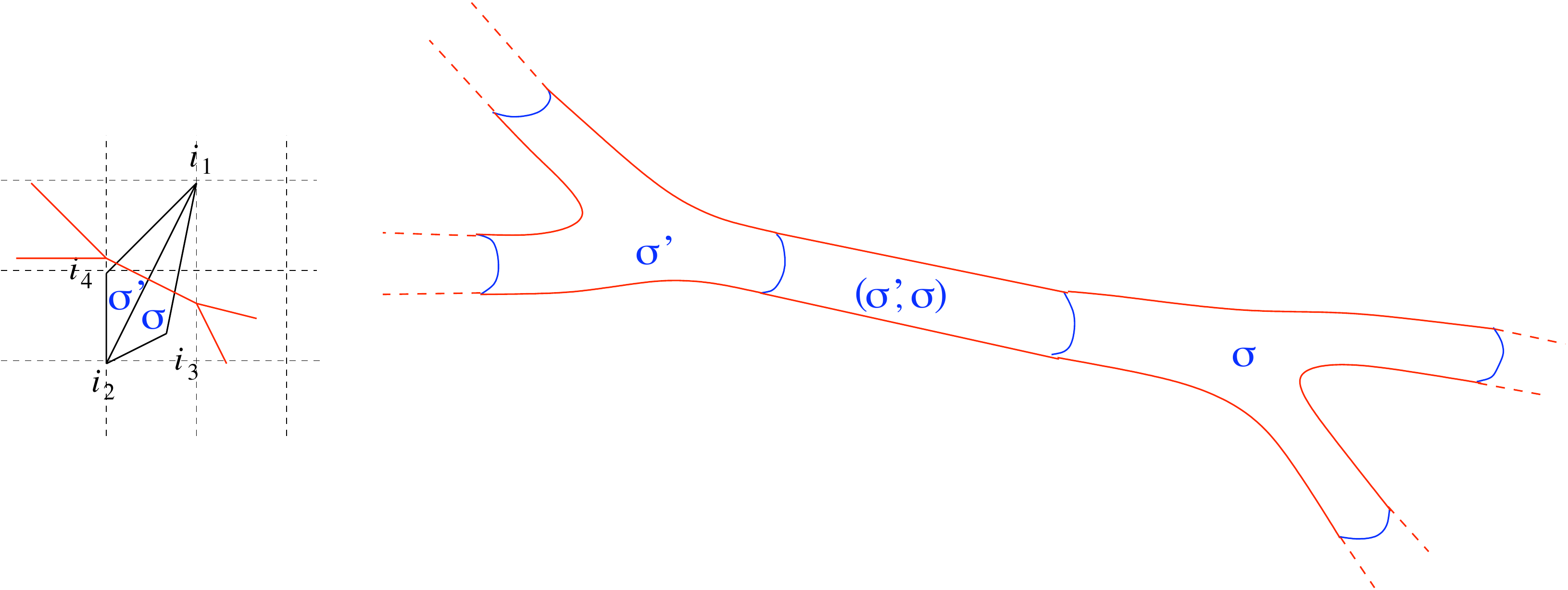}  $$
\caption{The curve $\curve$ is a union of cylinders $\curve_{\sigma,\sigma'}$ corresponding to edges of the toric graph $\Upsilon_\CYX$ and of pairs of pants $\curve_\sigma$ corresponding to vertices of $\Upsilon_\CYX$. Its Bergman kernel can be obtained as a combination of the Bergman kernels of each pieces. Notice that the framing of the edge $\sigma,\sigma'$ is $\fram_{\sigma,\sigma'}=-\fram_{\sigma',\sigma} = \fram_{b,\sigma'}=-\fram_{b,\sigma} = \beta_{i_2}-\beta_{i_1}$.
\label{figcurveBergman}}
\end{figure}

Recall that the curve $\curve$ is a union of cylinders $\curve_{\sigma,\sigma'}$ and of pairs of pants $\curve_\sigma$, cf fig.\ref{figspcurvelocP2framed} and fig \ref{figcurveBergman}.

The following lemma allows to express the Bergman kernel of the full curve $\curve$ in terms of Bergman kernels of its pieces $\curve_\sigma$ and $\curve_{\sigma,\sigma'}$.
Recall that we write (for any curve $\spcurve$) that the Bergman kernel is the double derivative of the log of the prime form $E$:
\beq
B(\spcurve;x_1,x_2) = d_{x_1}\,\otimes\,d_{x_2}\,\,\ln{E(\spcurve;x_1,x_2)}.
\eeq

We have:
\bl \label{lemmaBergman}
Let $\sigma, \sigma'$ be two vertices. 
Let $x\in \curve_\sigma$ and $x'\in\curve_{\sigma'}$, then we have
\bea
&& \ln E(\spcurve;\arond_\sigma+x,\arond_{\sigma'}+x') -   \delta_{\sigma,\sigma'} \ln E(\spcurve_\sigma;x,x') \cr
&=& \,\frac{1}{(2\pi i)^2}\oint_{x_2\in \d_{\sigma}\curve_{\sigma'}}\,\oint_{x_1\in \d_{\sigma'}\curve_\sigma}\, \ln E(\spcurve_\sigma;x,{x_1})\, \times \cr
&& \qquad \qquad \quad \qquad \qquad  \qquad \qquad \times B(\spcurve_{\sigma,\sigma'};x_1,{x_2+\arond_{\sigma'}-\arond_\sigma})\, \ln E(\spcurve_{\sigma'};x_2,x') \cr
&& +  \sum_{\sigma_1}\,\frac{1}{(2\pi i)^2}\oint_{x_2\in \d_{\sigma}\curve_{\sigma_1}}\,\oint_{x_1\in \d_{\sigma_1}\curve_\sigma}\, \ln E(\spcurve_\sigma;x,{x_1})\,B(\spcurve_{\sigma,\sigma_1};x_1,{x_2+\arond_{\sigma_1}-\arond_\sigma})\, \cr
&& \qquad \qquad \Big( \ln E(\spcurve;\arond_{\sigma_1}+x_2,\arond_{\sigma'}+x') - \delta_{\sigma_1,\sigma'}\,\ln E(\spcurve_{\sigma'};x_2,x')\Big) \cr
\eea

where
\beq
B(\spcurve_{\sigma,\sigma'};x_1,x_2)
 = A_{\sigma,\sigma'}\,\frac{1}{(\fram_{\sigma,\sigma'})^2}\,\,\frac{\ee{-\,\frac{x_1}{\fram_{\sigma,\sigma'}}}\,\ee{-\,\frac{x_2}{\fram_{\sigma,\sigma'}}}}{\left(\ee{-\,\frac{x_1}{\fram_{\sigma,\sigma'}}}-\ee{-\,\frac{x_2}{\fram_{\sigma,\sigma'}}}\right)^2}\,\,\,dx_1\otimes dx_2
\eeq
is the Bergman kernel on the cylinder $\curve_{\sigma,\sigma'}$, 
$A_{\sigma,\sigma'}$ is the adjacency matrix of the toric graph, i.e. $A_{\sigma,\sigma'}=1$ if $\sigma$ and $\sigma'$ are neighbors and $0$ otherwise,
and $\fram_{\sigma,\sigma'}$ is the framing of the edge $(\sigma,\sigma')$ as defined in def \ref{defframedges}.

\begin{figure}[t]
\centering
$$\includegraphics[width=10cm]{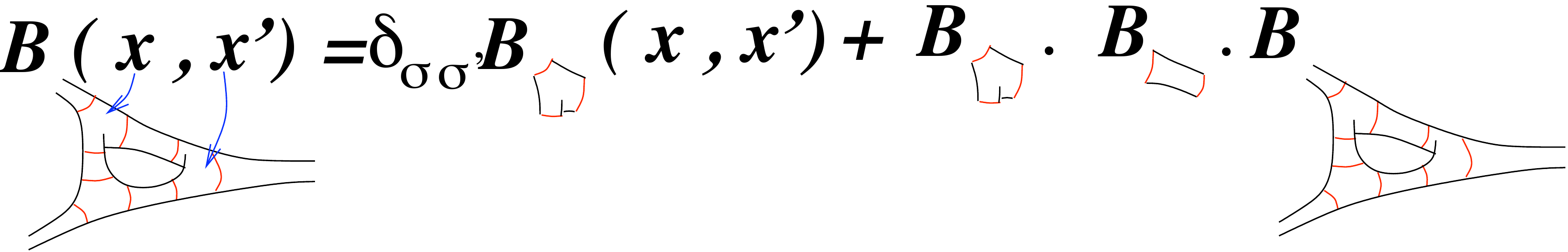}  $$
\caption{Lemma \ref{lemmaBergman} shows that the Bergman kernel of the full curve $\curve$, can be constructed by combining Bergman kernels of the pairs of pants $\curve_\sigma$ and Bergman kernels of cylinders $\curve_{\sigma,\sigma'}$. In some sense $B_{\rm total} = B_{\rm pant} + B_{\rm pant}\,B_{\rm cylinder}\,B_{\rm total}$.
\label{figBergmancombin}}
\end{figure}

\el
This lemma is illustrated in fig. \ref{figBergmancombin}.

\proof{
This lemma is proved in appendix \ref{applemmaBergman}. The proof is only complex analysis on $\curve$, it consists in writing Cauchy residue formula and moving the integration contours.}

\subsubsection{Renormalizing edges}

Now, it remains to compute the weights \eq{eqFeded} attached to edges of the graph decomposition of $W_{g,n}(\spcurve)$ through corollary \ref{corWnggraphsLhodgerenormtadpoles}, i.e. the integrals

\bea
\td {\cal F}_{\epsilon,d;\epsilon',d'} 
&=& \frac{1}{\fram_\epsilon\,\fram_{\epsilon'}\,(2\pi i)^2}\oint_{x\in \d_\epsilon \curve_{\sigma(\epsilon)}}\,\oint_{x'\in \d_{\epsilon'} \curve_{\sigma(\epsilon')}} \ee{- \frac{d}{\fram_\epsilon}\,x}\,\Big(B(\spcurve;x+\arond_{\sigma(\epsilon)},x'+\arond_{\sigma(\epsilon')}) \cr
&& -\delta_{\sigma(\epsilon),\sigma(\epsilon')} B(\spcurve_{\sigma(\epsilon)};x,x') \Big)\,\ee{- \frac{d'}{\fram_{\epsilon'}}\,x'} .\cr
\eea

From lemma \ref{lemmaBergman}  above, we prove that:
\bp\label{propedgestdFF}

The edge weight $\td F_{\epsilon,d;\epsilon',d'}$ satisfies:
\beq
 \td {\cal F}_{\epsilon,d;\epsilon',d'}  = {\cal F}_{\epsilon,d;\epsilon',d'}  + \sum_{\epsilon_1,d_1} \sum_{\epsilon_2,d_2} {\cal F}_{\epsilon,d;\epsilon_1,d_1}\,\, \td {\cal H}_{0,2,\sigma(\epsilon_1)}(d_1/\fram_{\epsilon_1};d_2/\fram_{\epsilon_2})\,\,\td {\cal F}_{\epsilon_2,d_2;\epsilon',d'}
\eeq
where
\beq
{\cal F}_{\epsilon,d;\epsilon',d'} =  \frac{d}{\fram_\epsilon^2}\,\,\ee{- \frac{d}{\fram_\epsilon}\,(\arond_{\sigma(\epsilon')}-\arond_{\sigma(\epsilon)})}\,\,A_{\epsilon,\epsilon'}\,\,\, \delta_{d,d'}
\eeq
and
\bea
\td {\cal H}_{0,2,\sigma}(k;k')
&=& {\cal H}_{0,2,\sigma}(k;k') \cr
&& + \frac{1}{2}\sum_{n=1}^\infty\frac{1}{n!}  
\Big< \CL_{\rm Hodge}(\fram_{a,\sigma})\,\CL_{\rm Hodge}(\fram_{b,\sigma})\,\CL_{\rm Hodge}(-\fram_{a,\sigma}-\fram_{b,\sigma}) \cr
&& \left(\sum_d R_{\sigma,d}\,\tau_d\right)^n
\,\,\frac{\gamma_{\fram_\sigma}(k)}{1-k\,\psi}
\,\,\frac{\gamma_{\fram_{\sigma'}}(k')}{1-k'\,\psi'} \Big>_{0,n+2}
\eea

and
\beq
{\cal H}_{0,2,\sigma}(k;k')
=
\,\frac{1}{k+k'}\,\,\gamma_{\fram_{\sigma}}(k)\,\gamma_{\fram_{\sigma'}}(k').
\eeq

\ep

This lemma means that:
\beq
\td {\cal F} = {\cal F} + {\cal F} \td {\cal H}_{0,2} {\cal F} + {\cal F} \td {\cal H}_{0,2}  {\cal F} \td {\cal H}_{0,2} {\cal F} + {\cal F} \td {\cal H}_{0,2} {\cal F} \td {\cal H}_{0,2} {\cal F} \td {\cal H}_{0,2} F+ \dots 
\eeq
which is illustrated as:
$$ \includegraphics[width=13cm]{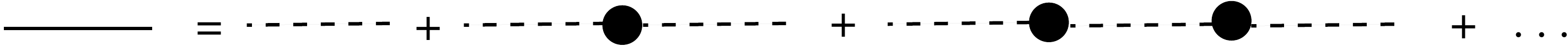} $$
i.e. ${\cal H}_{0,2,\sigma}$ can be viewed as the $(0,2)$ vertex, and ${\cal F}$ as an edge weight.

\proof{
From lemma \ref{lemmaBergman} in appendix \ref{applemmaBergman}, we see that we first have to compute:
\bea
{\cal F}_{\epsilon,d;\epsilon',d'} 
&=& \frac{1}{\fram_\epsilon\,\fram_{\epsilon'}\,(2\pi i)^4}
\,\oint_{x_1\in \d_{\sigma(\epsilon')}\curve_{\sigma(\epsilon)}}\, 
\oint_{x\in \d_\epsilon \curve_{\sigma(\epsilon)}}
\oint_{x_2\in \d_{\sigma(\epsilon)}\curve_{\sigma(\epsilon')}}
\,\oint_{x'\in \d_{\epsilon'} \curve_{\sigma(\epsilon')}}  
\,\, \frac{d}{\fram_{\epsilon}}\,\,\frac{d'}{\fram_{\epsilon'}}\,\,dx\,\,dx' \cr
&& \ee{- \frac{d}{\fram_\epsilon}\,x}\,\ln E(\spcurve_{\sigma(\epsilon)};x,{x_1})\,B(\spcurve_{(\sigma(\epsilon),\sigma(\epsilon'))};x_1,{x_2+\arond_{\sigma'}-\arond_\sigma})\,\cr
&& \qquad  \ln E(\spcurve_{\sigma(\epsilon')};x_2,x') 
\,\ee{- \frac{d'}{\fram_{\epsilon'}}\,x'} 
\eea
which is non-vanishing only if $\sigma(\epsilon)$ and $\sigma(\epsilon')$ are adjacent vertices in $\Upsilon_\CYX$.

Moreover, if $\epsilon$ (resp. $\epsilon'$) is not the half-edge linking $\sigma(\epsilon)$ to $\sigma(\epsilon')$ (resp. $\sigma(\epsilon')$ to $\sigma(\epsilon)$), we may push the integration contour for $x$  (resp. $x'$) towards the puncture of $\curve_{\sigma(\epsilon)}$ (resp. $\curve_{\sigma(\epsilon')}$) in the direction of the half-edge $\epsilon$ (resp. $\epsilon'$) without meeting any singularity, and thus the result vanishes.
${\cal F}_{\epsilon,d;\epsilon',d'} $ is thus proportional to the adjacency matrix $A_{\epsilon,\epsilon'}$ which is 1 if the half-edges $\epsilon$ and $\epsilon'$ form an edge of $\Upsilon_\CYX$, and 0 otherwise:
\bea
{\cal F}_{\epsilon,d;\epsilon',d'} 
&=& \frac{A_{\epsilon,\epsilon'}}{\fram_\epsilon\,\fram_{\epsilon'}\,(2\pi i)^4}
\,\oint_{x_1\in \d_{\epsilon}\curve_{\sigma(\epsilon)}}\, 
\oint_{x\in \d_\epsilon \curve_{\sigma(\epsilon)}}
\oint_{x_2\in \d_{\epsilon'}\curve_{\sigma(\epsilon')}}
\,\oint_{x'\in \d_{\epsilon'} \curve_{\sigma(\epsilon')}}  
\,\, \frac{d}{\fram_{\epsilon}}\,\,\frac{d'}{\fram_{\epsilon'}}\,\,dx\,\,dx' \cr
&& \ee{- \frac{d}{\fram_\epsilon}\,x}\,\ln E(\spcurve_{\sigma(\epsilon)};x,{x_1})\,B(\spcurve_{(\sigma(\epsilon),\sigma(\epsilon'))};x_1,{x_2+\arond_{\sigma'}-\arond_\sigma})\,\cr
&& \qquad  \ln E(\spcurve_{\sigma(\epsilon')};x_2,x') 
\,\ee{- \frac{d'}{\fram_{\epsilon'}}\,x'} .
\eea

So, from now on, we assume that $(\epsilon,\epsilon')$ is the edge  linking $\sigma(\epsilon)$ to $\sigma(\epsilon')$.

$$\includegraphics[width=10cm]{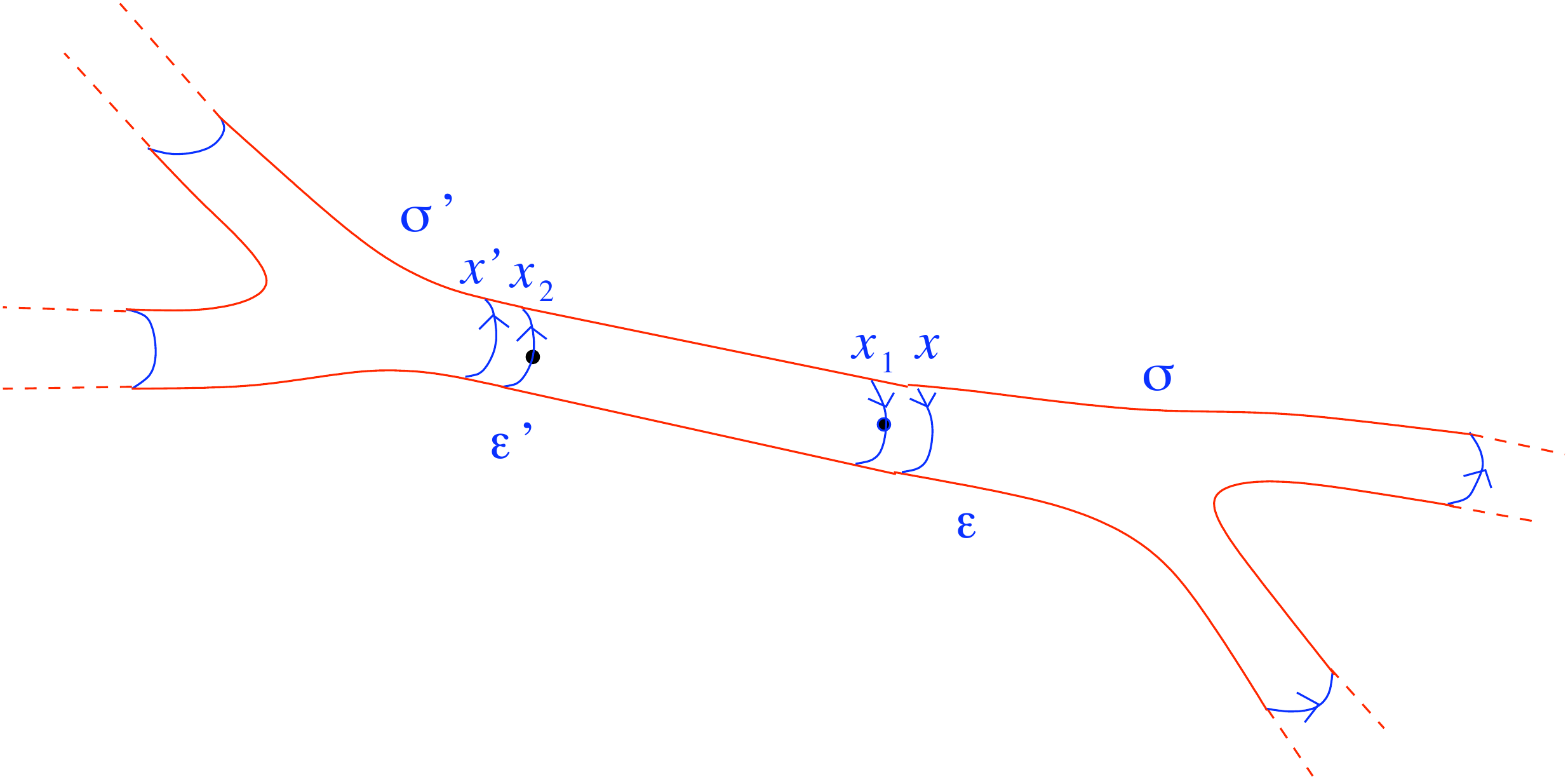}  $$

We can push the integration contour of $x$ through that of $x_1$ and send it to the puncture of $\curve_{\sigma(\epsilon)}$ where it vanishes, we only pick a residue at $x=x_1$, and similarly for $x'$ and $x_2$, and we get:
\bea
{\cal F}_{\epsilon,d;\epsilon',d'} 
&=& \frac{A_{\epsilon,\epsilon'}}{\fram_\epsilon\,\fram_{\epsilon'}\,(2\pi i)^2}
\,\oint_{x_1\in \d_{\epsilon}\curve_{\sigma(\epsilon)}}\, 
\oint_{x_2\in \d_{\epsilon'}\curve_{\sigma(\epsilon')}}  \cr
&& \ee{- \frac{d}{\fram_\epsilon}\,x_1}\,\,B(\spcurve_{(\epsilon,\epsilon')};x_1,{x_2+\arond_{\sigma'}-\arond_\sigma})\, 
\,\ee{- \frac{d'}{\fram_{\epsilon'}}\,x_2} .\cr
\eea
Similarly, on the cylinder $\curve_{(\epsilon,\epsilon')}$, we can push the integration contour of $x_1$ through that of $x_2$ and send it to the puncture where it vanishes, we only pick a residue at $x_1=x_2+\arond_{\sigma'}-\arond_\sigma$, and we get:
\bea
{\cal F}_{\epsilon,d;\epsilon',d'} 
&=& \frac{d}{\fram_\epsilon}\,\,\frac{A_{\epsilon,\epsilon'}}{\fram_\epsilon\,\fram_{\epsilon'}\,2\pi i}
\oint_{x_2\in \d_{\epsilon'}\curve_{\sigma(\epsilon')}}   \,\,\ee{- \frac{d}{\fram_\epsilon}\,(x_2+\arond_{\sigma'}-\arond_\sigma)}\,\,
\,\ee{- \frac{d'}{\fram_{\epsilon'}}\,x_2}\,\,dx_2 .\cr
\eea
Then, notice that if $A_{\epsilon,\epsilon'}\neq 0$, this means that $(\epsilon,\epsilon')$ is an edge of the toric graph and thus $\fram_{\epsilon'}=-\fram_\epsilon$. Consider the variable $z=\ee{-x_2/\fram_\epsilon}=\ee{x_2/\fram_{\epsilon'}}$, we thus have
\bea
{\cal F}_{\epsilon,d;\epsilon',d'} 
&=& \frac{-d}{\fram_\epsilon^3}\,\,\ee{- \frac{d}{\fram_\epsilon}\,(\arond_{\sigma'}-\arond_\sigma)}\,\,\frac{A_{\epsilon,\epsilon'}}{2\pi i}
\oint_{x_2\in \d_{\epsilon'}\curve_{\sigma(\epsilon')}}   \,\,\ee{- \frac{d-d'}{\fram_\epsilon}\,x_2}\,\,dx_2 \cr
&=& \frac{d}{\fram_\epsilon^3}\,\,\ee{- \frac{d}{\fram_\epsilon}\,(\arond_{\sigma'}-\arond_\sigma)}\,\,{A_{\epsilon,\epsilon'}}
\Res_{z\to 0}   \,\,z^{d-d'}\,\,\fram_\epsilon\,\,\frac{dz}{z} \cr
&=& \frac{d}{\fram_\epsilon^2}\,\,\ee{- \frac{d}{\fram_\epsilon}\,(\arond_{\sigma'}-\arond_\sigma)}\,\,A_{\epsilon,\epsilon'}\,\,\, \delta_{d,d'}.
\eea

Then, using lemma \ref{lemmaBergman}, we have:
\bea
&& \td {\cal F}_{\epsilon,d;\epsilon',d'}  - {\cal F}_{\epsilon,d;\epsilon',d'} \cr
&=& \frac{1}{\fram_\epsilon\,\fram_{\epsilon'}\,(2\pi i)^4}\,\sum_{\sigma_1}
\,\oint_{x_1\in \d_{\sigma_1}\curve_{\sigma(\epsilon)}}\, 
\oint_{x\in \d_\epsilon \curve_{\sigma(\epsilon)}}
\oint_{x_2\in \d_{\sigma(\epsilon)}\curve_{\sigma_1}}
\,\oint_{x'\in \d_{\epsilon'} \curve_{\sigma(\epsilon')}}  
\,\, \frac{d}{\fram_{\epsilon}}\,\,\frac{d'}{\fram_{\epsilon'}}\,\,dx\,\,dx' \cr
&& \ee{- \frac{d}{\fram_\epsilon}\,x}\,\ln E(\spcurve_{\sigma(\epsilon)};x,{x_1})\,B(\spcurve_{(\sigma(\epsilon),\sigma_1)};x_1,{x_2+\arond_{\sigma_1}-\arond_{\sigma(\epsilon)}})\, \cr
& & \Big(\ln E(\spcurve;\arond_{\sigma_1}+x_2,\arond_{\sigma(\epsilon')}+x') -\delta_{\sigma_1,\sigma(\epsilon')} \ln E(\spcurve_{\sigma(\epsilon')};x_2,x')\Big)
\,\ee{- \frac{d'}{\fram_{\epsilon'}}\,x'} \cr
\eea
where the sum vanishes if $\sigma_1$ is not a neighbor of $\sigma(\epsilon)$.

Again, if $\epsilon$ is not the half-edge linking $\sigma(\epsilon)$ to $\sigma_1$, we can push the integration contour of $x$ towards the puncture of $\curve_{\sigma(\epsilon)}$ without meeting any singularity and the result vanishes.
In other words, $\sigma_1$ has to be chosen as the vertex on the other side of the half-edge $\epsilon$.

We can then push the integration contour of $x$ through that of $x_1$ and send it to the puncture of $\curve_{\sigma(\epsilon)}$ where it vanishes, we only pick a residue at $x=x_1$
\bea
&& \td {\cal F}_{\epsilon,d;\epsilon',d'}  - {\cal F}_{\epsilon,d;\epsilon',d'} \cr
&=& \frac{1}{\fram_\epsilon\,\fram_{\epsilon'}\,(2\pi i)^3}\,
\,\oint_{x_1\in \d_{\epsilon}\curve_{\sigma(\epsilon)}}\, 
\oint_{x_2\in \d_{\sigma(\epsilon)}\curve_{\sigma_1}}
\,\oint_{x'\in \d_{\epsilon'} \curve_{\sigma(\epsilon')}}  
\,\, \frac{d'}{\fram_{\epsilon'}}\,\,dx' \cr
&& \ee{- \frac{d}{\fram_\epsilon}\,x_1}\,\,B(\spcurve_{(\sigma(\epsilon),\sigma_1)};x_1,{x_2+\arond_{\sigma_1}-\arond_{\sigma(\epsilon)}})\, \cr
& & \Big(\ln E(\spcurve;\arond_{\sigma_1}+x_2,\arond_{\sigma(\epsilon')}+x') -\delta_{\sigma_1,\sigma(\epsilon')} \ln E(\spcurve_{\sigma(\epsilon')};x_2,x')\Big)
\,\ee{- \frac{d'}{\fram_{\epsilon'}}\,x'} \cr
\eea
then we can push the integration contour of $x_1$ through that of $x_2$ and send it to the puncture where it vanishes, we only pick a residue at $x_1=x_2+\arond_{\sigma_1}-\arond_{\sigma(\epsilon)}$
\bea
&& \td {\cal F}_{\epsilon,d;\epsilon',d'}  - {\cal F}_{\epsilon,d;\epsilon',d'} \cr
&=& \frac{1}{\fram_\epsilon\,\fram_{\epsilon'}\,(2\pi i)^2}\,
\oint_{x_2\in \d_{\sigma(\epsilon)}\curve_{\sigma_1}}
\,\oint_{x'\in \d_{\epsilon'} \curve_{\sigma(\epsilon')}}  
\,\,\frac{d}{\fram_{\epsilon}}\,\,\frac{d'}{\fram_{\epsilon'}}\,\,dx_2\,\,dx' \,\,\, \ee{- \frac{d}{\fram_\epsilon}\,(x_2+\arond_{\sigma_1}-\arond_{\sigma(\epsilon)})} \cr
&& \Big(\ln E(\spcurve;\arond_{\sigma_1}+x_2,\arond_{\sigma(\epsilon')}+x') -\delta_{\sigma_1,\sigma(\epsilon')} \ln E(\spcurve_{\sigma(\epsilon')};x_2,x')\Big)
\,\ee{- \frac{d'}{\fram_{\epsilon'}}\,x'} \cr
&=&  \sum_{\epsilon_1,d_1} F_{\epsilon,d;\epsilon_1,d_1}
 \frac{1}{\,\fram_{\epsilon'}\,(2\pi i)^2}\, 
\oint_{x_2\in \d_{\epsilon_1}\curve_{\sigma(\epsilon_1)}}
\,\oint_{x'\in \d_{\epsilon'} \curve_{\sigma(\epsilon')}}  
\,\,\frac{d'}{\fram_{\epsilon'}}\,\,dx_2\,\,dx' \,\,\, \ee{ \frac{d_1}{\fram_{\epsilon_1}}\,x_2} \cr
&& \Big(\ln E(\spcurve;\arond_{\sigma(\epsilon_1)}+x_2,\arond_{\sigma(\epsilon')}+x') -\delta_{\sigma(\epsilon_1),\sigma(\epsilon')} \ln E(\spcurve_{\sigma(\epsilon')};x_2,x')\Big)
\,\ee{- \frac{d'}{\fram_{\epsilon'}}\,x'} \cr
\eea
where we have used that ${\cal F}_{\epsilon,d;\epsilon_1,d_1}$ is non-vanishing only if $\epsilon_1$ is the half edge linking $\sigma_1$ to $\sigma(\epsilon)$, and thus $\fram_{\epsilon_1}=-\fram_\epsilon$.
One would be tempted to identify this last integral with $\td {\cal F}_{\epsilon_1,-d_1;\epsilon',d'}$, but it is not possible because of the wrong sign of $d_1$ in the exponential.

Instead, we insert another integral:
\bea\label{eqinsertW02tdF}
&& \td {\cal F}_{\epsilon,d;\epsilon',d'}  - {\cal F}_{\epsilon,d;\epsilon',d'} \cr
&=&  - \sum_{\epsilon_1,d_1} F_{\epsilon,d;\epsilon_1,d_1}
 \frac{\fram_{\epsilon_1}}{d_1\,\fram_{\epsilon'}\,(2\pi i)^2}\, 
\oint_{x_2\in \d_{\epsilon_1}\curve_{\sigma(\epsilon_1)}}
\,\oint_{x'\in \d_{\epsilon'} \curve_{\sigma(\epsilon')}}  
\Res_{x_3\to x_2} 
\,\,\frac{d'}{\fram_{\epsilon'}}\,\,dx' \,\,\, \cr
&& \,\,\ee{ \frac{d_1}{\fram_{\epsilon_1}}\,x_2}\,\, 
B(\spcurve_{\sigma(\epsilon_1)}; x_2,x_3)
\Big(\ln E(\spcurve;\arond_{\sigma(\epsilon_1)}+x_3,\arond_{\sigma(\epsilon')}+x') \cr
&& -\delta_{\sigma(\epsilon_1),\sigma(\epsilon')} \ln E(\spcurve_{\sigma(\epsilon')};x_3,x')\Big)
\,\ee{- \frac{d'}{\fram_{\epsilon'}}\,x'} \cr
\eea
and we deform the integration contour of $x_3$ (i.e. a small circle around $x_2$) into a pair of circles around the cylinder, one on each side of the integration contour of $x_2$.
$$\includegraphics[height=5.5cm]{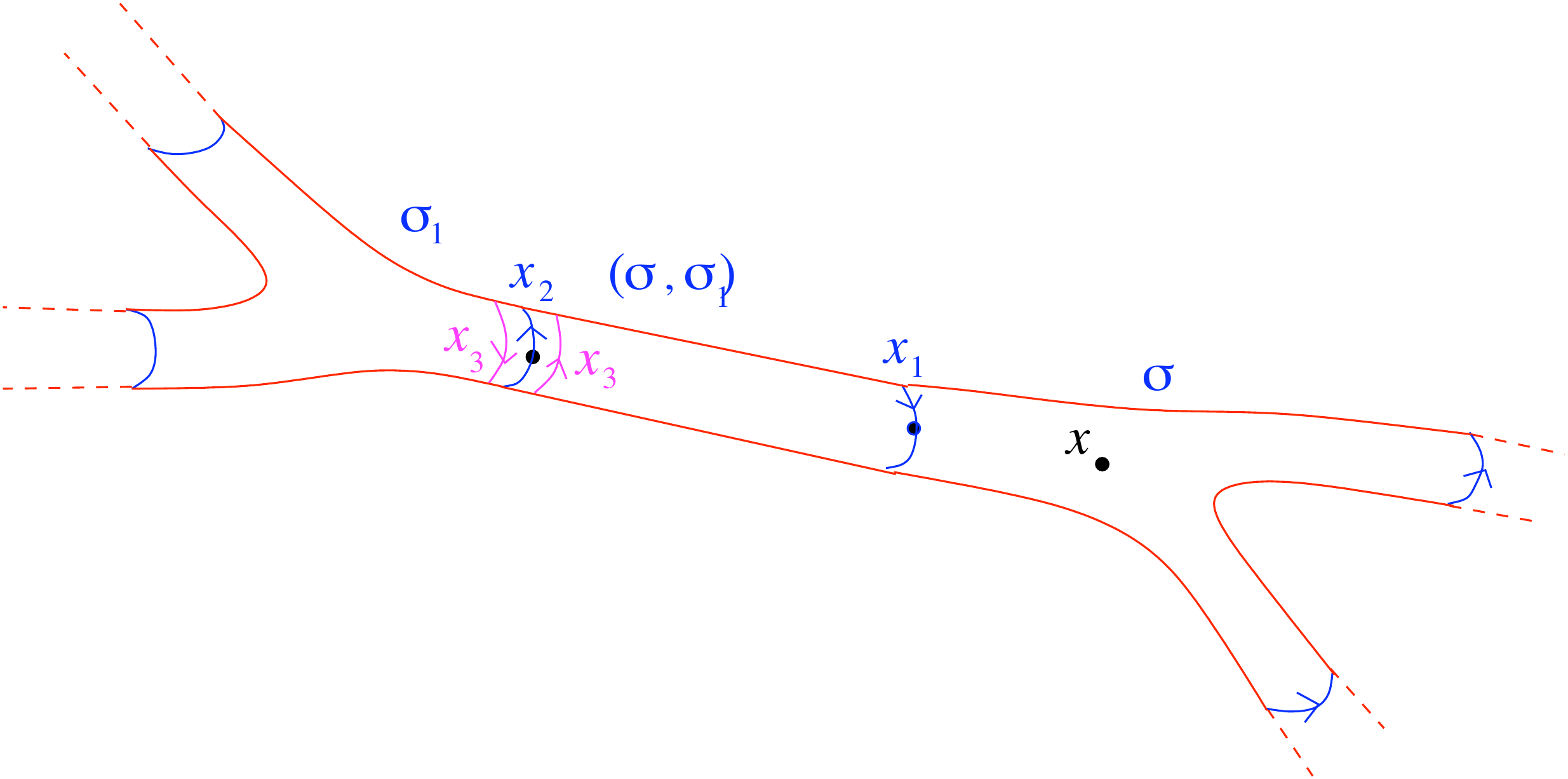}  $$

Then, we push the $x_3$ circle which is inside $\curve_{\sigma_1}$ through the pair of pants, and thus we deform the integration contour for $x_3$ into $ \d\curve_{\sigma_1}$:
$$ \includegraphics[width=10cm]{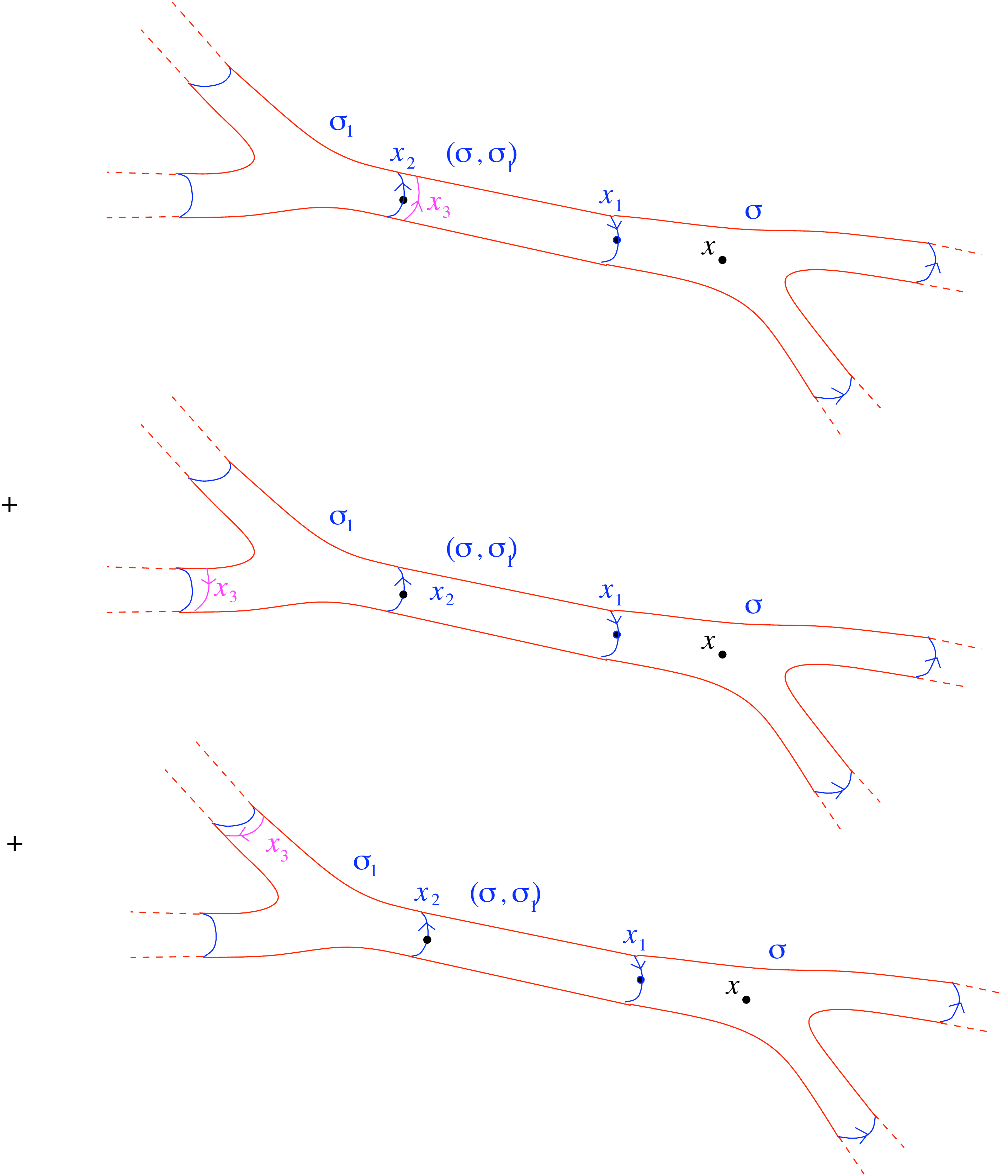} $$
Eventually we have
\bea
&& \td {\cal F}_{\epsilon,d;\epsilon',d'}  - {\cal F}_{\epsilon,d;\epsilon',d'} \cr
&=& -  \sum_{\epsilon_1,d_1} F_{\epsilon,d;\epsilon_1,d_1}
\,\,\sum_{\epsilon''} \frac{\fram_{\epsilon_1}}{d_1\,\fram_{\epsilon'}\,\,(2\pi i)^3}\, 
\,\oint_{x3\in \d_{\epsilon''} \curve_{\sigma(\epsilon_1)}}   
\,\oint_{x_2\in \d_{\epsilon_1}\curve_{\sigma(\epsilon_1)}}
\,\oint_{x'\in \d_{\epsilon'} \curve_{\sigma(\epsilon')}}  
\,\,\frac{d'}{\fram_{\epsilon'}}\,\,dx' \,\,\, \cr
&& \,\,\ee{ \frac{d_1}{\fram_{\epsilon_1}}\,x_2}
\,\,B(\spcurve_{\sigma(\epsilon_1)}; x_2,x_3)
\Big(\ln E(\spcurve;\arond_{\sigma(\epsilon_1)}+x_3,\arond_{\sigma(\epsilon')}+x') \cr
&& -\delta_{\sigma(\epsilon_1),\sigma(\epsilon')} \ln E(\spcurve_{\sigma(\epsilon')};x_3,x')\Big)
\,\ee{- \frac{d'}{\fram_{\epsilon'}}\,x'} \cr
\eea
where $\sum_{\epsilon''}$ means the sum over the 3 boundaries of $\curve_{\sigma_1}$, and where $x_2$ is integrated in the interior of $\curve_{\sigma_1}$.

Then, notice that the following integral (whose integration contour leaves $x_3$ near one of the punctures):
\beq
\,\oint_{x_2\in \d_{\epsilon_1}\curve_{\sigma(\epsilon_1)}}
 \,\,\ee{ \frac{d_1}{\fram_{\epsilon_1}}\,x_2}
\,\,B(\spcurve_{\sigma(\epsilon_1)}; x_2,x_3)
\eeq
is an analytic function of $\ee{-x_3/\fram_{\epsilon''}}$ when $x_3$ approaches the puncture $\epsilon''$. Therefore we may expand it as:
\bea
&& -\frac{\fram_{\epsilon_1}\,\fram_{\epsilon''}}{d_1\,\,2i\pi}\,\oint_{x_2\in \d_{\epsilon_1}\curve_{\sigma(\epsilon_1)}}
 \,\,\ee{ \frac{d_1}{\fram_{\epsilon_1}}\,x_2}
\,\,B(\spcurve_{\sigma(\epsilon_1)}; x_2,x_3) \cr
&\sim& \sum_{d''=0}^\infty \td {\cal H}_{0,2,\sigma(\epsilon_1)}(d_1/\fram_{\epsilon_1};d''/\fram_{\epsilon''})
\,\,\ee{-\,\frac{d''}{\fram_{\epsilon''}}\,x_3}\,\,\frac{d''}{\fram_{\epsilon''}}\,dx_3\,\, ,
\eea
whose coefficients $\td {\cal H}_{0,2,\sigma}(d/\fram_\epsilon;d'/\fram_{\epsilon'})$ can be determined by computing a residue at the puncture $\epsilon'$, i.e. a contour integral around $\d_{\epsilon'}\curve_\sigma$:
\beq
\td {\cal H}_{0,2,\sigma}(d/\fram_\epsilon;d'/\fram_{\epsilon'})
= -\,
\frac{\fram_\epsilon\,\fram_{\epsilon'}}{d\,d'}\,\, \frac{1}{(2i\pi)^2}\,\oint_{x\in \d_{\epsilon}\curve_{\sigma}}\,\oint_{x'\in \d_{\epsilon'}\curve_{\sigma}}
 \,\,\ee{ \frac{d}{\fram_{\epsilon}}\,x}
\,\,B(\spcurve_{\sigma}; x,x')
 \,\,\ee{ \frac{d'}{\fram_{\epsilon'}}\,x'} .
\eeq
According to corollary \ref{corWgnMgn}, we have
\bea
- \td {\cal H}_{0,2,\sigma}(k;k')
&=& {\cal H}_{0,2,\sigma}(k;k') \cr
&& + \,\, \frac{1}{2}\sum_{n=1}^\infty\frac{1}{n!}  
\Big< \CL_{\rm Hodge}(\fram_{a,\sigma})\,\CL_{\rm Hodge}(\fram_{b,\sigma})\,\CL_{\rm Hodge}(-\fram_{a,\sigma}-\fram_{b,\sigma}) \cr
&& \left(\sum_d R_{\sigma,d}\,\tau_d\right)^n
\,\,\frac{\gamma_{\fram_\sigma}(k)}{1-k\,\psi}
\,\,\frac{\gamma_{\fram_{\sigma'}}(k')}{1-k'\,\psi'} \Big>_{0,n+2}
\eea
where
\beq
{\cal H}_{0,2,\sigma}(d/\fram_\epsilon;d'/\fram_{\epsilon'})= \,
\frac{\fram_{\epsilon}\,\fram_{\epsilon'}}{d\,d'}\,\,  \frac{1}{(2i\pi)^2}\,\oint_{x\in \d_{\epsilon}\curve_{\sigma}}\,\oint_{x'\in \d_{\epsilon'}\curve_{\sigma}}
 \,\,\ee{ \frac{d}{\fram_{\epsilon}}\,x}
\,\,\Brond_{\fram_\sigma}(x,x')
 \,\,\ee{ \frac{d'}{\fram_{\epsilon'}}\,x'} 
\eeq
can be computed explicitely:
\beq
{\cal H}_{0,2,\sigma}(k;k')= \frac{1}{k+k'}\,\,\gamma_{\fram_\sigma}(k)\,\gamma_{\fram_\sigma}(k')
\eeq
and   can be included into the sum by formally writing:
\beq
\Big< \CL_{\rm Hodge}(\fram_{a,\sigma})\,\CL_{\rm Hodge}(\fram_{b,\sigma})\,\CL_{\rm Hodge}(-\fram_{a,\sigma}-\fram_{b,\sigma}) 
\,\,\frac{1}{1-k\,\psi}
\,\,\frac{1}{1-k'\,\psi'} \Big>_{0,2} "="\,\, \frac{2}{k+k'}.
\eeq

\medskip
Finally we have:
\beq
 \td {\cal F}_{\epsilon,d;\epsilon',d'}  = {\cal F}_{\epsilon,d;\epsilon',d'}  + \sum_{\epsilon_1,d_1} \sum_{\epsilon_2,d_2} {\cal F}_{\epsilon,d;\epsilon_1,d_1}\,\, \tilde {\cal H}_{0,2,\sigma(\epsilon_1)}(\epsilon_1,d_1;\epsilon_2,d_2)\,\,\td {\cal F}_{\epsilon_2,d_2;\epsilon',d'}
\eeq

which proves the proposition.
}

\subsubsection{External legs}

Then, we need to compute the weight of external legs given by \eq{defdJsigmad}:

\bp
If $x\in \curve_{\epsilon_i}$ is on an open edge, we have
\beq
d\td J_{\epsilon,d}(x)
= d J_{\epsilon,d}(x)  
 + \sum_{\epsilon',d',\epsilon'',d''}\, {\cal F}_{\epsilon,d;\epsilon',d'}\,
\td {\cal H}_{0,2,\sigma(\epsilon')}(d'/\fram_{\epsilon'};d''/\fram_{\epsilon''}) \,\, 
d\td J_{\epsilon'',d''}(x) 
\eeq
where
\beq
d J_{\epsilon,d}(x) =
\delta_{\epsilon,\epsilon_i}\,\ee{-\frac{d}{\fram_\epsilon}(x-\arond_{\sigma(\epsilon)})}\,\,\frac{d}{\fram_\epsilon^2}\,\,dx .
\eeq

\ep

\proof{
Assume that $x\in \curve_{\epsilon_i}$.
We have from \eq{defdJsigmad}
\beq
d\td J_{\epsilon,d}(x)
= \frac{1}{\fram_\epsilon\,\,2\pi i}\, \oint_{x'\in \d_\epsilon\curve_{\sigma(\epsilon)}} \ee{-\frac{d}{\fram_\epsilon}(x'-\arond_{\sigma(\epsilon)})}\,\,B(\spcurve;x',x) .
\eeq
Let $\epsilon'$ be the other side of the edge of $\epsilon$ (i.e. such that $A_{\epsilon,\epsilon'}=1$, and then we have $\fram_{\epsilon'}=-\fram_\epsilon$). Let us move the integration contour to the other end of the cylinder, by doing so, we may pick a residue at $x=x'$ in the  case where $x'$ lies on the cylinder. We thus have
\bea
d\td J_{\epsilon,d}(x)
&=&  \delta_{\epsilon,\epsilon_i}\,\ee{-\frac{d}{\fram_\epsilon}(x-\arond_{\sigma(\epsilon)})}\,\,\frac{d}{\fram_\epsilon^2}\,\,dx \cr
&& + \sum_{\epsilon'}\, \frac{A_{\epsilon,\epsilon'}}{\fram_{\epsilon'}}\,\, \ee{-\,\frac{d}{\fram_\epsilon}\,(\arond_{\sigma(\epsilon')}-\arond_{\sigma(\epsilon)} )}\,\,\frac{1}{2\pi i}\, \oint_{x'\in \d_{\epsilon'}\curve_{\sigma(\epsilon')}} \ee{+\frac{d}{\fram_{\epsilon'}}(x'-\arond_{\sigma(\epsilon')})}\,\,B(\spcurve;x',x). \cr
\eea
Since we have the wrong sign for the exponential, we insert another integral like in \eqref{eqinsertW02tdF}:
\bea
d\td J_{\epsilon,d}(x)
&=&  \delta_{\epsilon,\epsilon_i}\,\ee{-\frac{d}{\fram_\epsilon}(x-\arond_{\sigma(\epsilon)})}\,\,\frac{d}{\fram_\epsilon^2}\,\,dx \cr
&& - \sum_{\epsilon',d'}\, {\cal F}_{\epsilon,d;\epsilon',d'}\,\frac{\fram_{\epsilon'}}{d'}\,\, \frac{1}{2\pi i}\, \oint_{x'\in \d_{\epsilon'}\curve_{\sigma(\epsilon')}} \Res_{x''\to x'} \cr
&&  \ee{+\frac{d}{\fram_{\epsilon'}}(x'-\arond_{\sigma(\epsilon')})}\,\,dS(\spcurve_{\sigma(\epsilon')},x'-\arond_{\sigma(\epsilon')},x''-\arond_{\sigma(\epsilon')}) B(\spcurve;x'',x) \cr
&=&  \delta_{\epsilon,\epsilon''}\,\ee{-\frac{d}{\fram_\epsilon}(x-\arond_{\sigma(\epsilon)})}\,\,\frac{d}{\fram_\epsilon^2}\,\,dx \cr
&& + \sum_{\epsilon',d',\epsilon''}\, {\cal F}_{\epsilon,d;\epsilon',d'}\,\frac{\fram_{\epsilon'}}{d'}\,\, \frac{1}{(2\pi i)^2}\, \oint_{x'\in \d_{\epsilon'}\curve_{\sigma(\epsilon')}} \oint_{x''\in \d_{\epsilon''}\curve_{\sigma(\epsilon')}} \cr
&&  \ee{+\frac{d'}{\fram_{\epsilon'}}(x'-\arond_{\sigma(\epsilon')})}\,\,dS(\spcurve_{\sigma(\epsilon')},x'-\arond_{\sigma(\epsilon')},x''-\arond_{\sigma(\epsilon')}) B(\spcurve;x'',x). \cr
\eea
Then we use
\beq
 -\frac{\fram_{\epsilon'}\,\fram_{\epsilon''}}{d'\,\,2i\pi}\,\oint_{x'\in \d_{\epsilon'}\curve_{\sigma(\epsilon')}}
 \,\,\ee{ \frac{d'}{\fram_{\epsilon'}}\,x'}
\,\,dS(\spcurve_{\sigma(\epsilon')}; x',x'') 
= -\sum_{d''=0}^\infty \td {\cal H}_{0,2,\sigma(\epsilon')}(d'/\fram_{\epsilon'};d''/\fram_{\epsilon''})
\,\,\ee{-\,\frac{d''}{\fram_{\epsilon''}}\,x''}.
\eeq
That gives
\bea
d\td J_{\epsilon,d}(x)
&=&  \delta_{\epsilon,\epsilon_i}\,\ee{-\frac{d}{\fram_\epsilon}(x-\arond_{\sigma(\epsilon)})}\,\,\frac{d}{\fram_\epsilon^2}\,\,dx \cr
&& - \sum_{\epsilon',d',\epsilon'',d''}\, {\cal F}_{\epsilon,d;\epsilon',d'}\,
\td H_{0,2,\sigma(\epsilon')}(d'/\fram_{\epsilon'};d''/\fram_{\epsilon''})\,\, 
d\td J_{\epsilon'',d''}(x). \cr
\eea

}

In other words, we may replace the edge weight $\td {\cal F}$ by ${\cal F}$, by introducing a 2-valent vertex $\td {\cal H}_{0,2}$.

We thus have:

\bt\label{thWngtdHtdFH02}
If $x_j\in \curve_{\epsilon_j}$ where $\epsilon_j$ is an half edge of $\Upsilon_\CYX$ (not necessarily corresponding to a non-compact edge, neither necessarily the one on which the brane $L$ ends), for  $(g,n) \in \mathbb{N}^2 \backslash \{(0,0),(1,0)\}$, we have
\bea
&& W_{g,n}(\spcurve;x_1,\dots,x_n)  \cr
&=& 2^{3g-3+n}\sum_{G\in {\cal G}^{{\rm stable}+(0,2)}_{g,n}} \frac{1}{\#{\rm Aut}(G)} \quad
 \prod_{v={\rm vertices}} 
 \frac{\td {\cal H}_{g_v,n_v,\sigma_v}(\{d_{h}/\fram_{\epsilon_h} \}_{h\in E_v})}{\prod_{h\in E_v} \fram_{\epsilon_h}}
 \cr
&& \prod_{e=(h_+,h_-)={\rm closed\,edges}} {\cal F}_{\epsilon_{h_+},d_{h_+};\epsilon_{h_-},d_{h_-} } 
\quad \qquad \prod_{j=1}^n  \ee{- \frac{d_{j}}{\fram_{\epsilon_j}}\,(x_j-\arond_{\sigma(\epsilon_j)})}\,\, \frac{d_{j}}{\fram_{\epsilon_j}}\,d x_j \cr
\eea
where the sum is over stable graphs with possibly $(0,2)$ vertices (every vertex $v$ is such that $2-2g_v-n_v<1$),
with
\bea
\td {\cal H}_{g,n,\sigma}(k_1,\dots,k_n) 
&=&  {\ee{\hat t_{\sigma,0}(2g-2+n)}}\,\int_{\overline\modsp_{g,n}} \CL_{\rm Hodge}(\fram_{a,\sigma})\CL_{\rm Hodge}(\fram_{b,\sigma})\CL_{\rm Hodge}(-\fram_{a,\sigma}-\fram_{b,\sigma}) \cr
&& \ee{l_{1*}\sum_d \td R_{\sigma_v,d}\tau_d} \quad \prod_{i=1}^n \,\,\frac{\gamma_{\fram_\sigma}(k_i)}{1-k_i\,\psi_i} ,
\eea
\bea
\td {\cal H}_{0,2,\sigma}(k_1,k_2) 
&=& {\cal H}_{0,2,\sigma}(k_1,k_2) \cr
&& +  \sum_{k=1}^\infty \frac{1}{k!}\, \int_{\overline\modsp_{0,k+2}} \CL_{\rm Hodge}(\fram_{a,\sigma})\CL_{\rm Hodge}(\fram_{b,\sigma})\CL_{\rm Hodge}(-\fram_{a,\sigma}-\fram_{b,\sigma}) \cr
&& \quad \qquad \qquad  \left(\sum_d \td R_{\sigma_v,d}\tau_d\right)^k \quad  \,\,\frac{\gamma_{\fram_\sigma}(k_1)}{1-k_1\,\psi_1} \,\,\frac{\gamma_{\fram_\sigma}(k_2)}{1-k_2\,\psi_2} 
\eea
and
\beq
{\cal F}_{\epsilon,d;\epsilon',d'} =  d\,\,\ee{- \frac{d}{\fram_\epsilon}\,(\arond_{\sigma'}-\arond_\sigma)}\,\,A_{\epsilon,\epsilon'}\,\,\, \delta_{d,d'}.
\eeq
\et

\proof{
The sum over all $(0,2)$ vertices can be performed due to
$$
\td {\cal F} = {\cal F} + {\cal F} \td {\cal H}_{0,2} {\cal F} + {\cal F} \td {\cal H}_{0,2}  {\cal F} \td {\cal H}_{0,2} {\cal F} + {\cal F} \td {\cal H}_{0,2} {\cal F} \td {\cal H}_{0,2} {\cal F} \td {\cal H}_{0,2} F+ \dots 
$$
which is illustrated as:
$$ \includegraphics[width=13cm]{propagrenorm.pdf} $$
and it exactly reproduces the left hand side.
This is a usual trick used in combinatorics of graphs.}

\subsubsection{Renormalized disc amplitude $R_{\sigma,d}$}

\bp
The vertex weights can be renormalized by
\bea
&& \td {\cal H}_{g,n,\sigma}(d_1,\dots,d_n) \cr
&=& {\cal H}_{g,n,\sigma}(d_1,\dots,d_n) \cr
&& + \sum_{k=1}^\infty \frac{1}{k!} \sum_{\epsilon_{n+1},\dots,\epsilon_{n+k}} \sum_{d_{n+1},\dots,d_{n+k}} {\cal H}_{g,n+k,\sigma}(d_1,\dots,d_{n+k})\,\,\prod_{i=n+1}^{n+k} {\cal F}_{\epsilon,d_i;\epsilon_i,d_i} C_{\sigma(\epsilon'_i)}(d_i) \cr
\eea
and
\bea
C_{\sigma}(d) &=&  {\cal H}_{0,1,\sigma}(d)   \cr
&& + \sum_{k=1}^\infty \frac{1}{k!} \sum_{\epsilon_{2},\dots,\epsilon_{k+1}} \sum_{d_{2},\dots,d_{k+1}} {\cal H}_{0,k+1,\sigma}(d_1,d_2,\dots,d_{k+1})\,\,\,\prod_{i=2}^{k+1} {\cal F}_{\epsilon,d_i;\epsilon_i,d_i} C_{\sigma(\epsilon'_i)}(d_i)
\eea
where
\beq
{\cal H}_{0,1,\sigma}(k) = \frac{\ee{-\hat t_{\fram_\sigma,0}}}{4\,k^2}\,\,\, \gamma_{\fram_{\sigma}}(k)
\eeq
and, for $2-2g-n<0$:
\bea
{\cal H}_{g,n,\sigma}(k_1,\dots,k_n) 
&=&  \frac{2^{3g-3+n}}{\ee{(2-2g-n)\hat t_{\fram_\sigma,0}}} \Big< \CL_{\rm Hodge}(\fram_{a,\sigma})\,\CL_{\rm Hodge}(\fram_{b,\sigma})\,\CL_{\rm Hodge}(-\fram_{a,\sigma}-\fram_{b,\sigma})  \cr
&& \prod_{i=1}^n\,\, \frac{\gamma_{\fram_\sigma}(k_i)}{1-k_i \,\psi_i}\,\, \Big>_{g,n}.
\eea

\ep

\proof{

From  lemma \ref{lemmarenorm01}, we have
\beq
R_{\sigma,d} 
=  \frac{-2\,\ee{\hat t_{\fram_\sigma,0}}}{2\pi i} 
\oint_{\d\curve_\sigma}\,\, \td\xirond_{\fram_\sigma,d}(x)\,\,(y(x+\arond_\sigma)-\brond_\sigma-\yrond_{\fram_\sigma}(x))\,dx .
\eeq
We first decompose $\d\curve_\sigma$ into its 3 circles $\cup_\epsilon \d_\epsilon\curve_\sigma$, and on each $\d_\epsilon\curve_\sigma$ we use the expansion \eqref{eqexpxirondz0} for $\td\xirond_{\fram_\sigma,d}(x)$. This
 implies
\beq
\sum_d \psi^d R_{\sigma,d} 
=
2\,\ee{\hat t_{\fram_{\sigma(\epsilon)},0}}\,\sum_\epsilon \sum_k\,\,\frac{\gamma_\sigma(k/\fram_\epsilon)}{1-\frac{k}{\fram_\epsilon}\psi} \hat R_{\epsilon}(k/\fram_\epsilon)
\eeq
where
\beq
\hat R_{\epsilon}(k/\fram_\epsilon)
=\frac{1}{\fram_\epsilon\,2\pi i}\,
\oint_{\d_\epsilon\curve_{\sigma(\epsilon)}}\,\, \ee{-\,\frac{k}{\fram_\epsilon}\,(x-\arond_{\sigma(\epsilon)})}\,\,(y(x)-\brond_{\sigma(\epsilon)}-\yrond_{\fram_{\sigma(\epsilon)}}(x-\arond_{\sigma(\epsilon)}))\,dx .
\eeq

When $k=0$, we have to compute
\beq
\oint_{\d_\epsilon\curve_{\sigma(\epsilon)}}\,\, \,(y(x)-\brond_{\sigma(\epsilon)}-\yrond_{\fram_{\sigma(\epsilon)}}(x-\arond_{\sigma(\epsilon)}))\,dx =0
\eeq
which vanishes (order by order in the $Q$ expansion) due to \eqref{eqyyrondexpQk}.
So, let us assume $k\neq 0$ and integrate by parts:
\beq
\hat R_{\epsilon}(k/\fram_\epsilon)
=  \frac{1}{k}\,\,\frac{1}{2\pi i}\,
\oint_{\d_\epsilon\curve_{\sigma(\epsilon)}}\,\, \ee{-\,\frac{k}{\fram_\epsilon}\,(x-\arond_{\sigma(\epsilon)})}\,\,(dy(x)-d\yrond_{\fram_{\sigma(\epsilon)}}(x-\arond_{\sigma(\epsilon)})) .
\eeq
Using the parametrization of def. \ref{defspcurverond} one can compute explicitely
\beq
\oint_{\d_\epsilon\curve_{\sigma(\epsilon)}}\,\, \ee{-\,\frac{k}{\fram_\epsilon}\,x}\,\,d\yrond_{\fram_{\sigma(\epsilon)}}(x) = 0,
\eeq
and thus
\beq
\hat R_{\epsilon}(k/\fram_\epsilon)
=  \frac{1}{k}\,\,\frac{1}{2\pi i}\,
\oint_{\d_\epsilon\curve_{\sigma(\epsilon)}}\,\, \ee{-\,\frac{k}{\fram_\epsilon}\,(x-\arond_{\sigma(\epsilon)})}\,\,dy(x) .
\eeq
Then, let us move the integration contour through the cylinder $\curve_\epsilon$. Let us call $\epsilon'$ the other half-edge of the cylinder (i.e. $A_{\epsilon,\epsilon'}=1$, and in that case $\fram_{\epsilon'}=-\fram_\epsilon$):
\beq
\hat R_{\epsilon}(k/\fram_\epsilon)
= -\, \frac{1}{k}\,\,\frac{1}{2\pi i}\, \sum_{\epsilon'}\,A_{\epsilon,\epsilon'}\,\ee{-\,\frac{k}{\fram_{\epsilon}}\,(\arond_{\sigma(\epsilon')}-\arond_{\sigma(\epsilon)})}
\oint_{\d_{\epsilon'}\curve_{\sigma(\epsilon')}}\,\, \ee{+\,\frac{k}{\fram_{\epsilon'}}\,(x-\arond_{\sigma(\epsilon')})}\,\,dy(x)
\eeq
which we can write
\beq
\hat R_{\epsilon}(k/\fram_\epsilon)
= \sum_{\epsilon',k'}\,F_{\epsilon,k;\epsilon',k'}\,  C_{\epsilon'}(k'/\fram_{\epsilon'})  
\eeq
with
\bea
C_{\epsilon}(k/\fram_{\epsilon}) 
&=& -\frac{\fram_{\epsilon}^2}{k^2}\,\,\frac{1}{2\pi i}\, 
\oint_{\d_{\epsilon}\curve_{\sigma(\epsilon)}}\,\, \ee{+\,\frac{k}{\fram_\epsilon}\,(x-\arond_{\sigma(\epsilon)})}\,\,dy(x) \cr
&=& -\frac{\fram_{\epsilon}^2}{k^2}\,\,\frac{1}{2\pi i}\, 
\oint_{\d_{\epsilon}\curve_{\sigma(\epsilon)}}\,\, \ee{+\,\frac{k}{\fram_\epsilon}\,x}\,\,d\Phi(x)  -\frac{\fram_{\epsilon}^2}{k^2}\,\,\frac{1}{2\pi i}\, 
\oint_{\d_{\epsilon}\curve_{\sigma(\epsilon)}}\,\, \ee{+\,\frac{k}{\fram_\epsilon}\,x}\,\,d\yrond_{\fram_{\sigma(\epsilon)}}(x) \cr
\eea
where we have introduced $\Phi(x)$ such that
\beq
d\Phi(x) = (y(x)-\brond_\sigma- \yrond_{\fram_\sigma}(x-\arond_\sigma))\,dx.
\eeq

Then, let us use \eq{eqW01HodgeMgnR} of corollary \ref{corWgnMgn}: 
\bea
 d\Phi(x) 
&=&   \frac{1}{2\pi i}\oint_{x'\in\d\curve_\sigma} \Brond_{\fram_\sigma}(x,x')\,\Phi(x')
- \frac{\ee{-\hat t_{\fram_\sigma,0}}}{4}\, \sum_{k_1}\,\, \sum_{n=2}^\infty \frac{1}{n!}  \cr
&&   \Big< \CL_{\rm Hodge}(\fram_{a,\sigma})\,\CL_{\rm Hodge}(\fram_{b,\sigma})\,\CL_{\rm Hodge}(-\fram_{a,\sigma}-\fram_{b,\sigma}) \cr
  && \left(\sum_d  R_{\sigma,d}\tau_d\right)^n\,\, \frac{\gamma_{\fram_\sigma}(k_1/\fram_{\epsilon})}{1-\frac{k_1}{\fram_{\epsilon}} \,\psi_1}\,\, \Big>_{0,n+1} \,
 \, \frac{k_1}{(\fram_{\epsilon})^2} \ee{- \,\frac{k_1}{\fram_{\epsilon}}\,x}\,dx . \cr
\eea

We thus get that:
\bea
C_{\epsilon}(k/\fram_{\epsilon}) 
&=& -\frac{\fram_\epsilon}{k}\,\frac{1}{(2\pi i)^2}\,  \oint_{x\in\d_{\epsilon}\curve_{\sigma(\epsilon)}}\,\,\oint_{x'\in\d\curve_\sigma} \,\, \ee{+\,\frac{k}{\fram_\epsilon}\,x}\,\,\Brond_{\fram_\sigma}(x,x')\,\Phi(x') \cr
&& +\frac{\ee{-\hat t_{\fram_\sigma,0}}}{4}\, \,  \,\, \sum_{n=2}^\infty \frac{1}{n!}  
   \Big< \CL_{\rm Hodge}(\fram_{a,\sigma})\,\CL_{\rm Hodge}(\fram_{b,\sigma})\,\CL_{\rm Hodge}(-\fram_{a,\sigma}-\fram_{b,\sigma}) \cr
&& \left(\sum_d  R_{\sigma,d}\tau_d\right)^n\,\, \frac{\gamma_{\fram_\sigma}(k/\fram_{\epsilon})}{1-\frac{k}{\fram_{\epsilon}} \,\psi_1}\,\, \Big>_{0,n+1} . \,
  \cr
\eea

For $2-2g-n<0$ we define:
\bea
{\cal H}_{g,n,\sigma}(k_1,\dots,k_n) 
&=&  \frac{2^{3g-3+n}}{\ee{(2-2g-n)\hat t_{\fram_\sigma,0}}} \Big< \CL_{\rm Hodge}(\fram_{a,\sigma})\,\CL_{\rm Hodge}(\fram_{b,\sigma})\,\CL_{\rm Hodge}(-\fram_{a,\sigma}-\fram_{b,\sigma})  \cr
&& \prod_{i=1}^n\,\, \frac{\gamma_{\fram_\sigma}(k_i)}{1-k_i \,\psi_i}\,\, \Big>_{g,n}
\eea
and 
\beq
{\cal H}_{0,2,\sigma}(k/\fram_\epsilon,k'/\fram_{\epsilon'}) 
= -\frac{\fram_\epsilon\,\fram_{\epsilon'}}{k\,k'}\,\frac{1}{(2\pi i)^2}\,  \oint_{x\in\d_{\epsilon}\curve_{\sigma(\epsilon)}}\,\,\oint_{x'\in\d\curve_\sigma} \,\, \ee{+\,\frac{k}{\fram_\epsilon}\,x}\,\,\Brond_{\fram_\sigma}(x,x')\,\ee{+\,\frac{k'}{\fram_{\epsilon'}}\,x'} 
\eeq
and
\beq
H_{0,1,\sigma}(k) = \frac{\ee{-\hat t_{\fram_\sigma,0}}}{4\,k^2}\,\,\, \gamma_{\fram_{\sigma}}(k) .
\eeq

This amounts to write virtually:
\beq
\Big< \CL_{\rm Hodge}(\fram_{a,\sigma})\,\CL_{\rm Hodge}(\fram_{b,\sigma})\,\CL_{\rm Hodge}(-\fram_{a,\sigma}-\fram_{b,\sigma})  \frac{1}{1-k \,\psi}\,\, \Big>_{0,1} "=" \frac{1}{k^2}.
\eeq

This gives that
\bea
C_\sigma(k/\fram_\epsilon) 
&=& {\cal H}_{0,1,\sigma}(k/\fram_\epsilon)  + \sum_{n=1}^\infty \frac{1}{n!}
\sum_{\epsilon'_i,\,i=1,\dots,n}\,\sum_{d'_i,\, i=1,\dots,n} \cr
&& \,{\cal H}_{0,n+1,\sigma}(k/\epsilon,k_1/\epsilon_1,\dots,k_n/\epsilon_n) \prod_{i=1}^n {\cal F}_{\epsilon_i,k_i;\epsilon'_i,k'_i}\, C_{\sigma(\epsilon'_i)}(k'_i/\epsilon'_i) .\cr
\eea

%

%

%

}

\subsection{Finishing the proof}

So far we had from theorem \ref{thWngtdHtdF} that,
if $x_j\in \curve_{\epsilon_j}$ where $\epsilon_j$ is an half edge of $\Upsilon_\CYX$ (not necessarily corresponding to a non-compact edge, neither necessarily the one on which the brane $L$ ends), we have
\bea
&& W_{g,n}(\spcurve;x_1,\dots,x_n)  \cr
&=& \sum_{G\in {\cal G}^{\rm stable}_{g,n}} \frac{1}{\#{\rm Aut}(G)} \quad
 \prod_{v={\rm vertices}} 
\td {\cal H}_{g_v,n_v,\sigma_v}(\{d_{h}/\fram_{\epsilon_h} \}_{h\in E_v})
 \cr
&& \prod_{e=(h_+,h_-)={\rm closed\,edges}} \td {\cal F}_{\epsilon_{h_+},d_{h_+};\epsilon_{h_-},d_{h_-} } 
\qquad \qquad \prod_{j=1,\, h_j={\rm open\, half-edges}}^n  d\td J_{\epsilon_{h_j},d_{h_j}}(x_j) \cr
\eea
where the quantities $\td {\cal H}_{g,n,\sigma}$ and $\td {\cal F}_{\epsilon,d;\epsilon',d'}$ are directly computed from the spectral curve, and
where the sum is only over stable graphs (every vertex $v$ is such that $2-2g_v-n_v<0$).

However, we have just found that we have for $(g,n)\neq (0,1)$:
\bea
&& \td {\cal H}_{g,n,\sigma}(k_1/\fram_{\epsilon_1},\dots,k_n/\fram_{\epsilon_n}) \cr
&=& {\cal H}_{g,n,\sigma}(k_1/\fram_{\epsilon_1},\dots,k_n/\fram_{\epsilon_n})  \cr
&& + \sum_{m=1}^\infty \frac{1}{m!}
\sum_{\epsilon'_i,\,i=n+1,\dots,n+m}\,\sum_{d'_i,\, i=n+1,\dots,n+m} \cr
&& \,{\cal H}_{g,n+m,\sigma}(k_1/\epsilon_1,\dots,k_{n+m}/\epsilon_{n+m}) \prod_{i=n+1}^{n+m} {\cal F}_{\epsilon_i,k_i;\epsilon'_i,k'_i}\, C_{\sigma(\epsilon'_i)}(k'_i/\epsilon'_i) \cr
\eea
and for $(g,n)=(0,1)$
\bea
C_\sigma(k/\fram_\epsilon) 
&=& {\cal H}_{0,1,\sigma}(k/\fram_\epsilon)  + \sum_{n=1}^\infty \frac{1}{n!}
\sum_{\epsilon'_i,\,i=1,\dots,n}\,\sum_{d'_i,\, i=1,\dots,n} \cr
&& \,{\cal H}_{0,n+1,\sigma}(k/\epsilon,k_1/\epsilon_1,\dots,k_n/\epsilon_n) \prod_{i=1}^n {\cal F}_{\epsilon_i,k_i;\epsilon'_i,k'_i}\, C_{\sigma(\epsilon'_i)}(k'_i/\epsilon'_i) \cr
\eea
and we have
\beq
\td {\cal F}_{\epsilon,k;\epsilon',k'} = {\cal F}_{\epsilon,k;\epsilon',k'} + \sum_{\epsilon'',\epsilon''',k'',k'''} {\cal F}_{\epsilon,k;\epsilon'',k''} \td {\cal H}_{0,2,\sigma(\epsilon'')}(k''/\fram_{\epsilon''},k'''/\fram_{\epsilon'''})\,\td {\cal F}_{\epsilon''',k''';\epsilon',k'}
\eeq
and
\beq
d\td J_{\epsilon,k}(x)= dJ_{\epsilon,k}(x) + \sum_{\epsilon',\epsilon'',k',k''} \td {\cal F}_{\epsilon,k;\epsilon,k'}\,\td {\cal H}_{0,2,\sigma(\epsilon')}(k'/\fram_{\epsilon'},k''/\fram_{\epsilon''}) \,dJ_{\epsilon'',k''}(x)
\eeq

\begin{figure}[t]
\centering
$$\includegraphics[height=2.7cm]{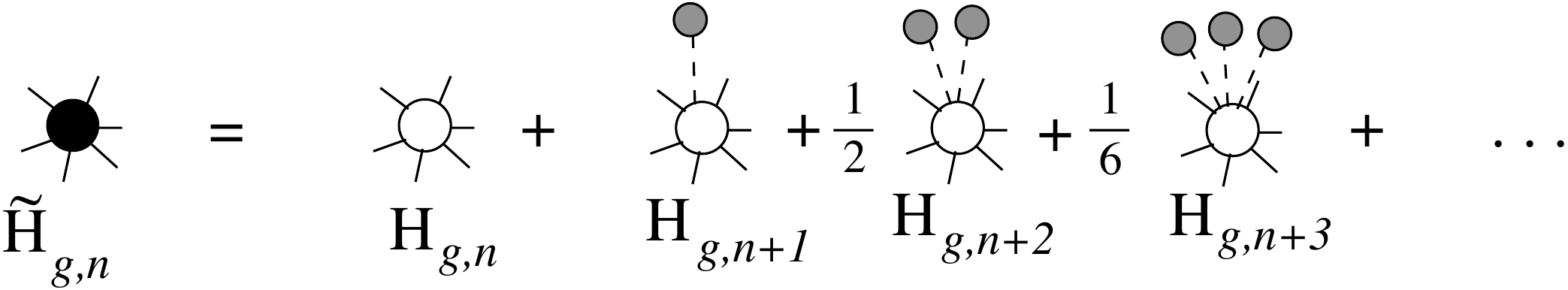} $$
$${}$$
$$\includegraphics[height=2.7cm]{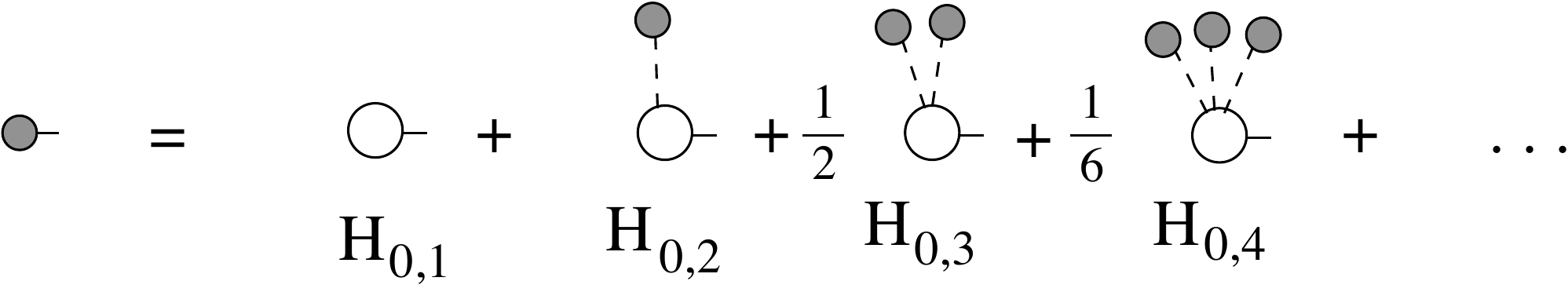} $$
$${}$$
$$ \includegraphics[width=13cm]{propagrenorm.pdf} $$
\caption{The vertices weights $\td {\cal H}_{g,n}$ are obtained by gluing $C$ in all possible ways, and $C$ must be chosen such that $\td {\cal H}_{0,1}=C$.
The weight  for the propagator is $\td {\cal F} = {\cal F} + {\cal F} \td {\cal H}_{0,2} {\cal F} + {\cal F} \td {\cal H}_{0,2} {\cal F} \td {\cal H}_{0,2} {\cal F} + \dots = {\cal F} + {\cal F} \td {\cal H}_{0,2} \td {\cal F}$} 
\label{figtadpolerenorm2}
\end{figure}

All this is sufficient to prove that:

\bt\label{thBmodelWgnrenorm}

For $(g,n) \in \mathbb{N}^2 \backslash \{(0,0),(1,0)\}$, we have:
\bea
&& W_{g,n}(\spcurve;x_1,\dots,x_n)  \cr
&=& \sum_{G\in {\cal G}_{g,n}} \frac{1}{\#{\rm Aut}(G)} \quad
 \prod_{v={\rm vertices}} 
 {\cal H}_{g_v,n_v,\sigma_v}(\{d_{h}/\fram_{\epsilon_h} \}_{h\in E_v})
 \cr
&& \prod_{e=(h_+,h_-)={\rm closed\,edges}}  {\cal F}_{\epsilon_{h_+},d_{h_+};\epsilon_{h_-},d_{h_-} } 
\qquad \qquad \prod_{j=1,\, h_j={\rm open\, half-edges}}^n  dJ_{\epsilon_{h_j},d_{h_j}}(x_j) \cr
\eea
where now the sum is over all graphs (not only stable ones).
Moreover we have:

\bea
{\cal H}_{g,n,\sigma}(k_1,\dots,k_n) 
&=&  \frac{2^{3g-3+n}}{\ee{(2-2g-n)\hat t_{\fram_\sigma,0}}} \Big< \CL_{\rm Hodge}(\fram_{a,\sigma})\,\CL_{\rm Hodge}(\fram_{b,\sigma})\,\CL_{\rm Hodge}(-\fram_{a,\sigma}-\fram_{b,\sigma})  \cr
&& \prod_{i=1}^n\,\, \frac{\gamma_{\fram_\sigma}(k_i)}{1-k_i \,\psi_i}\,\, \Big>_{g,n}
\eea
where for $(g,n)=(0,2)$ and $(0,1)$ we have defined
\beq
\Big< \CL_{\rm Hodge}(\fram_{a,\sigma})\,\CL_{\rm Hodge}(\fram_{b,\sigma})\,\CL_{\rm Hodge}(-\fram_{a,\sigma}-\fram_{b,\sigma})  \frac{1}{1-k \,\psi}\,\, \Big>_{0,1} \stackrel{{\rm def}}{=}\,\, \frac{1}{k^2}
\eeq
\beq
\Big< \CL_{\rm Hodge}(\fram_{a,\sigma})\,\CL_{\rm Hodge}(\fram_{b,\sigma})\,\CL_{\rm Hodge}(-\fram_{a,\sigma}-\fram_{b,\sigma})  \frac{1}{1-k \,\psi_1}\frac{1}{1-k' \,\psi_2}\,\, \Big>_{0,2} \stackrel{{\rm def}}{=}\,\, \,\frac{2}{k+k'}
\eeq
and
\beq
{\cal F}_{\epsilon,d;\epsilon',d'} = A_{\epsilon,\epsilon'}\,\delta_{d,d'}\, \frac{d}{\fram_\epsilon^2}\,\,\ee{-\frac{d}{\fram_\epsilon}\,(\arond_{\sigma'}-\arond_\sigma)}
\eeq
\beq
dJ_{\epsilon,k}(x) = \,\ee{-\,\frac{k}{\fram_\epsilon}\,(x-\arond_{\sigma(\epsilon)})}\,\,\frac{k}{\fram_\epsilon^2}\,dx
\eeq

\et

\proof{The proof is best represented graphically, this is fig. \ref{figtadpolerenorm2}.}

One can check that this expression coincides with the localization formula for Gromov--Witten invariants.
This concludes the proof of the BKMP conjecture:

\bt
The BKMP conjecture holds true.
In other words,  the invariants $W_{g,n}$ of the mirror curve $\spcurve$ do coincide with the Gromov--Witten invariants:
\beq
\forall (g,n) \in \mathbb{N}^2 \backslash \{(0,0),(1,0)\} \, , \; W_{g,n}(\spcurve;x_1,\dots,x_n) = {\cal W}_{g,n}(\CYX,x_1,\dots,x_n)\,\,dx_1\otimes \dots \otimes dx_n.
\eeq
\et

\section{Conclusion}

We have obtained theorem \ref{thBmodelWgnrenorm} using only properties of the topological recursion (mostly combinatorics of graphs and complex analysis on the spectral curve), and it is remarkable that what we obtain is exactly the localization formula of Gromov-Witten invariants.

Our proof is thus a proof which works mostly on the B-model side. The main ingredients are 
localization, tropical limit, special geometry, graph combinatorics and complex analysis on $\curve$.

\bigskip

En route we have seen that the B-model formula continues to make sense when the boundaries are not all on the same brane, each boundary can be chosen on a different brane, and also the brane needs not be on a non-compact edge of the toric graph, it can be on any half-edge.

We hope that the present proof may shed some new light on the A-model side geometry.

\section*{Acknowledgments}
We would like to thank G. Borot, V. Bouchard, A. Brini, R. Kashaev, A. Kashani Poor, C. Kozcaz, M. Liu, M. Mari\~no, M. Mulase, S. Shadrin, A. Szenes, J. Zhou 
for useful and fruitful discussions on this subject.
The work of B. E.  is partly supported by the ANR project GranMa ``Grandes Matrices Al\'{e}atoires" ANR-08-BLAN-0311-01, by the European Science Foundation through the Misgam program, by the Quebec government with the FQRNT. B. E. thanks the CERN, the university of Geneva and Stas Smirnov for their hospitality. The work of N.O. is partly supported by the FCT through the fellowship
SFRH/BPD/70371/2010. N.O. would like to thank the CERN and the KdV Institute for their hospitality while part of this work was carried out.


\setcounter{section}{0}
\appendix{}
\setcounter{section}{0}

\section{Invariants of spectral curves and topological recursion\label{secappdefWgn}}

\subsection{Spectral curves}

A spectral curve is in fact the data of a plane curve with some additional structure. We set:

\bd[Spectral curve]
a spectral curve $\spcurve=(\curve,x,y,B)$, is the data of:

$\bullet$ a Riemann surface $\curve$, not necessarily compact,

$\bullet$ two analytical functions $x:\curve\to\mathbb C$, $y:\curve\to\mathbb C$,

$\bullet$ a Bergman kernel $B$, i.e. a symmetric 2nd kind bilinear meromorphic differential, having a double pole on the diagonal and no other pole, and normalized (in any local coordinate $z$) as:
\beq
B(z_1,z_2) \mathop{{\sim}}_{z_2\to z_1} \frac{dz_1\otimes dz_2}{(z_1-z_2)^2} + {\rm analytical}.
\eeq

\bigskip

Moreover, the spectral curve $\spcurve$ is called regular if the  1-form $dx$ has a finite number of zeroes on $\curve$, denoted $\{\alpha_1, \dots,\alpha_\bpt\}$, and they are simple zeroes, and $dy$ doesn't vanish at the zeroes of $dx$. In other words, locally near a branchpoint $\alpha$, $y$ behaves like a square root of $x$:
\beq\label{bpysqrtxapp}
y(z) \mathop{{\sim}}_{z\to \alpha} y(\alpha) + y'(a)\,\sqrt{x(z)-a} + O(x(z)-a) \qquad , \,\, y'(a)\neq 0
\eeq
and where $a=x(\alpha)$ is the $x$--projection of the branchpoint $\alpha$:
\beq
x(\alpha)=a.
\eeq

\ed

\subsection{Invariants}\label{secdefomng}

In \cite{EOFg}, it was defined how to associate to a regular spectral curve $\spcurve$, an infinite sequence of symmetric multilinear meromorphic forms $\om_n^{(g)}\in T^*(\curve)\otimes\dots\otimes T^*(\curve)$, and a sequence of complex numbers $F_g(\spcurve)\in\mathbb C$.
The definition is given by a recursion, often called "{\bf topological recursion}", which we recall:

\bd[Invariants $\om_{g,n}(\spcurve)$]
Let $\spcurve=(\curve,x,y,B)$ be a regular spectral curve.
Let $\alpha_1,\dots,\alpha_\bpt$ be its branchpoints (zeroes of $dx$ in $\curve$), and $a_i=x(\alpha_i)$.
We define
\beq
\om_{0,1}(\spcurve;z) = y(z)\,dx(z),
\eeq
\beq
\om_{0,2}(\spcurve;z_1,z_2) = B(z_1,z_2),
\eeq
and for $2g-2+(n+1)>0$:
\bea
\om_{g,n+1}(\spcurve;z_1,\dots,z_n,z_{n+1})
&=& \sum_{i=1}^\bpt \Res_{z\to \alpha_i} K(z_{n+1},z)\,\Big[ \om_{g-1,n+2}(z,\bar z,z_1,\dots,z_n) \cr
&& \qquad + \sum_{h=0}^g\sum'_{I\uplus J = \{z_1,\dots,z_n\}} \om_{h,1+\#I}(z,I)\,\om_{g-h,1+\#J}(z,J) \Big] \cr
\eea
where the prime in $\sum_h\sum_{I\uplus J}'$ means that we exclude from the sum the terms $(h=0,I=\emptyset)$ and $(h=g,J=\emptyset)$, 
and
where $\bar z$ means the other branch of the square-root in \eqref{bpysqrtx} near a branchpoint $\alpha_i$, i.e. if $z$ is in the vicinity of $\alpha_i$, $\bar z\neq z$ is the other point in the vicinity of $\alpha_i$ such that
\beq
x(\bar z)=x(z),
\eeq
and thus $y(\bar z) \sim y(\alpha) - y'(a)\sqrt{x(z)-a}$.
The recursion kernel $K(z_{n+1},z)$ is defined as
\beq
K(z_{n+1},z) = \frac{\int_{z'=\bar z}^z B(z_{n+1},z')}{2(y(z)-y(\bar z))\,dx(z)}
\eeq
$K$ is a 1-form in $z_{n+1}$ defined on $\curve$ with a simple pole at $z_{n+1}=z$ and at $z_{n+1}=\bar z$, and in $z$ it is the inverse of a 1-form, defined only locally near branchpoints, and it has a simple pole at $z=\alpha_i$.

\medskip

Using the $x(z)$ coordinate instead of $z$, we define
\beq
W_{g,n}(\spcurve;x(z_1),\dots,x(z_n)) = \om_{g,n}(z_1,\dots,z_n).
\eeq

\medskip

We also define for $g\geq 2$:
\beq
F_g(\spcurve)=\om_{g,0}(\spcurve) = \frac{1}{2-2g}\,\sum_{i=1}^\bpt \Res_{z\to \alpha_i}\,\,\om_{g,1}(\spcurve;z)\,\,\left(\int_{z'=\alpha_i}^z\,y(z')dx(z')\right).
\eeq

\ed

With this definition, $F_g(\spcurve)\in\mathbb C$ is a complex number associated to $\spcurve$, sometimes called the $g^{\rm th}$ symplectic invariant of $\spcurve$, and $\om_{g,n}(\spcurve;z_1,\dots,z_n)$ is a symmetric multilinear differential $\in T^*(\curve)\otimes\dots\otimes T^*(\curve)$, sometimes called the $n^{\rm th}$ descendant of $F_g$. Very often we denote $F_g=\om_{g,0}$.
If $2-2g-n<0$, $\om_{g,n}$ is called stable, and otherwise unstable, the only unstable cases are $F_0, F_1, \om_{0,1},\om_{0,2}$.
 For $2-2g-n<0$, $\om_{g,n}$ has poles only at branchpoints (when some $z_k$ tends to a branchpoint $a_i$), without residues, and the degrees of the poles are $\leq 6g+2n-4$.
In the $x$ variables, $W_{g,n}$ are multivalued functions of the $x_i$'s and their singular behavior near $x_i\to a_j$ are half integer power singularities:
\beq
W_{g,n}(\spcurve;x_1,\dots,x_n) \sim O\left((x_i-a_j)^{-d_{i,j}-\frac{1}{2}}\right)\,dx_i
\eeq
where
\beq
d_{i,j} \leq 3g-3+n
\eeq

\smallskip

We shall not write here the definition of $F_0$ and $F_1$, see \cite{EOFg}, since we shall not use them here.

\medskip

Those invariants $F_g$ and $\om_{g,n}$'s have many fascinating properties, in particular related to integrability, to modular forms, and to special geometry, and we refer the reader to \cite{EOFg, EOreview}.


\section{Intersection numbers}\label{apppsikappa}

Since many of our formula involve intersection numbers in moduli spaces of curves,  let us introduce basic concepts. We refer the reader to \cite{} for deeper description.

\subsection{Definitions}

\begin{figure}[t]
\centering
\includegraphics[height=5cm]{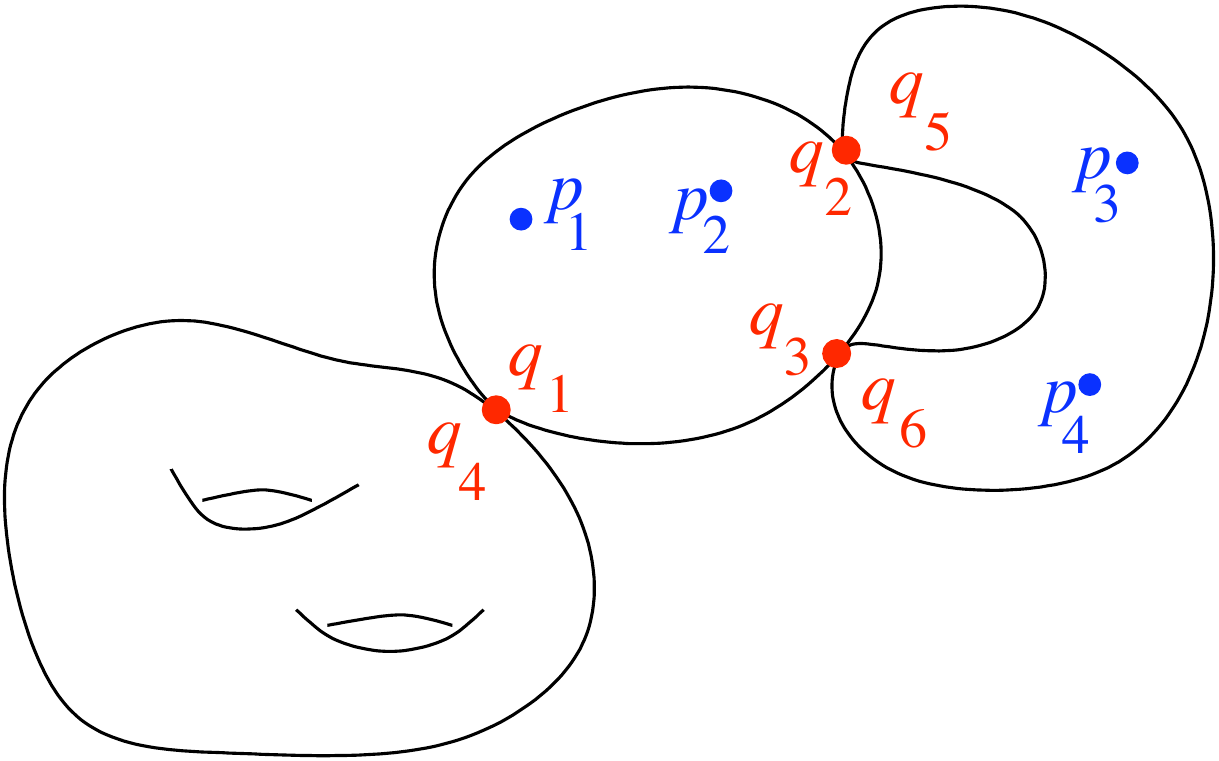} 
\caption{A stable curve in $\ovl\modsp_{g,n}$ can be smooth or nodal. Here we have an example in $\ovl\modsp_{3,4}$ of a stable curve of genus $g=3$, with $n=4$ marked points $p_1,\dots, p_4$, and made of 3 components, glued by 3 nodal points. Each nodal point is a pair of marked points $(q_i,q_j)$. Each component is a smooth Riemann surface of some genus $g_i$, and with $n_i$ marked or nodal points. Stability means that for each component $\chi_i=2-2g_i-n_i<0$. Here, one component has genus 2 and 1 nodal point $q_4$ so $\chi=-3$, another component is a sphere with 2 marked points $p_1,p_2$ and 3 nodal points $q_1,q_2,q_3$ i.e. $\chi=-3$, and the last component is a sphere with 2 marked points $p_3,p_4$ and 2 nodal points $q_5,q_6$ so $\chi=-2$. The total Euler characteristics is $\chi=-3-3-2=-8$ which indeed corresponds to $2-2g-n$ for a Riemann surface of genus $g=3$ with $n=4$ marked points.\label{fignodalmap1}
} 
\end{figure}

Let ${\cal M}_{g,n}$ be the moduli space of complex curves of genus $g$ with $n$ marked points.
It is a complex orbifold (manifold quotiented by a group of symmetries), of dimension
\beq
{\rm dim}\,{\cal M}_{g,n}=d_{g,n}=3g-3+n.
\eeq
Each element $(\Sigma,p_1,\dots,p_n)\in {\cal M}_{g,n}$ is a smooth complex curve $\Sigma$ of genus $g$ with $n$ smooth distinct marked points $p_1,\dots,p_n$.
$\modsp_{g,n}$ is not compact because the limit of a family of smooth curves may be non--smooth, some cycles may shrink, or some marked points may collapse in the limit.
The Deligne--Mumford compactification $\overline\modsp_{g,n}$ of $\modsp_{g,n}$ also contains stable nodal curves of genus $g$ with $n$ distinct smooth marked points (a nodal curve is a set of smooth curves glued at nodal points, and thus nodal points are equivalent to pairs of marked points, and stability means that each punctured component curve has an Euler characteristics $<0$), see fig \ref{fignodalmap1}.
$\overline\modsp_{g,n}$ is then a compact space.

Let ${\cal L}_i$ be the cotangent bundle at the marked point $p_i$, i.e. the bundle over $\overline{\cal M}_{g,n}$ whose fiber is the cotangent space $T^*(p_i)$ of $\Sigma$ at $p_i$.
It is customary to denote its first Chern class:
\beq
\psi_i=\psi(p_i)=c_1({\cal L}_i).
\eeq
$\psi_i$ is (the cohomology equivalence class modulo exact forms, of) a 2-form on $\ovl{\cal M}_{g,n}$.
Since $\dim_{\mathbb R} \ovl\modsp_{g,n}= 2 \dim_{\mathbb C} \ovl\modsp_{g,n}= 6g-6+2n$, it makes sense to compute the integral of the exterior product of $3g-3+n\,$ 2-forms, i.e. to compute the "intersection number"
\bd
\bea
\left<\tau_{d_1}\dots \tau_{d_n}\right>_{g,n}  &:=& \left<\psi_1^{d_1}\dots \psi_n^{d_n}\right>_{g,n} \cr
&:=& 
\left\{\begin{array}{ll}
\displaystyle \int_{\overline{\cal M}_{g,n}}\psi_1^{d_1}\dots \psi_n^{d_n}  \quad & {\rm if}\, {\displaystyle \sum_i} d_i=d_{g,n}=3g-3+n \cr
& \cr
0 & {\rm otherwise}
\end{array}\right. .\cr
\eea

\ed

\medskip

More interesting characteristic classes and intersection numbers are defined as follows.
Let (we follow the notations of \cite{Liu2009}, and refer the reader to it for details)
$$
\pi:\overline{\cal M}_{g,n+1}\to \overline{\cal M}_{g,n}
$$
be the forgetful morphism (which forgets the last marked point), and let $\sigma_1,\dots,\sigma_n$ 
be the canonical sections of $\pi$, and $D_1,\dots,D_n$ be the corresponding divisors in $\overline{\cal M}_{g,n+1}$. Let $\om_\pi$ be the relative dualizing sheaf.
We consider the following tautological classes on $\overline{\cal M}_{g,n}$:

$\bullet$ The $\psi_i$ classes (which are 2-forms), already introduced above:
$$ \psi_i = c_1(\sigma_i^*(\om_\pi)) $$
It is customary to use Witten's notation:
\beq\label{deftaud}
\psi_i^{d_i}=\tau_{d_i}.
\eeq

$\bullet$ The Mumford $\kappa_k$ classes \cite{Mumford1983, Arbarello1996}:
$$ \kappa_k = \pi_*(c_1(\om_\pi(\sum_i D_i))^{k+1} ) .$$
$\kappa_k$ is a $2k$--form. 
$\kappa_0$ is the Euler class, and in $\ovl{\cal M}_{g,n}$, we have
$$
\kappa_0=-\chi_{g,n}=2g-2+n.
$$
$\kappa_1$ is known as the Weil-Petersson form since it is given by $2\pi^2\kappa_1=\sum_i dl_i\wedge d\theta_i$ in the Fenchel-Nielsen coordinates $(l_i,\theta_i)$ in Teichm\"uller space \cite{Wolpert1983}.

In some sense, $\kappa$ classes are the remnants of the $\psi$ classes of (clusters of) forgotten points.
There is the formula \cite{Arbarello1996}:
\beq
\pi_* \psi_1^{d_1}\dots\psi_n^{d_n}\,\psi_{n+1}^{k+1}\, = \psi_1^{d_1}\dots\psi_n^{d_n}\,\kappa_k
\eeq
\beq
\pi_*\pi_* \psi_1^{d_1}\dots\psi_n^{d_n}\,\psi_{n+1}^{k+1}\,\psi_{n+2}^{k'+1}\, = \psi_1^{d_1}\dots\psi_n^{d_n}\,(\kappa_k\,\kappa_{k'}+\kappa_{k+k'})
\eeq
and so on...

\medskip
$\bullet$ The Hodge class  $\Lambda(\alpha)=1+\sum_{k=1}^g \,(-1)^k\,\alpha^{-k}\,c_k(\mathbb E)$ where $c_k(\mathbb E)$ is the $k^{\rm th}$ Chern class of the Hodge bundle $\mathbb E=\pi_*(\om_\pi)$.
Mumford's formula \cite{Mumford1983, FaberC.1998} says that
\beq\label{eqmumfordhodge}
\Lambda_{\rm Hodge}(\alpha)= \ee{-\sum_{k\geq 1} {\frac{B_{2k}\,\alpha^{1-2k}}{2k(2k-1)}\,\,\left(\kappa_{2k-1}-\sum_i \psi_i^{2k-1}+\frac{1}{2}\sum_\delta \sum_j (-1)^j\,\,l_{\delta*} \psi_j\,\psi'^{2k-2-j}\right)}}
\eeq
where $\Ber_{k}$ is the $k^{\rm th}$ Bernoulli number, $\delta$ a boundary divisor (i.e. a cycle which can be pinched so that the pinched curve is a stable nodal curve, i.e. replacing the pinched cycle by a pair of marked points, all components have a strictly negative Euler characteristics), and $l_{\delta*}$ is the natural inclusion into the moduli spaces of each connected component. In other words $\sum_\delta l_{\delta*}$ adds a nodal point in all possible stable ways, i.e. it adds two marked points, and $\psi$ and $\psi'$ are their $\psi$ classes. 

\bigskip
In fact, all tautological classes in $\overline{\cal M}_{g,n}$ can be expressed in terms of $\psi$-classes or their pull back or push forward from some $\overline{\cal M}_{h,m}$ \cite{Bertram2006}.
Faber's conjecture \cite{FaberC.1998} (partly proved in \cite{Mulase2006} and \cite{Liu2009}) proposes an efficient method to compute intersection numbers of $\psi, \kappa$ and Hodge classes.

\section{Bergman kernel of a spectral curve\label{applemmaBergman}}

\begin{figure}[t]
\centering
$$\includegraphics[width=14cm]{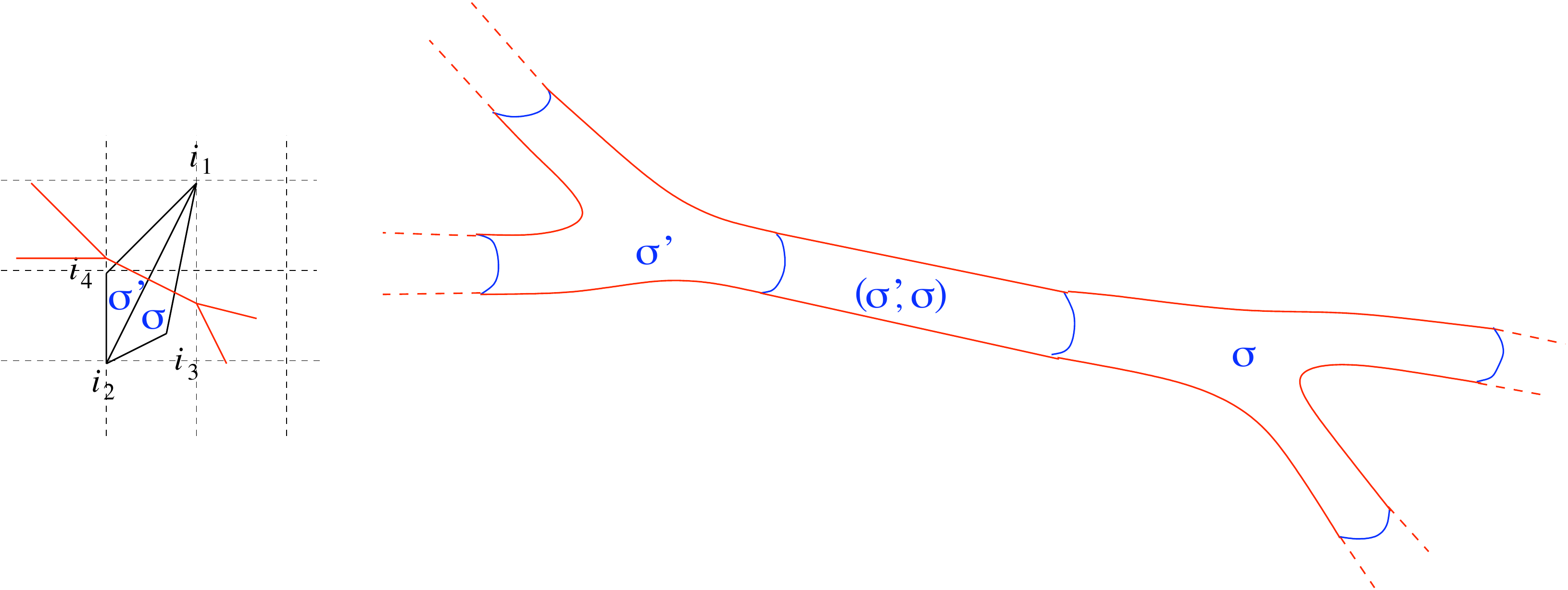}  $$
\caption{The curve $\curve$ is a union of cylinders $\curve_{\sigma,\sigma'}$ corresponding to edges of the toric graph $\Upsilon_\CYX$ and of pairs of pants $\curve_\sigma$ corresponding to vertices of $\Upsilon_\CYX$. Its Bergman kernel can be obtained as a combination of the Bergman kernels of each pieces. Notice that the framing of the edge $\sigma,\sigma'$ is $\fram_{\sigma,\sigma'}=-\fram_{\sigma',\sigma} = \fram_{b,\sigma'}=-\fram_{b,\sigma} = \beta_{i_2}-\beta_{i_1}$.
}
\end{figure}

{\bf Lemma \ref{lemmaBergman}}
{\em
Let $\sigma, \sigma'$ be two vertices. 
Let $x\in \curve_\sigma$ and $x'\in\curve_{\sigma'}$, then we have
\bea
&& \left(2 i \pi \right)^2 \,  \left[ \ln E(\spcurve;\arond_\sigma+x,\arond_{\sigma'}+x') -   \delta_{\sigma,\sigma'} \ln E(\spcurve_\sigma;x,x')\right] \cr
&=& \, \oint_{x_2\in \d_{\sigma}\curve_{\sigma'}}\,\oint_{x_1\in \d_{\sigma'}\curve_\sigma}\, \ln E(\spcurve_\sigma;x,{x_1})\,B(\spcurve_{\sigma,\sigma'};x_1,{x_2+\arond_{\sigma'}-\arond_\sigma})\, \ln E(\spcurve_{\sigma'};x_2,x') \cr
&& +  \sum_{\sigma_1}\,\oint_{x_2\in \d_{\sigma}\curve_{\sigma_1}}\,\oint_{x_1\in \d_{\sigma_1}\curve_\sigma}\, \ln E(\spcurve_\sigma;x,{x_1})\,B(\spcurve_{\sigma,\sigma_1};x_1,{x_2+\arond_{\sigma_1}-\arond_\sigma})\, \cr
&& \qquad \qquad \Big( \ln E(\spcurve;\arond_{\sigma_1}+x_2,\arond_{\sigma'}+x') - \delta_{\sigma_1,\sigma'}\,\ln E(\spcurve_{\sigma'};x_2,x')\Big) \cr
\eea

where
\beq
B(\spcurve_{\sigma,\sigma'};x_1,x_2)
 = A_{\sigma,\sigma'}\,\frac{1}{(\fram_{\sigma,\sigma'})^2}\,\,\frac{\ee{-\,\frac{x_1}{\fram_{\sigma,\sigma'}}}\,\ee{-\,\frac{x_2}{\fram_{\sigma,\sigma'}}}}{\left(\ee{-\,\frac{x_1}{\fram_{\sigma,\sigma'}}}-\ee{-\,\frac{x_2}{\fram_{\sigma,\sigma'}}}\right)^2}\,\,\,dx_1\otimes dx_2
\eeq
is the Bergman kernel on the cylinder $\curve_{\sigma,\sigma'}$, 
$A_{\sigma,\sigma'}$ is the adjacency matrix of the toric graph, i.e. $A_{\sigma,\sigma'}=1$ if $\sigma$ and $\sigma'$ are neighbors and $0$ otherwise,
and $\fram_{\sigma,\sigma'}$ is the framing of the edge $(\sigma,\sigma')$ as defined in def \ref{defframedges}.

}

\proof{

For some choice of a basepoint $o$, we define the Cauchy kernel
\beq
dS(\spcurve;x,{x'}) = \int_{x''=o}^{x'}\, B(\spcurve;x,x'')
\eeq
Notice that $\curve$ is not simply connected, neither is any pair of pants $\curve_\sigma$ nor any cylinder $\curve_{\sigma,\sigma'}$, so the integral from $o$ to $x'$ might depend on the integration contour.
However,  since we have normalized our Bergman kernels on $\acycle$ cycles (cf section \ref{secmirrormap}), and $\acycle$-cycles surround cylinders, we see that $dS(\spcurve;x,{x'})$ is globally well defined on any pair of pants $\curve_\sigma$ or on any cylinder $\curve_{\sigma,\sigma'}$ (it would be ill defined on a subdomain of $\curve$ containing a $\bcycle$-cycle because the $\bcycle$-cycle integral of $B$ does'nt vanish).

Locally near $x'=x$ it behaves like
\beq
dS(\spcurve;x,{x'}) \sim \frac{dx}{x-x'} + {\rm analytical}.
\eeq

This can thus be used to write Cauchy formula, and we write:
\bea
B(\spcurve;\arond_\sigma+x,\arond_{\sigma'}+x') 
=-\,\Res_{x_1\to x} dS(\spcurve_\sigma;x,{x_1})\, B(\spcurve;\arond_\sigma+x_1,\arond_{\sigma'}+x') 
\eea
and by moving the integration contour (a small circle around $x$) to the boundaries of $\curve_\sigma$ (we pick a pole at $x_1=x'$ in the case $x'\in \curve_\sigma$, i.e. in the case $\sigma'=\sigma$), we get:
\bea
&& B(\spcurve;\arond_\sigma+x,\arond_{\sigma'}+x') -   \delta_{\sigma,\sigma'} B(\spcurve_\sigma,x,x') \cr
&=&  -\,\sum_{\sigma_1}\,\frac{1}{2\pi i}\oint_{x_1\in \d_{\sigma_1}\curve_\sigma} dS(\spcurve_\sigma;x,{x_1})\, B(\spcurve;\arond_\sigma+x_1,\arond_{\sigma'}+x') \cr
\eea
(we choose to orient the boundaries of $\curve_\sigma$ such that the surface lies on the left of its boundaries).
$$\includegraphics[height=3cm]{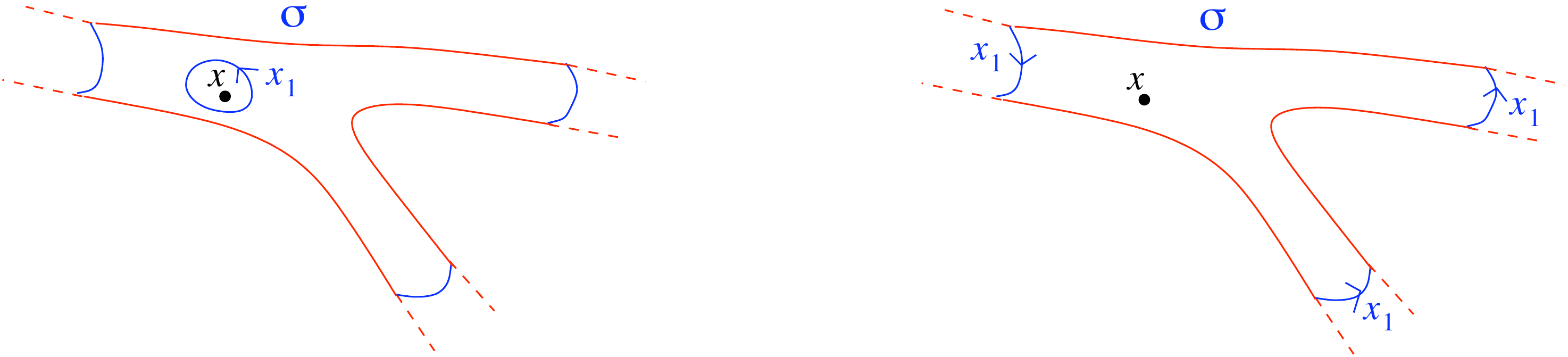}  $$
Then since $x_1\in \curve_{\sigma,\sigma_1}$, we write Cauchy formula again with the Cauchy kernel $dS(\spcurve_{\sigma,\sigma_1};x_1,{x_2})$ of $\curve_{\sigma,\sigma_1}$, i.e.
\bea
&& B(\spcurve;\arond_\sigma+x,\arond_{\sigma'}+x') -   \delta_{\sigma,\sigma'} B(\spcurve_\sigma,x,x') \cr
&=&  -\,\sum_{\sigma_1}\,\frac{1}{2\pi i}\oint_{x_1\in \d_{\sigma_1}\curve_\sigma} dS(\spcurve_\sigma;x,{x_1})\, B(\spcurve;\arond_\sigma+x_1,\arond_{\sigma'}+x') \cr
&=&  \sum_{\sigma_1}\,\frac{1}{2\pi i}\oint_{x_1\in \d_{\sigma_1}\curve_\sigma} \Res_{x_2\to x_1} dS(\spcurve_\sigma;x,{x_1})\,dS(\spcurve_{\sigma,\sigma_1};x_1,{x_2})\, B(\spcurve;\arond_\sigma+x_2,\arond_{\sigma'}+x') \cr
\eea
and again, moving the integration contour (a small circle around $x_1$) to a pair of circles around the cylinder $\curve_{\sigma,\sigma_1}$, we get:
\bea
&& B(\spcurve;\arond_\sigma+x,\arond_{\sigma'}+x') -   \delta_{\sigma,\sigma'} B(\spcurve_\sigma,x,x') \cr
&=&   - \sum_{\sigma_1}\,\frac{1}{(2\pi i)^2}\oint_{x_1\in \d_{\sigma_1}\curve_\sigma}\,\oint_{x_2\in \d_{\sigma_1}\curve_\sigma} dS(\spcurve_\sigma;x,{x_1})\,dS(\spcurve_{\sigma,\sigma_1};x_1,{x_2})\, \cr
&& \qquad B(\spcurve;\arond_\sigma+x_2,\arond_{\sigma'}+x') \cr
&&  -\sum_{\sigma_1}\,\frac{1}{(2\pi i)^2}\oint_{x_2\in \d_{\sigma}\curve_{\sigma_1}}\,\oint_{x_1\in \d_{\sigma_1}\curve_\sigma}\, dS(\spcurve_\sigma;x,{x_1})\,dS(\spcurve_{\sigma,\sigma_1};x_1,{x_2})\, \cr
&& \qquad B(\spcurve;\arond_\sigma+x_2,\arond_{\sigma'}+x') 
\eea
$$\includegraphics[height=3.5cm]{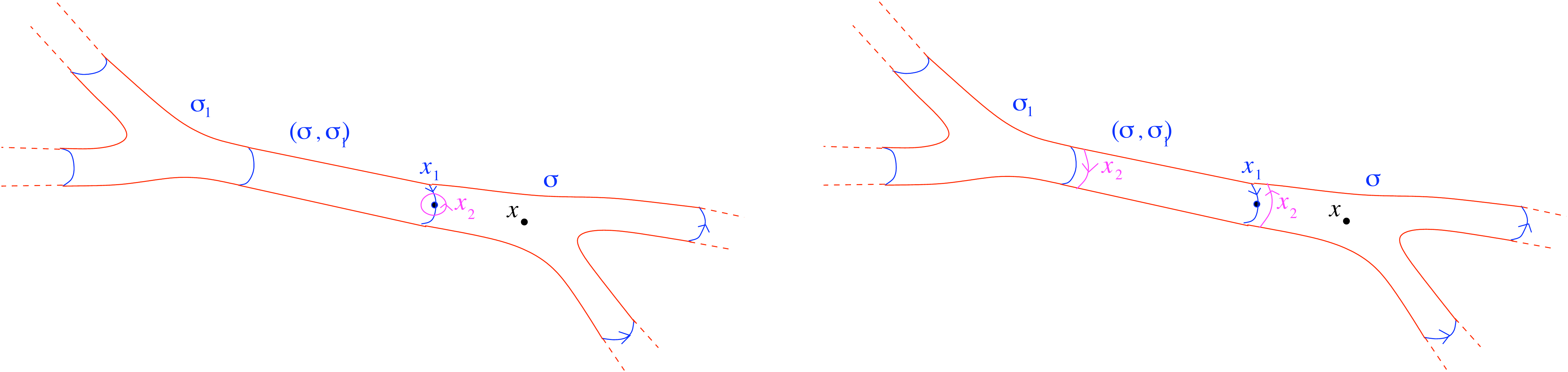}  $$

For the first line, notice that both $\curve_\sigma$ and $\curve_{\sigma,\sigma_1}$ can be realized as the complex projective plane $\mathbb C P^1$, and $\curve_{\sigma,\sigma_1}$ can be realized as a subset of $\curve_\sigma$ (a disc around one of the punctures of $\curve_\sigma=\mathbb C P^1\setminus\{0,1,\infty\}$) and we can send $x_1\to \infty$ (i.e. to the puncture) without picking any singularity (because both $B(\spcurve_\sigma;x,x_1)$ and $B(\spcurve_{\sigma,\sigma_1};x_1,x_2)$ are analytical at the punctures, they have poles only at coinciding points), i.e.
\beq
\oint_{x_1\in \d_{\sigma_1}\curve_\sigma}\,\oint_{x_2\in \d_{\sigma_1}\curve_\sigma} dS(\spcurve_\sigma;x,{x_1})\,dS(\spcurve_{\sigma,\sigma_1};x_1,{x_2})\, B(\spcurve;\arond_\sigma+x_2,\arond_{\sigma'}+x')  = 0
\eeq
$$\includegraphics[height=4cm]{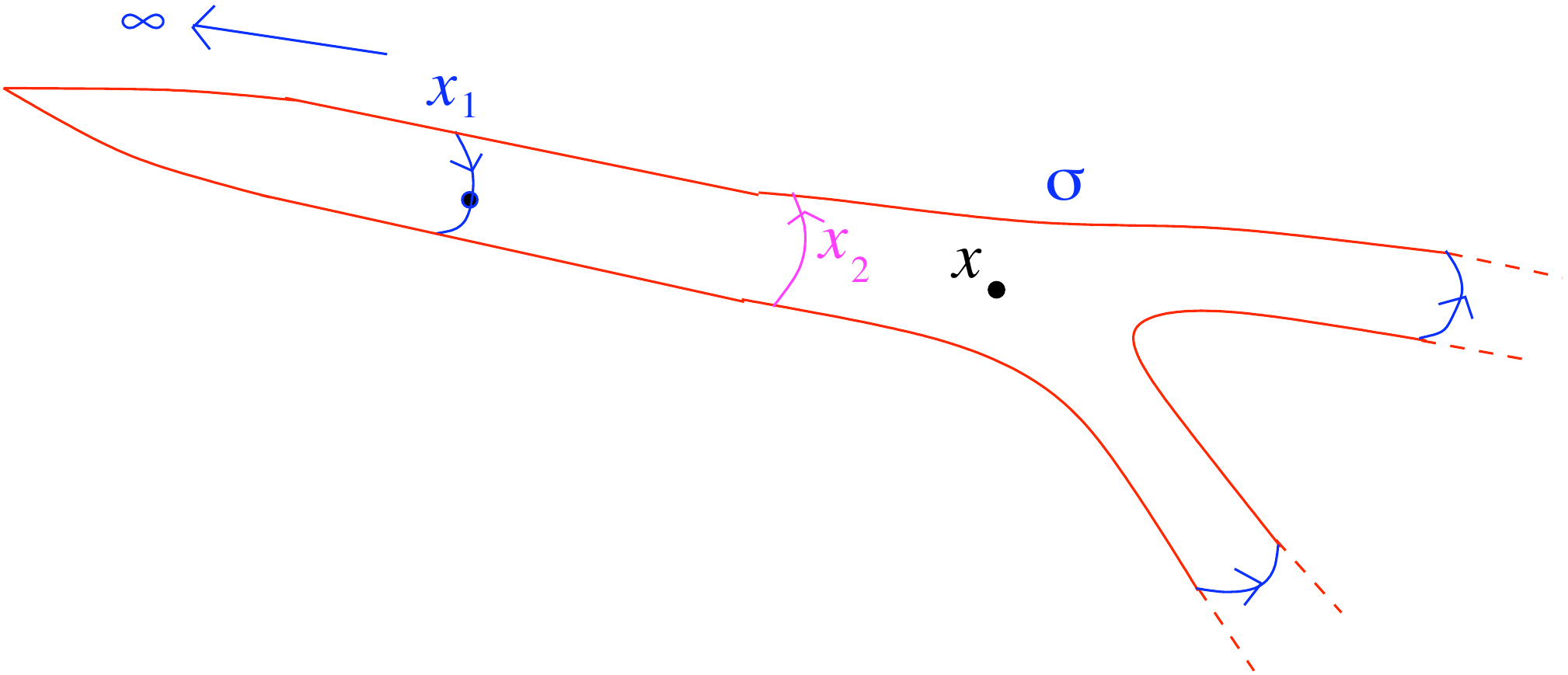}  $$

Therefore, after changing the variable $x_2$:
\bea
&& B(\spcurve;\arond_\sigma+x,\arond_{\sigma'}+x') -   \delta_{\sigma,\sigma'} B(\spcurve_\sigma,x,x') \cr
&=& -\,\sum_{\sigma_1}\,\frac{1}{(2\pi i)^2}\oint_{x_2\in \d_{\sigma}\curve_{\sigma_1}}\,\oint_{x_1\in \d_{\sigma_1}\curve_\sigma}\, dS(\spcurve_\sigma;x,{x_1})\,dS(\spcurve_{\sigma,\sigma_1};x_1,{x_2+\arond_{\sigma_1}-\arond_\sigma})\, \cr
&& \qquad \qquad \qquad \qquad B(\spcurve;\arond_{\sigma_1}+x_2,\arond_{\sigma'}+x') 
\eea
$$\includegraphics[height=5.5cm]{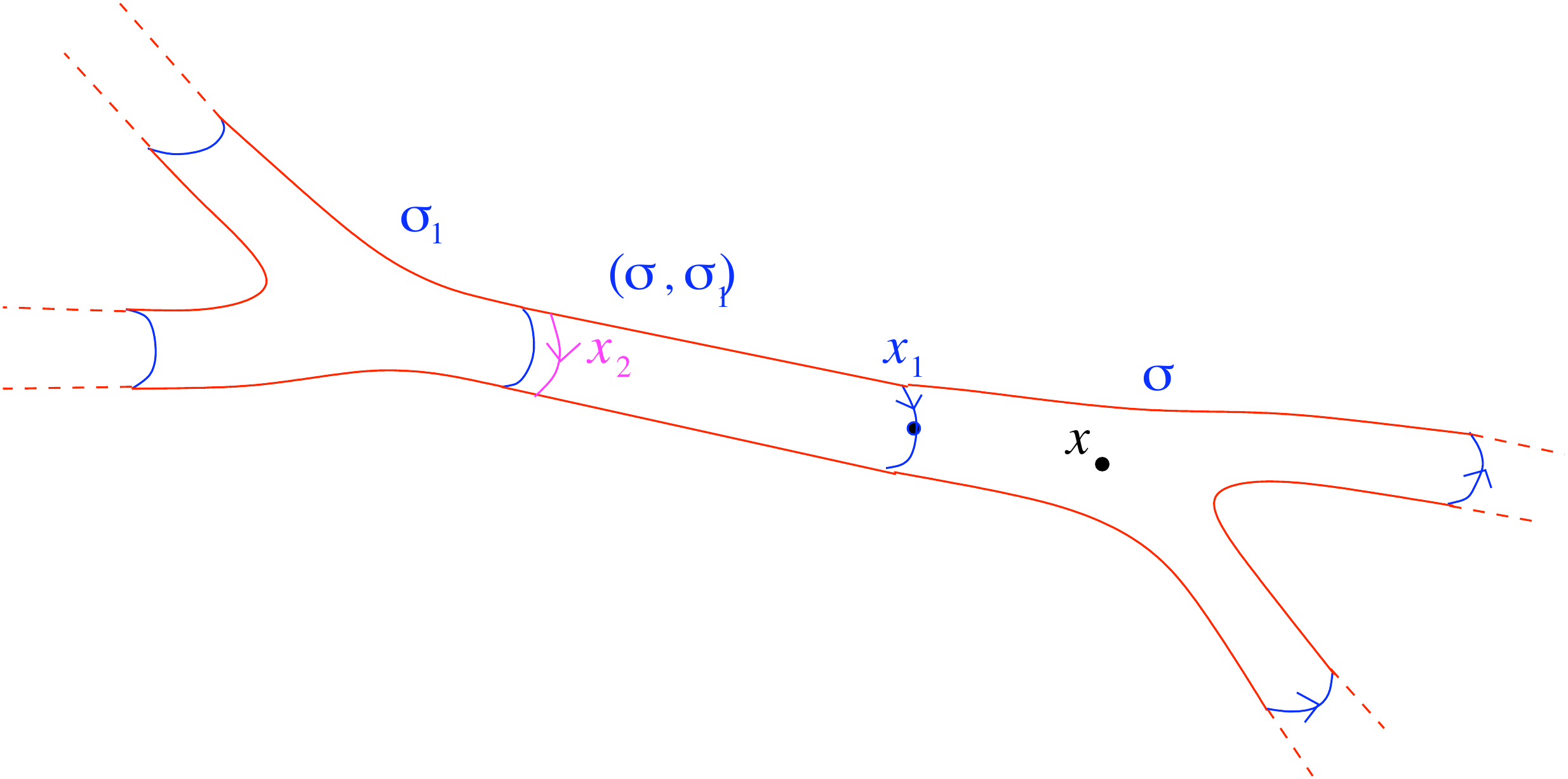}  $$
we may rewrite it as:
\beq
\begin{array}{rcl}
&& B(\spcurve;\arond_\sigma+x,\arond_{\sigma'}+x') -   \delta_{\sigma,\sigma'} B(\spcurve_\sigma,x,x') \cr
&=& - \,\frac{1}{(2\pi i)^2}\oint_{x_2\in \d_{\sigma}\curve_{\sigma'}}\,\oint_{x_1\in \d_{\sigma'}\curve_\sigma}\, dS(\spcurve_\sigma;x,{x_1})\,dS(\spcurve_{\sigma,\sigma'};x_1,{x_2+\arond_{\sigma'}-\arond_\sigma})\, B(\spcurve_{\sigma'};x_2,x') \cr
&& -  \sum_{\sigma_1}\,\frac{1}{(2\pi i)^2}\oint_{x_2\in \d_{\sigma}\curve_{\sigma_1}}\,\oint_{x_1\in \d_{\sigma_1}\curve_\sigma}\, dS(\spcurve_\sigma;x,{x_1})\,dS(\spcurve_{\sigma,\sigma_1};x_1,{x_2+\arond_{\sigma_1}-\arond_\sigma})\, \cr
&& \qquad \qquad \Big( B(\spcurve;\arond_{\sigma_1}+x_2,\arond_{\sigma'}+x') - \delta_{\sigma_1,\sigma'}\,B(\spcurve_{\sigma'};x_2,x')\Big) \cr
\end{array}
\eeq

and integrating:
\beq
\begin{array}{rcl}
&& \left(2i \pi\right)^2 \, \left[\ln E(\spcurve;\arond_\sigma+x,\arond_{\sigma'}+x') -   \delta_{\sigma,\sigma'} \ln E(\spcurve_\sigma,x,x')\right] \cr
&=&  \,
\oint_{x_2\in \d_{\sigma}\curve_{\sigma'}}\,\oint_{x_1\in \d_{\sigma'}\curve_\sigma}\, \ln E(\spcurve_\sigma;x,{x_1})\,B(\spcurve_{\sigma,\sigma'};x_1,{x_2+\arond_{\sigma'}-\arond_\sigma})\, \ln E(\spcurve_{\sigma'};x_2,x') \cr
&& +  \sum_{\sigma_1}\,
\oint_{x_2\in \d_{\sigma}\curve_{\sigma_1}}\,\oint_{x_1\in \d_{\sigma_1}\curve_\sigma}\, \ln E(\spcurve_\sigma;x,{x_1})\,B(\spcurve_{\sigma,\sigma_1};x_1,{x_2+\arond_{\sigma_1}-\arond_\sigma})\, \cr
&& \qquad \qquad \Big( \ln E(\spcurve;\arond_{\sigma_1}+x_2,\arond_{\sigma'}+x') - \delta_{\sigma_1,\sigma'}\,\ln E(\spcurve_{\sigma'};x_2,x')\Big) \cr
\end{array}
\eeq
which proves the lemma.
}

\section{Local description of the spectral curve near branchpoints\label{applaplacespcurvelocal}}

Let $\spcurve=(\curve,x,y,B)$ be a regular spectral curve,
let $\{\alpha_1,\alpha_2,\dots,\alpha_\bpt\}$ be the set of its branchpoints, i.e. the zeores of $dx$, and $a_i=x(\alpha_i)$.
We first need to set up notations.

\medskip

For each branchpoint $a_\sigma$ we define the steepest descent path $\gamma_\sigma$, as a connected arc on $\curve$ passing through $\alpha_\sigma$ such that
\beq
x(\gamma_{\sigma})-a_\sigma=\mathbb R_+ \, ,
\eeq
i.e. the vertical trajectory of $x$ going through $\alpha_\sigma$.
In a vicinity of $\alpha_\sigma$ the following quantity is a good local coordinate
\beq
\sqrt{x(z)-a_\sigma}.
\eeq

\subsection{Coefficients $\hat B_{\sigma,k;\sigma',l}$}

We expand the Bergman kernel in the vicinity of branchpoints in powers of the local coordinates $\sqrt{x(z)-a}$ as follows:
\bea\label{BTaylorexpaiaj}
B(z,z') 
&\displaystyle\mathop{{\sim}}_{\stackrel{z'\to \alpha_{\sigma'}}{z\to \alpha_\sigma}}&
\,\, \Big(\frac{\delta_{\sigma,\sigma'}}{(\sqrt{x(z)-a_\sigma} - \sqrt{x(z')-a_{\sigma'}})^2} \cr
&& \qquad  + \sum_{d,d'\geq 0} B_{\sigma,d;\sigma',d'}\,(x(z)-a_\sigma)^{d/2}\,(x(z')-a_{\sigma'})^{d'/2}\,\Big) \cr
&& \qquad\,\,\frac{ dx(z)\otimes dx(z')}{4\,\,\sqrt{x(z)-a_\sigma}\,\,\sqrt{x(z')-a_{\sigma'}}}
\eea
and then we define
\beq
\hat B_{\sigma,k;\sigma',k'} = (2k-1)!!\,(2l-1)!!\,\,2^{-k-l-1}\,\,B_{\sigma,2k;\sigma',2k'}.
\eeq
It is useful to notice that the generating function of these last quantities can also be defined through Laplace transform, we define:
\beq
\check B_{\sigma,\sigma'}(u,v)  =  \sum_{k,k'\geq 0}\, \hat B_{\sigma,k;\sigma',k'} u^{-k}\,v^{-l}, 
\eeq
which is given by the Laplace transform of the Bergman kernel
\beq
\check B_{\sigma,\sigma'}(u,v) = \delta_{\sigma,\sigma'}\,\frac{uv}{u+v} 
+ \frac{\sqrt{uv}\,\,\ee{u a_\sigma}\,\ee{v a_{\sigma'}}}{2\pi}\,\int_{z\in\gamma_{\sigma}}\,\int_{z'\in\gamma_{\sigma'}}\,B(z,z')\,\ee{-ux(z)}\,\ee{-vx(z')} 
\eeq
where the double integral is conveniently regularized when $\sigma=\sigma'$, so that $\check B_{\sigma,\sigma'}(u,v)$ is a power series of $u$ and $v$.

\subsection{Basis of differential forms $d\xi_{\sigma,d}(z)$}

We define the set of functions $\xi_{\sigma,d}(z)$ as follows:
\beq
d\xi_{\sigma,d}(z) = -\,(2d-1)!!\,2^{-d}\,\Res_{z'\to \alpha_\sigma}\, B(z,z')\,(x(z')-a_\sigma)^{-1/2-d}
\eeq
It is a meromorphic 1-form defined on $\curve$, with a pole only at $z=\alpha_\sigma$, of degree $2d+2$.

Namely, near $z\to \alpha_{\sigma'}$  it behaves like
\beq
\xi_{\sigma,d}(z) \mathop{{\sim}}_{z\to \alpha_{\sigma'}}\, \delta_{\sigma,\sigma'}\,\frac{(2d-1)!!}{2^d\,(x(z)-a_{\sigma'})^{1/2+d}}\,-\frac{(2d-1)!!}{2^d}\,\sum_k \frac{1}{k+1}\,B_{\sigma,2d;\sigma',k}\,(x(z)-a_{\sigma'})^{\frac{k+1}{2}}.
\eeq

These differential forms will play an important role because they give the behavior of the Bergman kernel $B$ near a branchpoint:
\beq\label{eqappBexpdxid}
B(z,z')-B(\bar z,z') \mathop{{\sim}}_{z\to a_\sigma}\,  -2\,\sum_{d\geq 0}\,\,\frac{2^d}{(2d-1)!!}\,\,\zeta_{a_\sigma}(z)^{2d}\,\,d\zeta_{a_\sigma}(z)\otimes d\xi_{a_\sigma,d}(z')
\eeq
where $\zeta_{a_\sigma}(z) = \sqrt{x(z)-a_\sigma}$.

\subsection{$f_{\sigma,\sigma'}(u)$}

Knowing $\xi_{\sigma',0}(z)$, it is useful to define its Laplace transform  along $\gamma_{\sigma}$ as
\bea
f_{\sigma,\sigma'}(u) 
&=& \, \frac{\sqrt u}{2\sqrt\pi}\,\ee{u a_\sigma}\,\, \int_{\gamma_{\sigma}}\,\ee{-u\,x}\,\,\xi_{\sigma',0}\,dx  \cr
&=& \, \frac{1}{2\sqrt{\pi\,u}}\,\ee{u a_\sigma}\,\, \int_{\gamma_{\sigma}}\,\ee{-u\,x}\,\,d\xi_{\sigma',0}  \cr
&=& \delta_{\sigma,\sigma'}- \sum_{k\geq 0} \frac{\hat B_{\sigma',0;\sigma,k}}{u^{k+1}} .
\eea

In \cite{eynclasses2}, it was proved that

\bl\label{lemmaBuv}
If $\curve$ is a compact Riemann surface and $dx$ is a meromorphic form on $\curve$ and $B$ is the fundamental form of the second kind normalized on $\acycle$-cycles, we have
\beq
\check B_{\sigma,\sigma'}(u,v) = \frac{uv}{u+v}\,\left( \delta_{\sigma,\sigma'}- \sum_{\sigma''=1}^\bpt f_{\sigma,\sigma''}(u)\,f_{\sigma',\sigma''}(v)\right)
\eeq
\el
so that all we need to compute is in fact $f_{\sigma,\sigma'}(u)$.

\subsection{The times $\hat t_{\sigma,k}$}

Finally we define the times $\hat t_{\sigma,k}$ at branchpoint $a_\sigma$ in terms of the local behavior of $y(z)$ by the Laplace transform of $ydx$ along $\gamma_{\sigma}$
\beq
\ee{-\hat t_{\sigma,0}}\,\ee{-g_{\sigma}(u)} 
= \frac{2\,u^{3/2}\,\ee{u a_\sigma}}{\sqrt{\pi}}\,\int_{z\in\gamma_{\sigma}}\,\, \ee{-ux(z)}\,y(z)\,dx(z)
= \frac{2\sqrt{u}\,\ee{u a_\sigma}}{\sqrt{\pi}}\,\int_{z\in\gamma_{\sigma}}\,\, \ee{-ux(z)}\,dy(z)
\eeq
The times $\hat t_{\sigma,k}$ are the coefficients of the expansion of $g(u)$ at large $u$:
\beq
g_{\sigma}(u) = \sum_{k\geq 1} \hat t_{\sigma,k} u^{-k}.
\eeq
Notice that the time $\hat t_{\sigma,0}$ is given by
\beq\label{eqt0yprime}
\ee{-\hat t_{\sigma,0}}  = \lim_{z\to \alpha_\sigma}\,\frac{y(z)-y(\bar z)}{\sqrt{x(z)-a_\sigma}} \stackrel{{\rm def}}{=} 2\,y'(a_\sigma).
\eeq

Here we add another lemma:
\bl\label{lemmagf}
If $\curve$ is a compact Riemann algebraic surface of equation 
\beq
H(\ee{-x},\ee{-y})=0
\eeq
where $H$ is a polynomial, 
and we assume that all poles of $dy$ are also poles of $dx$, and the Bergman kernel $B$ is the fundamental 2-form of the second kind normalized on $\acycle$-cycles (i.e. $\oint_{\acycle_i} B(.,z)=0$),
then we have
\beq
\ee{-g_\sigma(u)} = \sum_{\sigma'}\,\ee{\hat t_{\sigma,0}-\hat t_{\sigma',0}}\,f_{\sigma,\sigma'}(u).
\eeq

\el

\proof{
Since $\ee{-x}$ and $\ee{-y}$ are meromorphic functions, their logarithmic derivatives are meromorphic forms, i.e. $dx$ and $dy$ are meromorphic forms, and thus 
$$
\frac{dy}{dx}
$$
is a meromorphic function.

It has poles at all the zeroes of $dx$, namely
\beq
\frac{dy(z)}{dx(z)} \mathop{{\sim}}_{z\to \alpha_\sigma}\,\, \frac{y'(\alpha_\sigma)}{2\,\sqrt{x(z)-x(\alpha_\sigma)}} \,\,\mathop{{\sim}}_{z\to \alpha_\sigma}\,\, \frac{y'(\alpha_\sigma)}{2}\,\xi_{\sigma,0}(z).
\eeq
Since $dx$ and $dy$ are logarithmic derivatives of meromorphic functions, they can only have simple poles, and we assumed that all poles of $dy$ must also be poles of $dx$, therefore $dy/dx$ has no pole at the poles of $dy$.
The only poles of $dy/dx$ are thus the $\alpha_\sigma$, and therefore $dy/dx - \sum_\sigma y'(\alpha_\sigma) \xi_{\sigma,0}/2$ has no pole pole. 
Taking the differential once again says that
\beq
d\,\frac{dy}{dx} - \frac{1}{2}\sum_\sigma y'(\alpha_\sigma)\,d\xi_{\sigma,0}
\eeq
is a holomorphic differential without poles, therefore it can be written
\beq
d\,\frac{dy}{dx} = \frac{1}{2}\sum_\sigma y'(\alpha_\sigma)\,d\xi_{\sigma,0} + \sum_{i=1}^\genus c_i\,du_i 
\eeq
where $c_i$ are some coefficients determined by
\beq
c_i = -\frac{1}{2}\sum_\sigma y'(\alpha_\sigma)\,\oint_{\acycle_i}\,d\xi_{\sigma,0} =0
\eeq
(they vanish because $d\xi_{\sigma,0}$ is normalized on $\acycle_i$ like $B$).
This implies that
\beq
\frac{dy}{dx} = \frac{1}{2}\sum_\sigma y'(\alpha_\sigma)\,\xi_{\sigma,0} + C
\eeq
where $C$ is some integration constant.

The Laplace transforms are:
\beq
\ee{-\hat t_{\sigma,0}} \ee{-g_\sigma(u)} = \frac{2\sqrt{u}\,\ee{ua_\sigma}}{\sqrt\pi}\,\int_{\gamma_\sigma} \ee{-ux}\,\frac{dy}{dx}\,dx
\eeq
and
\beq
f_{\sigma,\sigma'}(u) = \frac{\sqrt{u}\,\ee{ua_\sigma}}{2\sqrt\pi}\,\int_{\gamma_\sigma} \ee{-ux}\,\xi_{\sigma',0}\,dx
\eeq
We thus obtain (notice that the constant term does'nt contribute because it is the integral of a total derivative):
\beq
\ee{-\hat t_{\sigma,0}} \ee{-g_\sigma(u)} = 2\sum_{\sigma'}\,y'(\alpha_{\sigma'})\,f_{\sigma,\sigma'}(u)
\eeq
Notice that at large $u$ we have $\ee{-g_\sigma(u)}\to 1$ and $f_{\sigma,\sigma'}(u)\to \delta_{\sigma,\sigma'}$, therefore we recover the relatonship \eq{eqt0yprime}
\beq
\ee{-\hat t_{\sigma,0}} = 2 y'(\alpha_{\sigma})
\eeq
and finally as announced
\beq
\ee{-g_\sigma(u)} = \sum_{\sigma'}\,\ee{\hat t_{\sigma,0}-\hat t_{\sigma',0}}\,f_{\sigma,\sigma'}(u).
\eeq

}






%

\subsection{More Lemmas}

All the following formulae are easy to prove and we write them for bookkeeping purpose:
\beq
\frac{d \xi_{a,d}(z)}{dx(z)} = -\xi_{a,d+1}(z) - \sum_{a'}\,\hat B_{a,d;a',0}\,\xi_{a',0}(z)
\eeq
\beq
\left.\frac{\d \xi_{a,d}(z)}{\d Q}\right|_{x(z)} = \frac{\d x(a)}{\d Q}\,\, \xi_{a,d+1}(z) + \sum_{a'}\,\hat B_{a,d;a',0}\,\frac{\d x(a')}{\d Q}\,\,\xi_{a',0}(z)
\eeq

i.e.
\beq
\left.\frac{\d \xi_{a,d}(z)}{\d Q}\right|_{x(z)}+\frac{\d x(a)}{\d Q}\,\, \frac{d\xi_{a,d}(z)}{dx(z)} =  \sum_{a'\neq a}\,\hat B_{a,d;a',0}\,\left(\frac{\d x(a')}{\d Q}-\frac{\d x(a)}{\d Q}\right)\,\,\xi_{a',0}(z)
\eeq

\beq
\frac{\d f_{a'',a}(u)}{\d Q} 
= u\,\left(\frac{\d x(a'')}{\d Q}-\frac{\d x(a)}{\d Q}\right)\,f_{a'',a}(u)
+  \sum_{a'\neq a}\,\hat B_{a,0;a',0}\,\left(\frac{\d x(a')}{\d Q}-\frac{\d x(a)}{\d Q}\right)\,\,f_{a'',a'}(u)
\eeq

\beq
\frac{\d \hat B_{a,0;a'',k}}{\d Q} 
= \left(\frac{\d x(a'')}{\d Q}-\frac{\d x(a)}{\d Q}\right)\,\hat B_{a,0;a'',k+1}
+  \sum_{a'\neq a}\,\hat B_{a,0;a',0}\,\left(\frac{\d x(a')}{\d Q}-\frac{\d x(a)}{\d Q}\right)\,\,\hat B_{a',0;a'',k}
\eeq

%
%
%

\section{Invariants of the topological vertex\label{appvertex}}

\bt\label{thWngvertexapp}["Mari\~no--Vafa formula"]
 For $(g,n) \in \mathbb{N}^2 \backslash \{(0,0),(1,0)\}$, we have:
\bea
&& W_{g,n}(\spcurverond_\fram;x_1,\dots,x_n) \cr
&=& \frac{2^{3g-3+n}}{\ee{\hat t_{\fram,0}(2-2g-n)}}\,\sum_{\{d_i\}}\Big\langle \CL_{\rm Hodge}(\fram_a)\,\CL_{\rm Hodge}(\fram_b)\,\CL_{\rm Hodge}(-\fram_a-\fram_b) \prod_{i=1}^n  \tau_{d_i} \Big\rangle_{g,n}\,\,\prod_{i=1}^n d\td\xirond_{\fram,d_i}(x_i) \cr
\eea
where, if $x$ lies near the puncture of $\mathbb P^1\setminus\{0,1,\infty\}$ (i.e. $z=0$, $1$ or $\infty$) corresponding to the half-edge $\epsilon$, (whose framing is $\fram_{\epsilon}=\fram_b$, $\fram_a$ or $-\fram_a-\fram_b$ respectively):
\beq
\td\xirond_{\fram_\epsilon,d}(x) = (-1)^d\,\left(\frac{{\mathrm d}}{{\mathrm d}x}\right)^d\,\, \xirond_{\fram_\epsilon,0}(x)
= \sum_k \frac{k^{d+1}}{\fram_{\epsilon_i}^{d+2}}\,\,\gamma_\fram(k/\fram_{\epsilon})\,\,\,\ee{-\,\frac{k}{\fram_{\epsilon}}\,x}\,dx.
\eeq

Thus, if $x_i$ lies near the puncture corresponding to the half-edge $\epsilon_i$, (whose framing is $\fram_{\epsilon}$):
\bea
&& W_{g,n}(\spcurverond_\fram;x_1,\dots,x_n) \cr
&=& \frac{2^{3g-3+n}}{\ee{\hat t_{\fram,0}(2-2g-n)}}\,\sum_{\{k_i\}}\Big\langle \CL_{\rm Hodge}(\fram_a)\,\CL_{\rm Hodge}(\fram_b)\,\CL_{\rm Hodge}(-\fram_a-\fram_b) \prod_{i=1}^n  \frac{1}{1-\frac{k_i}{\fram_{\epsilon_i}}\,\psi_i} \Big\rangle_{g,n} \cr
&& \qquad \,\,\prod_{i=1}^n \frac{k_i}{\fram_{\epsilon_i}^2}\,\,\gamma_\fram(k_i/\fram_{\epsilon_i})\,\,\,\ee{-\,\frac{k_i}{\fram_{\epsilon_i}}\,x_i}\,dx_i \cr
\eea
where the sum carries over positive integers $(k_1,\dots,k_n)\in \mathbb Z_+^n$.

\et

\proof{This is a mere application of theorem \ref{WgnCL1bp}, and is fully proved in \cite{eynclasses1}, or alternatively, it can be seen as a consequence of the proof of BKMP for the framed vertex \cite{ChenLin2009, ZhouJian2009}.

According to theorem \ref{WgnCL1bp}
the invariants of $\spcurverond_{\fram}$ are
\bea
W^{g,n}(\spcurverond_\fram;x_1,\dots,x_n) 
&=& \,\frac{2^{3g-3+n}}{\ee{-\hat t_{\fram,0}(2g-2+n)}}\, \sum_{d_1,\dots,d_n} \left< \CL_{\spcurverond(\fram)}\,\prod_{i=1}^n \psi_i^{d_i}\right>_{\modsp_{g,n}} \,\,\prod_{i=1}^n d\xirond_{\fram,d}(x_i) \cr
\eea
where the classes $\CL_{\spcurverond(f)}$ and $d\xirond_{\fram,d}(x)$ are computed from the recipe given in  theorem \ref{WgnCL1bp}:

$\bullet$ the times $\hat t_{\fram,k}$: their generating function $\grond_\fram(u) = \sum_k \hat t_{\fram,k} u^{-k}$  is obtained by computing the Laplace transform of $ydx$:
\bea
\ee{-\hat t_{\fram,0}}\,\ee{-\grond_\fram(u)} 
&=& \frac{2u^{3/2}\,}{\sqrt\pi}\int_{\gamma} \ee{-u \xrond}\,\yrond d\xrond \cr
&=& \frac{2u^{1/2}\,}{\sqrt\pi}\int_{\gamma} \ee{-u \xrond}\,d\yrond \cr
&=& \frac{2u^{1/2}\,}{\sqrt\pi}\frac{(\fram_a+\fram_b)^{(\fram_a+\fram_b)u}}{\fram_a^{\fram_a u}\,\fram_b^{\fram_b u}}\int_0^1  \,\, z^{\fram_b u} (1-z)^{\fram_a u} \left(\frac{\fram_c}{1-z}-\frac{\fram_d}{z} \right)\,\, dz \cr
&=& \frac{2u^{1/2}\,}{\sqrt\pi}\frac{(\fram_a+\fram_b)^{(\fram_a+\fram_b)u}}{\fram_a^{\fram_a u}\,\fram_b^{\fram_b u}}\,
\left(
\fram_c\,\frac{\Gamma(\fram_b u+1)\Gamma(\fram_a u)}{\Gamma((\fram_a+\fram_b)u+1)}
-\fram_d\,\frac{\Gamma(\fram_b u)\Gamma(\fram_a u+1)}{\Gamma((\fram_a+\fram_b)u+1)}
\right)\cr
&=&  \frac{\fram_c\fram_b-\fram_d\fram_a}{\fram_a+\fram_b}\,\, \frac{2 \sqrt{u}\,(\fram_a+\fram_b)^{(\fram_a+\fram_b)u}}{\sqrt\pi\,\fram_a^{\fram_a u}\,\fram_b^{\fram_b u}}\,\,\,\frac{\Gamma(\fram_a u)\,\Gamma(\fram_b u)}{\Gamma((\fram_a+\fram_b)u)} \cr
&=& -\,\,\frac{2\sqrt{2}}{\sqrt{\fram_a\fram_b(\fram_a+\fram_b)}}\,\,\,\frac{\hat\Gamma(\fram_a u)\,\hat\Gamma(\fram_b u)}{\hat\Gamma((\fram_a+\fram_b)u)} \cr
&=& -\,\,\frac{2\sqrt{2}}{\sqrt{\fram_a\fram_b(\fram_a+\fram_b)}}\,\,\,\frac{1}{\sqrt{\pi\,u}\,\,\gamma_\fram(u)}
\eea
where $\gamma_\fram(u)$ was introduced in \eq{eqdefgamma} for the localization formula in theorem \ref{thlocalizationgraphs}:
\beq
\gamma_\fram(u) = \frac{1}{\sqrt{\pi\,u}}\,\, \frac{\hat\Gamma(u(\fram_a+\fram_b))}{\hat\Gamma(u\,\fram_a)\,\,\hat\Gamma(u\,\fram_b)}.
\eeq
That gives
\beq
\ee{-\hat t_{\fram,0}} = \frac{-\,\,2\sqrt{2}}{\sqrt{\fram_a\fram_b(\fram_a+\fram_b)}}
\eeq
and, using the Stirling expansion of the Log of the Gamma--function:
\beq
\grond_\fram(u) = \sum_{k\geq 1} u^{-k}\hat t_{\fram,k}  = \sum_{k\geq 1} u^{1-2k}\,\frac{\Ber_{2k}}{2k(2k-1)}\,((\fram_a+\fram_b)^{1-2k}-\fram_a^{1-2k}-\fram_b^{1-2k})
\eeq
where $\Ber_k$ is the $k^{\rm th}$ Bernoulli number.

$\bullet$ the functions $\xirond_{\fram,d}$ are found from \eq{defdxironddef1}. For $d=0$, \eq{defdxironddef1} implies that $\xirond_{\fram,0}(x)$ is a meromorphic function on $\curve_\sigma$, i.e. a rational function of the variable $z\in \mathbb P^1$, and with a simple pole at the branchpoint, and which behaves like:
\beq
\xirond_{\fram,0}(x) \mathop{{\sim}}_{x\to 0} \,\, \frac{1}{\sqrt{x}} + {\rm analytical}.
\eeq
We thus  easily find the unique rational fraction of $z$ having that property:
\beq
\xirond_{\fram,0}(x) 
= \sqrt{\frac{2 \fram_a \fram_b}{\fram_a+\fram_b}}\,\,\frac{1}{({\fram_a+\fram_b})z-\fram_b}
\qquad \quad {\rm where}\,\,x=\xrond_\fram(z).
\eeq
Its Laplace transform is
\bea\label{fdxirond}
f(u) 
&=& \frac{\sqrt{u}}{2\sqrt\pi}\,\int_{z=0}^1 \ee{-u\xrond(z)}\,\xirond_{\fram,0}(\xrond(z))\,d\xrond(z) \cr
&=& \sqrt{\frac{2 \fram_a \fram_b}{\fram_a+\fram_b}}\,\,\frac{\sqrt{u}}{2\,\sqrt\pi}\,\frac{(\fram_a+\fram_b)^{(\fram_a+\fram_b)u}}{\fram_a^{\fram_a u}\,\fram_b^{\fram_b u}}\,\int_{z=0}^1 z^{\fram_b u}(1-z)^{\fram_a u}\,\,\frac{dz}{z(1-z)} \cr
&=& \sqrt{\frac{ \fram_a \fram_b}{\fram_a+\fram_b}}\,\,\frac{\sqrt{u}}{\sqrt{2\pi}}\,\,\frac{(\fram_a+\fram_b)^{(\fram_a+\fram_b)u}}{\fram_a^{\fram_a u}\,\fram_b^{\fram_b u}}\,\,
\frac{\Gamma(\fram_a u)\Gamma(\fram_b u)}{\Gamma((\fram_a+\fram_b)u)} \cr
&=& \ee{-\grond_\fram(u)}
\eea
(we could also have obtained this result directly from lemma \ref{lemmagf} in the appendix \ref{applaplacespcurvelocal}).


\bl\label{lemmaappexpxirond0}
In general, if $\epsilon$ is one of the punctures $0,1,\infty$, we write  $\fram_\epsilon=\fram_a, \fram_b, -\fram_a-\fram_b$ respectively, and we have when $x$ approaches the puncture $\epsilon$:
\beq\label{eqappexpxirond0}
\xirond_{\fram,0}(x) = \frac{-1}{\fram_\epsilon} \,\sum_{k=0}^\infty \gamma_\fram(k/\fram_\epsilon)\,\,\,\ee{-\frac{k}{\fram_\epsilon}\,x}
\eeq

\el

\proof{
Let us prove it near the puncture $z=0$ in $\mathbb P^1\setminus \{0,1,\infty\}$ (the other cases are obtained in the same way), we have $z\sim \ee{-\xrond(z)/\fram_b} $, and thus the Taylor expansion of $\xirond_{\fram,0}$ into powers of $z$ near $z\to 0$ gives a Fourrier expansion in powers of $\ee{-k\,\xrond(z)/\fram_b}$:
\beq
\xirond_{\fram,0}(x)  \sim \sum_{k=0}^\infty c_k\,\,\ee{-k\,\frac{x}{\fram_b}},
\eeq
where the coefficients $c_k$ can be computed by a residue formula:
\bea
c_k 
&=& -\, \Res_{z\to 0}\,\, \ee{k\,\frac{x}{\fram_b}}\,\, \xirond_{\fram,0}(x)\,\, \frac{dx}{\fram_b}  \cr
&=& \sqrt{\frac{2 \fram_a \fram_b}{\fram_a+\fram_b}}\, \frac{\fram_a^{k\,\frac{\fram_a}{\fram_b}}\,\fram_b^{k} }{(\fram_a+\fram_b)^{k\,\frac{\fram_a+\fram_b}{\fram_b}}}\Res_{z\to 0}\,\, z^{-k}\,(1-z)^{-k\,\frac{\fram_a}{\fram_b}} \,\, \cr
&& \frac{1}{({\fram_a+\fram_b})z-\fram_b}\,\, \left( \frac{1}{z} + \frac{\fram_a}{\fram_b}\,\frac{1}{z-1}\right)\,dz \cr
&=& \frac{-1}{\fram_b}\,\sqrt{\frac{2 \fram_a \fram_b}{\fram_a+\fram_b}}\, \frac{\fram_a^{k\,\frac{\fram_a}{\fram_b}}\,\fram_b^{k} }{(\fram_a+\fram_b)^{k\,\frac{\fram_a+\fram_b}{\fram_b}}}\Res_{z\to 0}\,\, z^{-k-1}\,(1-z)^{-k\,\frac{\fram_a}{\fram_b}-1} \,\, dz \cr
&=& \frac{-(\fram_a+\fram_b)}{\fram_a\,\fram_b}\,\sqrt{\frac{2 \fram_a \fram_b}{\fram_a+\fram_b}}\, \frac{\fram_a^{k\,\frac{\fram_a}{\fram_b}}\,\fram_b^{k} }{(\fram_a+\fram_b)^{k\,\frac{\fram_a+\fram_b}{\fram_b}}}
\,\,\,\frac{\Gamma(k\,\frac{\fram_a+\fram_b}{\fram_b})}{k!\,\,\Gamma(k\,\frac{\fram_a}{\fram_b})}
\eea
In other words we have
\bea\label{eqexpxirondz0}
\xirond_{\fram,0}(x) 
&=& -\,\sqrt{\frac{2(\fram_a+\fram_b)}{\fram_a\,\fram_b}}\,\, \sum_{k=0}^\infty \frac{\fram_a^{k\,\frac{\fram_a}{\fram_b}}\,\,\fram_b^k}{(\fram_a+\fram_b)^{k\frac{\fram_a+\fram_b}{\fram_b}}}\,\,\frac{\Gamma(k\,\frac{\fram_a+\fram_b}{\fram_b})}{k!\,\,\Gamma(k\,\frac{\fram_a}{\fram_b})}
\,\,\,\ee{-\frac{k}{\fram_b}\,x}  \cr
& =& -\,\frac{1}{\fram_b}\,\,\sum_k \gamma_\fram(k/\fram_b)\,\,\ee{-\frac{k}{\fram_b}\,x} 
\eea

This could also have been deduced by doing the inverse Laplace transform of \eq{fdxirond} and see that
\beq\label{eqxirondinvLaplace}
\xirond_{\fram,0}(z) =
-\,\frac{1}{2\pi i}\,\,\oint \,du\,\,\frac{2\sqrt\pi}{\sqrt u}\,\,
f(u)\,\, \ee{u\,x}
\eeq
where the integration contour surrounds in the trigonometric direction all the points $-k/\fram_b$, $k\in \mathbb N$ (i.e. the poles of $\Gamma(\fram_b u)$), but not the points $-k/\fram_a$ (poles of $\Gamma(\fram_a u)$). Writing that this integral is the sum of residues at all poles of $\Gamma(\fram_b u)$ i.e. at $u=-k/\fram_b$ gives the expansion \eq{eqexpxirondz0}.

}


%


$\bullet$ The coefficients $\hat B_{k,l}$. Their generating function $\hat \Brond(u,v) = \sum_{k,l} \hat B_{k,l}\,u^{-k}\,v^{-l}$ is obtained from  lemma \ref{lemmaBuv} in the appendix, i.e.
\beq
\hat\Brond_{\fram}(u,v) = uv\,\,\frac{1-f(u)\,f(v)}{u+v}= uv\,\,\frac{1-\ee{-\grond_\fram(u)}\,\ee{-\grond_\fram(v)}}{u+v}.
\eeq
Then, using Mumford formula \cite{Mumford1983} for the Hodge class:
\beq
 \CL_{\rm Hodge}(\alpha) = \ee{-\sum_{k=1}^\infty \frac{\Ber_{2k}\,\alpha^{1-2k}}{2k(2k-1)}\,\,(\kappa_{2k-1}-\sum_{i=1}^n \psi_i^{2k-1} + \frac{1}{2}\sum_{\delta\in \d\modsp_{g,n}} \sum_{l=0}^{2k-2} (-1)^l\,l_{\delta^*}\tau_l\tau_{2k-2-l}) }
\eeq
one can show (done in \cite{eynclasses1}) that the class $\CL_{\spcurverond_\fram}$ defined in \eq{eqdefCLkappa1}, is a product of 3 Hodge classes:

\bt (proved in \cite{eynclasses1})\label{theqClassevertexf}
\beq\label{eqClassevertexf}
\CL_{\spcurverond(\fram)}\,\prod_{i=1}^n \ee{-\grond_\fram(1/\psi_i)} = \CL_{\rm Hodge}(\fram_a)\,\CL_{\rm Hodge}(\fram_b)\,\CL_{\rm Hodge}(-\fram_a-\fram_b)
\eeq
\et

\bigskip
$\bullet$ The forms $d\td\xirond_{\fram,d}(x)$.

In order to absorb the term $\ee{-\grond_\fram(1/\psi_i)}$ in theorem \ref{theqClassevertexf} above, we shall define $d\td\xirond_{\fram,d}(x)$ such that:
\beq
\sum_d \psi^d\,\, d\xirond_{\fram,d}(x) = \ee{-\grond_\fram(1/\psi)}\,\,\sum_d \psi^d\,\, d\td\xirond_{\fram,d}(x).
\eeq

For that purpose, let us start from \eq{eqappBexpdxid}, and Laplace transform:
\beq
\sum_d u^{-d} d\xirond_{\fram,d}(x) = -\,\frac{\sqrt{u}}{\sqrt\pi}\,\,\oint_\gamma \ee{-ux'}\,B(x',x)
\eeq
Doing another Laplace transform in $x$ implies
\beq
\begin{array}{rcl}
\frac{\sqrt{v}}{2\sqrt\pi}\,\,\sum_d u^{-d} \oint_\gamma \ee{-vx} d\xirond_{\fram,d}(x) 
&=& \frac{uv}{u+v}-\hat B(u,v) \cr
&=&  \frac{u\, v \,\,\ee{-\grond_\fram(u)}\,\ee{-\grond_\fram(v)}}{u+v} \cr
&=& -\,\sum_d u^{-d} (-v)^{d+1}\ee{-\grond_\fram(u)}\,\ee{-\grond_\fram(v)} \cr
&=& -\,\ee{-\grond_\fram(u)}\,\sum_d u^{-d} (-v)^{d+1}\frac{\sqrt{v}}{2\sqrt\pi}\,\,\oint_\gamma \xirond_{\fram,0}(x) \,\ee{-vx}\,dx\cr
&=& -\,\ee{-\grond_\fram(u)}\,\sum_d u^{-d} \frac{\sqrt{v}}{2\sqrt\pi}\,\,\oint_\gamma \xirond_{\fram,0}(x) \,(d/dx)^{d+1}\ee{-vx}\,dx\cr
&=& -\,\ee{-\grond_\fram(u)}\,\sum_d u^{-d} \frac{\sqrt{v}}{2\sqrt\pi}\,\,\oint_\gamma \ee{-vx} \,(-d/dx)^{d+1}\, \xirond_{\fram,0}(x)\,dx\cr
\end{array}
\eeq
In other words
\beq
\sum_d u^{-d}  \xirond_{\fram,d}(x) 
= \ee{-\grond_\fram(u)}\,\sum_d u^{-d}   \,\, (-d/dx)^{d}\,\xirond_{\fram,0}(x)
\eeq
We are thus led to define:
\beq
\td\xirond_{\fram,d}(x) = (-1)^d\, (d/dx)^d\,\xirond_{\fram,0}(x).
\eeq

\smallskip
Using lemma \ref{lemmaappexpxirond0}, we can expand $\td\xirond_{\fram,d}(x)$ as:
\beq\label{eqappexptdxirondd}
\td\xirond_{\fram,d}(x) =  \,\sum_{k=0}^\infty \gamma_\fram(k/\fram_\epsilon)\,\,\,\ee{-\frac{k}{\fram_\epsilon}\,x}\,\,\frac{-k^d}{\fram_\epsilon^{d+1}}
\eeq
and thus
\beq\label{eqappexptdxirondpsi}
\sum_d \psi^d \xirond_{\fram,d}(x) =\ee{-\grond_\fram(1/\psi)}\, \sum_d \psi^d \td\xirond_{\fram,d}(x) 
= \ee{-\grond_\fram(1/\psi)} \,\frac{-1}{\fram_\epsilon}\,\,\sum_{k=0}^\infty \gamma_\fram(k/\fram_\epsilon)\,\,\,\ee{-\frac{k}{\fram_\epsilon}\,x}\,\,\frac{1}{1-\frac{k}{\fram_\epsilon}\,\psi}\, 
\eeq

}

\bigskip

\subsection{Summary of some formulae for the topological vertex}

\beq
\ee{-g(u)} = \frac{(f+1)^{(f+1)u}\,\sqrt u}{f^{fu}\,\sqrt{2\pi}} \,\, \frac{\Gamma(u)\,\Gamma(fu)}{\Gamma((f+1)u)}
\eeq
\beq
\Brond(u,v) = uv\,\frac{1-\ee{-g(u)}\,\ee{-g(v)}}{u-v}
\eeq

Notice that $g(u)=-g(-u)$ and we can write in the large $u$ expansion:
\beq
\ee{-g(-u)} = \frac{(f+1)^{(f+1)u}\,\sqrt{2\pi}}{f^{fu}\,\sqrt{u}} \,\, \frac{\Gamma((f+1)u)}{\Gamma(u)\,\Gamma(fu)}
\eeq

\bigskip
We also have:

\beq
\td\xirond_{\fram,d}(x) = \frac{-1}{\fram_\epsilon}\,\sum_{k=0}^\infty \gamma_\fram(k/\fram_\epsilon)\,\ee{-\,\frac{k}{\fram_\epsilon}\,x}\,\,\left(k/\fram_\epsilon\right)^d
\eeq

\beq
d\yrond_\fram(x) = \frac{-1}{\fram_\epsilon}\,\frac{\ee{-\hat t_{\fram,0}}}{4}\,\sum_{k=0}^\infty \gamma_\fram(k/\fram_\epsilon)\,\ee{-\,\frac{k}{\fram_\epsilon}\,x}\,\,dx
\eeq

\beq
\Brond_\fram(x,x') = \frac{1}{\fram_\epsilon\,\fram_{\epsilon'}}\sum_{k,l}\, \frac{\gamma_\fram(k/\fram_\epsilon)\,\gamma_\fram(l/\fram_{\epsilon'})}{\frac{k}{\fram_\epsilon}+\frac{l}{\fram_{\epsilon'}}}\,\ee{-\,\frac{k}{\fram_\epsilon}\,x}\,\ee{-\,\frac{l}{\fram_{\epsilon'}}\,x'}\,\,\frac{k}{\fram_\epsilon}dx\,\frac{l}{\fram_{\epsilon'}}dx'
\eeq

\section{Proof of Lemma \ref{lemmarenorm01}\label{appprooflemmarenorm01}}

{\bf Lemma \ref{lemmarenorm01}}
{\em
If $2-2g-n< 0$
\bea
&& W_{g,n}(\spcurve_\sigma;x_1,\dots,x_n)  \cr
&=& \frac{2^{3g-3+n}}{\ee{\hat t_{\fram_\sigma,0}(2-2g-n)}}\sum_{k=0}^\infty \frac{1}{k!}\,\sum_{d_1,\dots,d_{n+k}} \,\,\prod_{i=1}^k R_{\sigma,d_{n+i}}\cr
&&   \left< \CL_{\rm Hodge}(\fram_{a,\sigma})\,\CL_{\rm Hodge}(\fram_{b,\sigma})\,\CL_{\rm Hodge}(-\fram_{a,\sigma}-\fram_{b,\sigma})\, \prod_{j=1}^{n+k} \tau_{d_j}\right>_{g,n+k} \quad
\prod_{j=1}^n\,d\td \xirond_{\fram_\sigma,d_j}(x_j)
 \cr
&=& \frac{2^{3g-3+n}}{\ee{\hat t_{\fram_\sigma,0}(2-2g-n)}}\, \sum_{d_1,\dots,d_n} \prod_{j=1}^n\,d\td\xirond_{\fram_\sigma,d_j}(x_j) \cr
&&  \left< \CL_{\rm Hodge}(\fram_{a,\sigma})\,\CL_{\rm Hodge}(\fram_{b,\sigma})\,\CL_{\rm Hodge}(-\fram_{a,\sigma}-\fram_{b,\sigma})\,\,\, \ee{l_{1*} \sum_d R_{\sigma,d}\tau_d}\, \prod_{j=1}^{n} \tau_{d_j}\right>_{g,n} \cr
\eea
where 
\beq
R_{\sigma,d} 
= \frac{-\,2\,\ee{\hat t_{\fram_\sigma,0}}}{2\pi i}\,\oint_{\d \curve_\sigma} \td\xirond_{\fram_\sigma,d}(x)\,\,(y(x+\arond_\sigma)-\brond_\sigma-\yrond_{\fram_\sigma}(x)) \,dx,
\eeq
where $\d\curve_\sigma$ is the boundary of $\curve_\sigma$, i.e. the union of three circles, oriented so that $\curve_\sigma$ lies on the left of $\d\curve_\sigma$.
In the second equality,
$l_{1*}$ denotes the natural inclusion of $\modsp_{g,n}\subset \modsp_{g,n+1}$ ,so that $\ee{l_{1*} \sum_d C_{\sigma,d}\psi^d}$ is just a short hand notation for the formula above.

\medskip 

And similarly for $(g,n)=(0,2)$:
\bea\label{defBsigmaapp}
 W_{0,2}(\spcurve_\sigma;x_1,x_2) 
 &=& B_\sigma(x_1,x_2)  = \Brond_{\fram_\sigma}(x_1-a_\sigma+\arond_\sigma,x_2-a_\sigma+\arond_\sigma)  \cr
 &=& \Brond_{\fram_\sigma}(x_1,x_2)+ \frac{1}{2}\sum_{k=1}^\infty \frac{1}{k!}\,\sum_{d_1,\dots,d_{k+2}}
 \,d\td\xirond_{\fram_\sigma,d_1}(x_1)\,d\td\xirond_{\fram_\sigma,d_2}(x_2)
\prod_{i=3}^{k+2} R_{\sigma,d_{i}}
 \cr
&&  \left< \CL_{\rm Hodge}(\fram_{a,\sigma})\,\CL_{\rm Hodge}(\fram_{b,\sigma})\,\CL_{\rm Hodge}(-\fram_{a,\sigma}-\fram_{b,\sigma})\, \prod_{j=1}^{k+2} \tau_{d_j}\right>_{0,k+2} \cr
\eea
and for $(g,n)=(0,1)$:
\bea
 W_{0,1}(\spcurve_\sigma;x_1) 
 &=& (y(x_1+\arond_\sigma)-\brond_\sigma)\,dx_1 \cr
&=& \yrond_{\fram_\sigma}(x_1)dx_1
+ \frac{1}{2\pi i}\oint_{\d \curve_\sigma} \Brond_{\fram_\sigma}(x_1,x')\,\Phi(x') \cr
&& +  \frac{\ee{-\hat t_{\fram_\sigma,0}}}{4}\sum_{k=2}^\infty \frac{1}{k!}\,\sum_{d_1,\dots,d_{k+1}} 
d\td\xirond_{\fram_\sigma,d_1}(x_1)
\prod_{i=2}^{k+1} R_{\sigma,d_{i}} \cr
&&  \left< \CL_{\rm Hodge}(\fram_{a,\sigma})\,\CL_{\rm Hodge}(\fram_{b,\sigma})\,\CL_{\rm Hodge}(-\fram_{a,\sigma}-\fram_{b,\sigma})\,\tau_{d_1} \prod_{j=2}^{k+1} \tau_{d_j}\right>_{0,k+1} \,\,
\eea


}

\proof{ 
The proof is based on the fact that almost by definition this lemma holds at the tropicla limit $t_j=+\infty$, and then, in order to  show that it holds for all $t_j$'s (in an open vicinity of $t_j=+\infty$), we prove using special geometry, that both sides obey the same differential equation.

\smallskip

First, notice that $W_{g,n}(\spcurverond(\fram_\sigma);x_1,\dots,x_{n})$ is a meromorphic form on $\curve_\sigma=\mathbb C\setminus\{0,1,\infty\}$, i.e. on $\mathbb CP^1$, which has poles only at the branchpoint $x_i=0$ without residue (when $2g-2+n>0$), and thus its primitive is a meromorphic function.
In other words
\beq
W_{g,n}(\spcurverond(\fram_\sigma);x_1,\dots,x_n)
= d_1\otimes \dots \otimes d_{n} \Phirond_{g,n}(\spcurverond(\fram_\sigma);x_1,\dots,x_n)
\eeq
and $\Phirond_{g,n}$ is an algebraic function of each $X_i=\ee{-x_i}$ for all $i=1,\dots,n$, having a square--root branchcut $[0,\infty[$.
In particular, it is analytical in a vicinity of $\d\curve_\sigma$ (see figure \ref{figCsigmaxplane}).

\medskip
Notice that, by definition of $\yrond_{\fram}$,  $y(x+\arond_\sigma)-\brond_\sigma-\yrond_{\fram_\sigma}(x)$ vanishes in the tropical limit $t_i\to +\infty$, i.e. at $Q_i=\ee{-t_i}=0$, and can be Taylor expanded in powers of $\mathbf Q=\{Q_i\}$ near $\mathbf Q=0$. 

Indeed, first observe that the coefficients $H_{i,j}$ of the algebraic equation $H_f(X,Y)=0$, have a Laurent expansion in powers of $\mathbf Q$. 
Since $\arond_\sigma$ and $\brond_\sigma$ are linear combinations of the $t_i$'s, i.e. $\ee{-\arond_\sigma}$ and $\ee{-\brond_\sigma}$ are product or ratios of the $Q_i$'s, we see that $Y(x+\arond_\sigma)\ee{\brond_\sigma}$ is an algebraic function of $\ee{-x}$, which has a Laurent expansion into powers of $Q_i$'s.
This implies that
\beq
\frac{\d}{\d Q_i}\,(y(x+\arond_\sigma)-\brond_\sigma-\yrond_{\fram_\sigma}(x))
= - \frac{1}{Y(x+\arond_\sigma)}\left(\frac{\d Y(x+\arond_\sigma)}{\d Q_i}+\frac{\d \arond_\sigma}{\d Q_i}\,\frac{\d Y(x+\arond_\sigma)}{\d x}\right) - \frac{\d \brond_\sigma}{\d Q_i}
\eeq
has a Laurent expansion in powers of $Q_i$'s, whose coefficients are algebraic functions of $\ee{-x}$. And since we know that $(y(x+\arond_\sigma)-\brond_\sigma-\yrond_{\fram_\sigma}(x))$ vanishes at $\mathbf Q=0$, we see that the Laurent expansion is in fact a Taylor expansion.
This implies that:
\beq\label{eqyyrondexpQk}
y(x+\arond_\sigma)-\brond_\sigma-\yrond_{\fram_\sigma}(x)
= \sum_{\mathbf k} {\mathbf Q}^k\, Y_{\mathbf k}(\ee{-x})
\eeq
where each $Y_{\mathbf k}(X)$ is an algebraic function of $X$ on $\curve_\sigma$.
$Y_{\mathbf k}$ maybe singular at $x=0$ (where $\yrond$ has a squareroot branchcut), or also at the punctures $0,1,\infty$ in the pair of pants $\curve_\sigma$.

Notice that due to the log, $y=-\ln Y$ was well defined only on $\curve$ cut along a tree, and similarly $\yrond_{\fram_\sigma}$ is also well defined only on $\curve_\sigma$ with some cuts, but the difference $y(x+\arond_\sigma)-\yrond_{\fram_\sigma}(x)$ has, order by order in $\mathbf Q$, no logarithmic cut.

In particular, order by order in powers of $\mathbf Q$, $(y(x+\arond_\sigma)-\brond_\sigma-\yrond_{\fram_\sigma}(x))$ is analytical in a vicinity of $\d\curve_\sigma$.
Therefore, the following integral makes sense (as a formal power series in $\mathbf Q$):
\beq\label{eqointPgnrondy}
\oint_{\d\curve_\sigma} \Phirond_{g,n}(\spcurverond(\fram_\sigma);x_1,\dots,x_{n})\,\, (y(x_n+\arond_\sigma)-\brond_\sigma-\yrond_{\fram_\sigma}(x_n))\,dx_n
\eeq
and it depends only on the homotopy class of the integration contour, i.e. it is invariant under small continuous deformations of the integration contour.
We can also integrate by parts and write it as:
\beq\label{eqointWgnrondPhi}
\oint_{\d\curve_\sigma} W_{g,n}(\spcurverond(\fram_\sigma);x_1,\dots,x_{n})\,\, \Phi(x_n)
\eeq
where
\beq
\frac{d\,\Phi(x)}{dx} = y(x+\arond_\sigma)-\brond_\sigma-\yrond_{\fram_\sigma}(x).
\eeq
Notice that a priori, $\Phi(x)$ seems to be defined only on a universal covering of $\curve_\sigma$, i.e. it is not necessarily an algebraic function of $\ee{-x}$.
However, the monodromies of $\Phi$ are the integrals
\beq
\oint (y(x+\arond_\sigma)-\brond_\sigma-\yrond_{\fram_\sigma}(x))\,dx
\eeq
which are linear combinations of the $t_i$'s (due to the mirror map \eq{mirrormap}).
This shows that
\beq
\frac{\d}{\d Q_i}\,\oint (y(x+\arond_\sigma)-\brond_\sigma-\yrond_{\fram_\sigma}(x))\,dx \propto \frac{1}{Q_i}
\eeq
and since $y(x+\arond_\sigma)-\brond_\sigma-\yrond_{\fram_\sigma}(x)$ is a power series of $\mathbf Q$ with only positive powers of the $Q_i$'s, this shows that in fact the monodromies of $\Phi$ must be independent of $\mathbf Q$, and since they vanish at $\mathbf Q=0$, they must vanish identically
(the fact that the monodromies of $\Phi$ vanish could also have been deduced directly from the tropical limit of the mirror map relationship).
This proves that $\Phi(x)$ is in fact, to each order in powers of $\mathbf Q$, an algebraic function of $\ee{-x}$.

\smallskip
Therefore, we may define define the following as a formal power series in $\mathbf Q$:
\bea
U_{g,n}(x_1,\dots,x_n) 
&=& W_{g,n}(\spcurverond(\fram_\sigma);x_1,\dots,x_n)  
 + \sum_{k=1}^\infty \frac{1}{(2\pi i)^k\,k!} \oint_{\d \curve_\sigma} \dots \oint_{\d \curve_\sigma}  \cr
 && W_{g,n+k}(\spcurverond(\fram_\sigma);x_1,\dots,x_n,x_{n+1},\dots,x_{n+k}) \,\prod_{i=1}^k \Phi(x_{n+i}) . \cr 
\eea

Taking a derivative with respect to any $Q_i$, we have
\bea
\frac{\d}{\d Q_i}\,U_{g,n}(x_1,\dots,x_n) 
&=&  \sum_{k=1}^\infty \frac{1}{(2\pi i)^k\,k!}\,\, k  \oint_{\d \curve_\sigma} \dots \oint_{\d \curve_\sigma}   \cr
&& W_{g,n+k}(\spcurverond(\fram_\sigma);x_1,\dots,x_{n+k}) \,\prod_{i=1}^{k-1} \Phi(x_{n+i})  \,\,\frac{\d}{\d Q_i}\, \Phi(x_{n+k})\cr
&=&  \sum_{k=0}^\infty \frac{1}{(2\pi i)^{k+1}\,k!}\,\, \oint_{\d \curve_\sigma} \dots \oint_{\d \curve_\sigma}  \cr
&& W_{g,n+k+1}(\spcurverond(\fram_\sigma);x_1,\dots,x_{n+k},x') \,\prod_{i=1}^{k} \Phi(x_{n+i}) \,\,\frac{\d}{\d Q_i}\, \Phi(x')\cr
&=&  \frac{1}{2\pi i}\oint_{\d \curve_\sigma}  U_{g,n+1}(x_1,\dots,x_n,x')  \,\,\frac{\d}{\d Q_i}\, \Phi(x')\cr
\eea
i.e.
\beq\label{maineqproofHngWng}
 \frac{\d}{\d Q_i}\,U_{g,n}(x_1,\dots,x_n) 
=  \frac{1}{2\pi i}\oint_{\d \curve_\sigma}  U_{g,n+1}(x_1,\dots,x_n,x')  \,\,\frac{\d}{\d Q_i}\, \Phi(x').Œ
\eeq

Then, notice that we have
\bea
\frac{\d}{\d Q_i}\,(y(x+\arond_\sigma)-\brond_\sigma)\,dx 
&=& \Res_{x'\to x}\, B_\sigma(x,x')\,\frac{\d \Phi(x')}{\d Q_i} \cr
&=& \frac{1}{2i\pi}\oint_{\d\curve_\sigma}\, B_\sigma(x,x')\,\frac{\d \Phi(x')}{\d Q_i}
\eea
indeed, the first equality comes from the fact that $B_\sigma$ has a double pole on the diagonal, and the second equality, holds order by order in powers of $\mathbf Q$, because to each order $\d\Phi/\d Q_i$ is an analytical function on $\curve_\sigma$ and thus one can move the integration contour.

\smallskip

From this, the "special geometry" property of spectral invariants (see \cite{EOFg} or appendix \ref{secappdefWgn}) implies that
 \beq\label{maineqproofHngWng2}
 \frac{\d}{\d Q_i}\,W_{g,n}(\spcurve_\sigma;x_1,\dots,x_n) 
=  \frac{1}{2\pi i}\oint_{\d \curve_\sigma}  W_{g,n+1}(\spcurve_\sigma;x_1,\dots,x_n,x')  \,\,\frac{\d\,\Phi(x')}{\d Q_i}\, 
\eeq

\medskip

We can now prove by recursion on the power of $\mathbf Q$, that:
\beq\label{Rechypothesis}
\left\{\begin{array}{l}
U_{g,n}(x_1,\dots,x_n) = W_{g,n}(\spcurve_\sigma;x_1,\dots,x_n)  \cr
\cr
U_{0,1}(x) = (y(x+\arond_\sigma)-\brond_\sigma)\,dx \cr
\cr
U_{0,2}(x,x') = B_\sigma(x,x')
\end{array}\right.
\eeq
This is clearly true when $\mathbf Q=0$, and if it is true to order ${\mathbf Q}^k$, then the right hand side of \eq{maineqproofHngWng} and \eq{maineqproofHngWng2} coincide to order $k$, and thus $\d W_{g,n}/\d Q_i$ and $\d U_{g,n}/\d Q_i$ coincide to order $k$, which implies that $W_{g,n}$ and $U_{g,n}$ coincide to order $k+1$.
We have thus proved the recursion hypothesis to order $k+1$, and thus it holds to all orders.

\smallskip

Then, when $2g-2+n+k>0$, use theorem \ref{thWngvertex}:
\bea
&& W_{g,n}(\spcurve_\sigma;x_1,\dots,x_n) \cr
&=& \sum_{d_1,\dots,d_n}\,\,\, \sum_k \frac{1}{k!} \,\, \sum_{d_{n+1},\dots,d_{n+k}}\cr
&& \frac{2^{3g-3+n+k}}{\ee{\hat t_{\fram_\sigma,0}(2-2g-n-k)}}\,\Big\langle \CL_{\rm Hodge}(\fram_{a,\sigma})\,\CL_{\rm Hodge}(\fram_{b,\sigma})\,\CL_{\rm Hodge}(-\fram_{a,\sigma}-\fram_{b,\sigma}) \prod_{i=1}^{n+k}  \tau_{d_i} \Big\rangle_{g,n+k} \cr
&& \prod_{i=1}^k \frac{1}{2\pi i}\oint_{x_{n+i}\in \d\curve_\sigma} \,\,\Phi(x_{n+i}) \,\,\,
d\,\td\xirond_{\sigma,d_{n+i}}(x_{n+i}) \qquad
 \qquad \,\,\prod_{i=1}^n d\,\td\xirond_{\sigma,d_{i}}(x_{i}) \cr
&=& \frac{2^{3g-3+n}}{\ee{\hat t_{\fram_\sigma,0}(2-2g-n)}}\,\sum_{d_1,\dots,d_n}\,\,\, \sum_k \frac{1}{k!} \,\, \sum_{d_{n+1},\dots,d_{n+k}}\cr
&& \Big\langle \CL_{\rm Hodge}(\fram_{a,\sigma})\,\CL_{\rm Hodge}(\fram_{b,\sigma})\,\CL_{\rm Hodge}(-\fram_{a,\sigma}-\fram_{b,\sigma}) \prod_{i=1}^{n+k}  \tau_{d_i} \Big\rangle_{g,n+k} \cr
&&  \,\,\,\prod_{i=1}^k R_{\sigma,d_{n+i}}
 \qquad \,\,\prod_{i=1}^n d\,\td\xirond_{\sigma,d_{i}}(x_{i}) 
\eea
where
\bea
R_{\sigma,d}
&=& \frac{2\,\ee{\hat t_{\fram_\sigma,0}}}{2\pi i}\oint_{x\in \d\curve_\sigma} \,\,\Phi(x) \,\,\,d\,\td\xirond_{\sigma,d}(x) \cr
&=& \frac{-\,2\,\ee{\hat t_{\fram_\sigma,0}}}{2\pi i}\oint_{x\in \d\curve_\sigma} \,\td\xirond_{\sigma,d}(x) \,\,\,(y(x+\arond_\sigma)-\brond_\sigma-\yrond_{\fram_\sigma}(x))\,dx
\eea

The cases $(g,n)=(0,1)$ and $(0,2)$ are obtained in the same way, except that we can't use theorem \ref{thWngvertex} for the first few values of $k$.

This proves the lemma.
}

\bibliographystyle{plain}


\end{document}